\documentclass[mathpazo]{cicp}

\usepackage{amssymb}
\usepackage{array}
\usepackage{latexsym}
\usepackage{graphicx}
\usepackage{caption}
\usepackage{titlesec}
\usepackage{mathrsfs}
\usepackage{amsmath}
\usepackage{epstopdf}
\usepackage{subfigure}
\usepackage{algorithm}
\usepackage{algorithmic}
\usepackage{setspace}
\usepackage{bm}
\usepackage{amsthm}
\usepackage{multicol}
\usepackage{multirow}
\usepackage{booktabs}
\usepackage{color,xcolor}
\usepackage{listings}
\usepackage{chemarrow}
\definecolor{codegreen}{rgb}{0,0.6,0}
\begin{document}

\title{A machine learning enhanced discontinuous Galerkin method for simulating transonic airfoil flow-fields}

\author[Y.W.Feng and L.L.Lv et.~al.]{Yiwei Feng\affil{1}\comma\corrauth,
		Lili Lv\affil{2}, Weixiong Yuan\affil{2}, Liang Xu\affil{1}, Tiegang Liu\affil{2}}
 \address{\affilnum{1}\ China Academy of Aerospace Aerodynamics, Beijing, 100074, P.R. China. \\
          \affilnum{2}\ School of Mathematical Sciences, Beihang University, Beijing 100191, P.R. China.
        }
 \emails{{\tt fengyw@buaa.edu.cn} (Y.W.~Feng)}

\begin{abstract}
Accurate and rapid prediction of flow-fields is crucial for aerodynamic design. This work proposes a discontinuous Galerkin method (DGM) whose performance enhances with increasing data, for rapid simulation of transonic flow around airfoils under various flow conditions. A lightweight and continuously updated data-driven model is built offline to predict the roughly correct flow-field, and the DGM is then utilized to refine the detailed flow structures and provide the corrected data. During the construction of the data-driven model, a zonal proper orthogonal decomposition (POD) method is designed to reduce the dimensionality of flow-field while preserving more near-wall flow features, and a weighted-distance radial basis function (RBF) is constructed to enhance the generalization capability of flow-field prediction. Numerical results demonstrate that the lightweight data-driven model can predict the flow-field around a wide range of airfoils at Mach numbers ranging from $0.7$ to $0.95$ and angles of attack from $-5^{\circ}$ to $5^{\circ}$ by learning from sparse data, and maintains high accuracy of the location and essential features of flow structures (such as shock waves). In addition, the machine learning (ML) enhanced DGM is able to significantly improve the computational efficiency and simulation robustness as compared to normal DGMs in simulating transonic inviscid/viscous airfoil flow-fields on arbitrary grids, and further enables rapid aerodynamic evaluation of numerous sample points during the surrogate-based aerodynamic optimization.
\end{abstract}

\ams{65M60, 68Q32}
\keywords{data-driven model, discontinuous Galerkin methods, CFD simulation of transonic flow-field, aerodynamic optimization}

\maketitle


\section{Introduction}
\label{sec:1}
\qquad Modern industrial aerodynamic design demands heightened design efficiency and exceptional aerodynamic performance of the final designs. In the conceptual design stage, surrogate-based optimization (SBO)~\cite{queipo2005surrogate, forrester2009recent, koziel2021recent} can achieve more comprehensive and superior aerodynamic performance in optimization results due to its ability to better explore the design space. However, when addressing high-dimensional design space, SBO is recognized extremely expensive due to the requirement of assessing quickly increasing sample or design points for sufficient coverage of the design space~\cite{shan2008survey}.

The primary expense within SBO for evaluating aerodynamic performance at specific design points primarily comes from using computational fluid dynamics (CFD) to simulate the flow-fields around continuously changing aerodynamic shapes under various flow conditions. Hence, the CFD solver needs to be robust, efficient but still accurate. Our previous works~\cite{feng2024adjoint,feng2024sbo} have demonstrated that high-order DGMs~\cite{fidkowski2005p, hartmann2010discontinuous} are capable of providing highly reliable aerodynamic quantities and sensitivities even on coarse grids. However, DGMs still needs improvements in terms of robustness and computational cost when treating intricate especially 3D shapes or large-scale computational grids.

Over the past few decades, the reduced-order models (ROMs)~\cite{benner2015survey}, particularly nonintrusive reduced-order models (NIROMs), provide attractive alternatives to alleviate the high computational cost of CFD simulation. The key idea of the NIROMs for flow-field prediction is to construct a map between the design variable and the reduced-dimensional flow-field based on CFD simulation data. Classical dimensionality reduction methods include POD, dynamic mode decomposition (DMD), and manifold learning~\cite{franz2014interpolation,li2024manifold}, etc. POD~\cite{chatterjee2000introduction} and DMD~\cite{alla2017nonlinear} utilize a low-dimensional linear space spanned by a set of orthogonal basis to approximate the high-dimensional space, and are widely used in fluid dynamics due to their simplicity and interpretability. After dimensionality reduction, appropriate interpolation or fitting model, such as RBF, Kriging interpolation, Gaussian regression, is constructed with a acceptable loss of accuracy to replace the original full-order model (CFD). NIROMs that combine POD/DMD with interpolation are easily implementable, conveniently updatable, and low computational cost. However, linear dimensionality reduction methods might lose intricate nonlinear characteristics of data. Additionally, for high-dimensional design variables encompassing both geometric and flow state parameters, classical interpolation mappings sometimes exhibit limited predictive performance beyond the interpolated dataset.

Recently, deep neural networks (DNNs) have been developed to accomplish the task of flow-field prediction. The DNN directly establish a map between high-dimensional geometry or grid and full-order flow-field through its exceptional nonlinear encoding and decoding capabilities. Typical DNNs used to predict flow-field include convolutional neural networks (CNNs)~\cite{duru2021cnnfoil}, U-Nets~\cite{chen2023towards}, generative adversarial networks (GANs)~\cite{haizhou2022generative}, attention-based nets~\cite{deng2023prediction, zuo2022fast}, PointNets~\cite{kashefi2021point}, graph convolutional networks (GCNs)~\cite{massegur2024graph}, etc. CNNs are widely used in the field of computer vision for feature extraction, and exhibit a remarkable capacity to reduce the number of training parameters as compared with fully connected neural networks. U-Nets enhance the feature extraction capability of deep learning models by adding more densely skip connections. Attention-based nets such as vision transformers (ViTs) utilize the self-attention mechanism to incorporate a large amount of global information and establish long-range relationships. PointNets and GCNs extend the flow-field prediction to arbitrary 3D geometries and unstructured grids through encoding respective geometric and grid topological information into the networks. DNN-based models can accomplish broader and more complicate flow-field prediction tasks, and demonstrate improved predictive performance compared to classical NIROMs~\cite{immordino2024steady}. However, DNNs usually possess complex structures with deep layers and numerous training DoFs to achieve generalized representation capabilities. This results in much higher computational costs during both the early training and updating stages and much more runtime memery as compared with NIROMs. Additionally, it is theoretically uncertain about the extrapolation or generalization capabilities of the models, making it difficult to directly used in engineering applications.

The CFD computation assisted by data-driven methods provides a promising perspective~\cite{sousa2024enhancing}. For flow-field prediction, data-driven methods are capable of rapidly providing predictions that closely resemble actual flow-field distribution, whereas CFD computations are more reliable but relatively costly. The integration of data-driven models with CFD holds promise in both enhancing the credibility of data-driven models and improving computational efficiency of CFD. For instance, Zuo et al.~\cite{zuo2024fast} utilized their FU-CBAM-Net to produce a better initial flow-field distribution for PHengLEI to accelerate convergence of steady subsonic viscous flow.

Leveraging the sparse data generated by the HODG platform~\cite{he2023hodg}, this work aims to develop a lightweight data-driven model based on the NIROM framework to accelerate the DGMs in simulating transonic airfoil flow-field with a wide range of shape variations under various flow conditions. The advantage is threefold: (a) the model is readily implementable, capable of being seamlessly integrated within the unified HODG framework; (b) the model is lightweight, thereby reducing the offline training costs and online runtime memory; (c) the model is easily updatable, allowing for more convenience to continually improve its predictive performance based on the expanded dataset, and facilitating its integration with SBO for online learning.

In order to enhance the fitting and predictive performance of the data-driven model while maintaining its lightweight nature, this work has made the following two efforts: (a) a zonal POD method is designed to preserve more near-wall flow features with a relatively few reduced dimensionality, thereby improving the interpolation/fitting performance of the model; (b) a weighted-distance RBF is constructed based on sensitivity analysis to balance the relationship between geometric and flow condition design variables, thereby enhancing the predictive performance of the model.

The remainder of this paper is organized as follows. Section~\ref{sec:2} describes the problem of steady-state transonic airfoil flow-field prediction. The overall framework of the machine learning (ML) enhanced DGM is detailed in Section~\ref{sec:3}, encompassing the generation of data samples, construction and updating of the data-driven models, as well as the synergy between data-driven models and DGMs. The predictive performance of the data-driven model is numerically analyzed in Section~\ref{sec:4}. The practical applications of the ML-enhanced DGM in viscous flow simulation on arbitrary shapes/grids and in in SBO for rapid aerodynamic evaluation are presented in Section~\ref{sec:5}. Concluding remarks and perspectives are made in Section~\ref{sec:6}.

\section{Problem description}
\label{sec:2}
\qquad The simulation of steady-state transonic flow around an airfoil can be mathematically described as a map from the airfoil geometry $\bm{D}$ and the flow condition $\bm{S}$ to the airfoil flow-field distribution $\bm{U}$.
\begin{equation}
 \bm{U}=\bm{U}(\bm{D},\bm{S}).
 \label{eq:problem}
\end{equation}

In~\eqref{eq:problem}, $\bm{D}$ is called the geometric design variable which describes the aerodynamic shape of an airfoil. For instance, $\bm{D}$ can be collection of the coordinates of control points on the lower and upper sides of the airfoil, i.e.,
\begin{equation*}
 \bm{D}=\left( x^{\text{low}}_{1}, y^{\text{low}}_{1},...,x^{\text{low}}_{n}, y^{\text{low}}_{n},x^{\text{up}}_{1}, y^{\text{up}}_{1},...,x^{\text{up}}_{n}, y^{\text{low}}_{n}\right).
\end{equation*}

In general, $\bm{D}$ is selected as the design variables' vector to describe the aerodynamic shape, and the lower and upper curves of the airfoil can be parameterized as,
\begin{equation}
 \left\{ (x,y)\in\mathbb{R}^2\big|\; \bm{P}(x,y,\bm{D})=\bm{0}, \; \bm{D}\in\mathcal{D}\subseteq \mathbb{R}^{d} \right\}.
 \label{eq:design-variable}
\end{equation}
Here, $\bm{P}(\cdot)$ is a specific shape parameterization method, and an explicit parameterization method can be expressed by $y^{\text{low/up}}=P(x,\bm{D})$. $\mathcal{D}$ denotes the geometric design space with the dimension of $d$.

In~\eqref{eq:problem}, $\bm{S}$ is called the freestream design variable which describes the flight state of an airfoil, i.e.,
\begin{equation}
 \bm{S} = \left( \text{AoA, Ma, Re},... \right).
 \label{eq:state-variable}
\end{equation}
Here, \text{AoA, Ma, Re} denote the angle of attach, Mach number and the Reynolds number, respectively.

In~\eqref{eq:problem}, $\bm{U}$ is the flow-field variables' vector to describe the spatial distribution of steady fluid (density, pressure, etc.) around the airfoil. Generally, $\bm{U}$ depends on the computational grid variable $\bm{X}$ which contains coordinates of grid vertices and their topological relationships, i.e.,
\begin{equation}
 \bm{X} = \big( x_1,\;y_1,...,\;x_{n_e},\;y_{n_e} \big).
 \label{eq:grid-variable}
\end{equation}
\begin{equation}
 \bm{U} = \big( \rho_1, u_1, v_1, p_1,...,\rho_{n_e}, u_{n_e}, v_{n_e}, p_{n_e} \big).
 \label{eq:flow-variable}
\end{equation}
Here, $n_e$ is the number of points or elements, and $(x_i,y_i)$ denotes the coordinate of the $i$-th point. $(\rho_i, u_i, v_i, p_i)$ denotes the fluid state at $(x_i,y_i)$, $\rho,p$ are density and pressure of the fluid, respectively, and $(u,v)$ is the fluid velocity vector.

The workflow for classical aerodynamic simulation and analysis is presented below,
\begin{equation}
 \bm{D} \xrightarrow[\quad 1 \quad]{} \bm{X} \xrightarrow[\quad 2 \quad]{+\bm{S}} \bm{U} \xrightarrow[\quad 3\quad]{} C
 \label{eq:CFD-workflow}
\end{equation}

1. Generation of computational grid $\bm{X}$ based on geometric variable $\bm{D}$. $\bm{X}$ includes the surface grid $\bm{X}_{\text{surf}}$ and the inner volume grid $\bm{X}_{\text{vol}}$.

2. Computation of flow-field $\bm{U}$ on grid $\bm{X}$ based on freestream $\bm{S}$. This step involves numerically solving the governing partial differential equations of fluid dynamics, which is commonly known as the classical CFD process.

3. Evaluation of aerodynamic quantities $C$ based on the flow-field $\bm{U}$. This is a downstream application of CFD, where the flow-field $\bm{U}$ can be used to evaluate aerodynamic parameters such as surface pressure distribution $C_p$ and lift/drag coefficients $C_l/C_d$, which are then utilized for aerodynamic analysis.

During the above phase~\eqref{eq:CFD-workflow}, CFD (step-2) is recognized the most time-intensive stage. The objective of this work is to utilize data-driven methodologies to accelerate the convergence process of the steady solution during CFD process.

\section{The ML-enhanced DGM}
\label{sec:3}
\qquad The idea underlying the ML-enhanced CFD method is to utilize high-precision CFD data of transonic inviscid flow around airfoils to establish a practical and lightweight data-driven model, and the flow-field predicted by the data-driven model serves as the initial flow-field distribution for CFD simulations, thereby accelerating the convergence of the steady solution.

\begin{figure}[htbp]
  \centering
  \includegraphics[width=13.5cm]{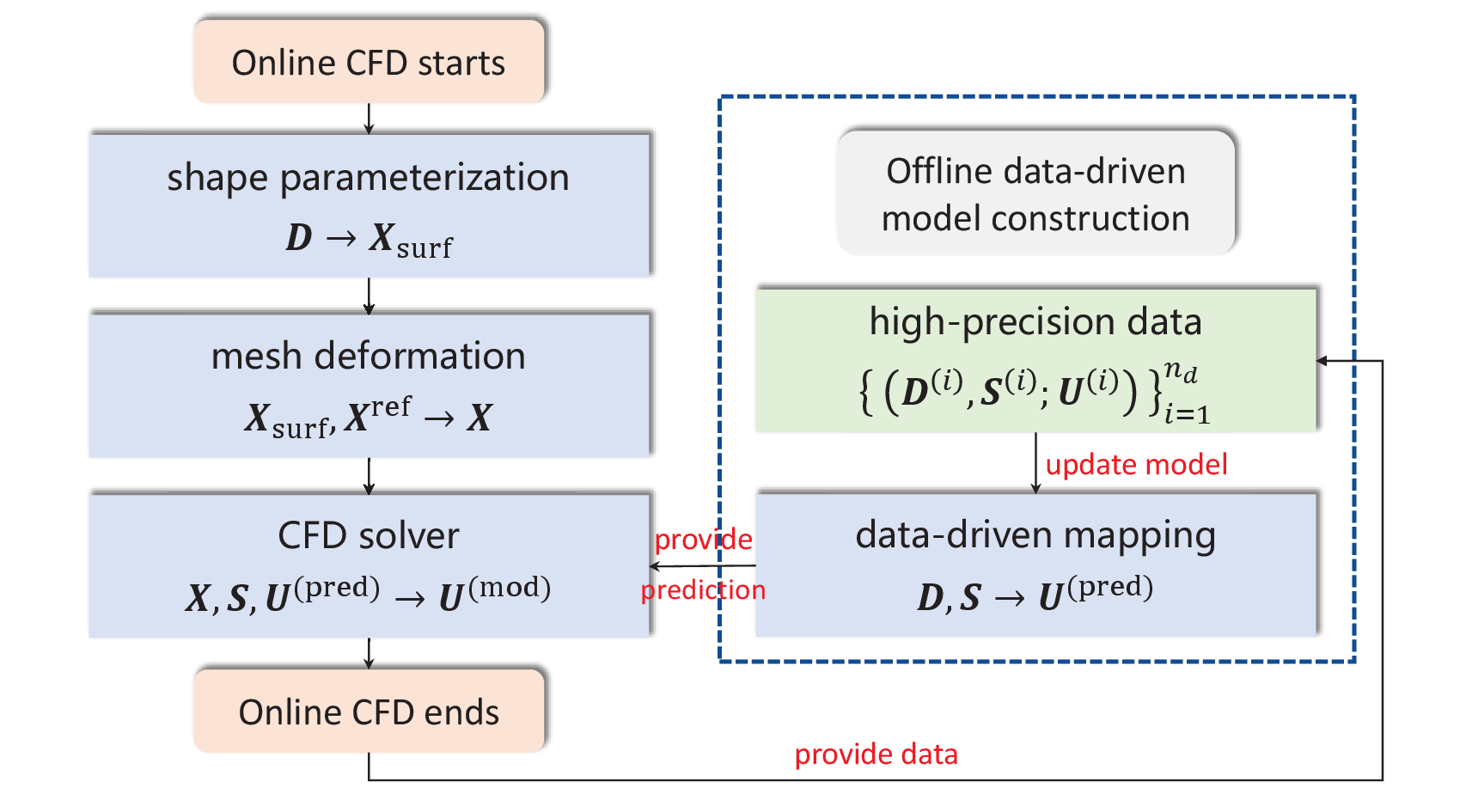}
  \caption{The workflow for the ML-enhanced CFD method.}
  \label{fig:AICFD}
\end{figure}

The data-assisted DGM in this work is developed and implemented using the open-sourced HODG platform~\cite{he2023hodg, feng2024adjoint}, which is a component-based framework of high-order discontinuous Galerkin methods (DGMs) for CFD simulation and aerodynamic shape optimization (ASO). The overall framework mainly comprises three modules: (a) geometry parameterization and mesh deformation, (b) offline data-driven model construction and updating, and (c) CFD solver with initial DoFs assignment. The entire workflow is summarized in Figure~\ref{fig:AICFD} and each module will be elaborated in the following sections.

It can be observed from Figure~\ref{fig:AICFD} that the online CFD simulation mainly contains the following three steps:\newline
\textbf{Step-1}: Mesh generation via deformation from reference grid: $\bm{D},\bm{X}^{\text{ref}}\to \bm{X}.$\newline
\textbf{Step-2}: Flowfield prediction via offline data-driven model: $\bm{D},\bm{S}\to \bm{U}^{\text{(pred)}}.$\newline
\textbf{Step-3}: Flowfield modification via high-order CFD method: $\bm{X},\bm{S},\bm{U}^{\text{(pred)}}\to\bm{U}^{\text{(mod)}}.$

Furthermore, the data produced by CFD can be utilized to enrich the data set and update the parameters within the offline data-driven model, thereby continuously enhancing its prediction capability.

\subsection{Geometry parameterization and mesh deformation}
\qquad Currently, the process of automatically generating computational grid based on geometric shapes still poses challenges, and the quality of the meshes sometimes fails to meet the computational requirements.

In order to stably provide high-quality computational grids for data-driven model and CFD, this work adopts the free-form deformation (FFD) geometry parameterization method \cite{sederberg1986free, hsu1992direct} and the radial basis function (RBF) based mesh deformation method \cite{de2007mesh} to acquire high-quality computational grids based on the baseline shape and reference grid. This work focuses on the prediction of transonic flow, and thus selects the RAE-2822 airfoil as the baseline shape, and adopts a C-type structured grid as the baseline grid, which comprises a total of 1536 quadrilateral elements.

The idea of the FFD method for geometry parameterization is to embed a baseline geometry together with its surface nodes $\bm{X}_{\text{surf}}$ into a box of flexible plastic which is known as the FFD box, and the internal geometry as well as its surface nodes are morphed when the FFD box is deformed. Specifically, the physical coordinate of a surface node $\bm{X}_{\text{surf}}=(x,y)\in\mathbb{R}^2$ is presented by
\begin{equation}
 \bm{X}_{\text{surf}}(u,v)=\sum_{i=0}^l \sum_{j=0}^m B_i^l(u) B_j^m(v) \bm{P}_{i,j}.
\label{eq:FFD}
\end{equation}
Here, $(u,v)\in [0,1]^2$ is the parametric coordinate of the surface node inside the FFD box, and $\bm{P}_{i,j}$ is the geometric design variable $\bm{D}$ which denotes the physical coordinates of the FFD box control points. $B_i^l(u),B_j^m(v)\in [0,1]$ are Bernstein polynomials of degree $l,m$ respectively, and the form is expressed by
$$
 B_i^l(u) = \frac{l!}{i! (l-i)!}u^i(1-u)^{l-i}.
$$

The local parametric coordinates $(u,v)$ of each surface node $\bm{X}_{\text{surf}}$ need to be calculated at the beginning, and remain unchanged during the morph of geometry. The movement of surface nodes $\Delta\bm{X}_{\text{surf}}$ is presented by the change of the design variable $\Delta \bm{P}_{i,j}$,
\begin{equation}
 \Delta\bm{X}_{\text{surf}}=\sum_{i=0}^l \sum_{j=0}^m B_i^l(u) B_j^m(v) \Delta \bm{P}_{i,j}.
 \label{eq:FFD2}
\end{equation}

\begin{figure}[htbp]
  \centering
  \includegraphics[width=7.5cm]{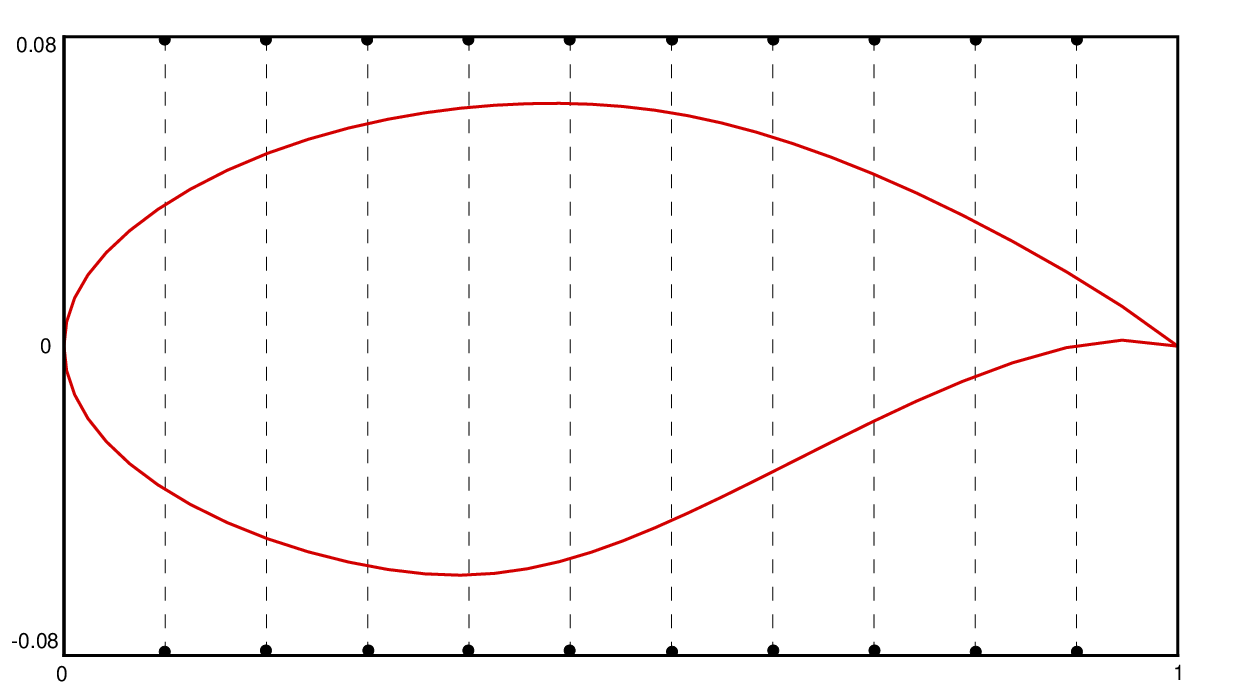}
  \caption{Diagram of the FFD box and FFD control points.}
  \label{fig:FFD-box}
\end{figure}

The initial FFD box used in this work is a rectangular frame (width $\times$ height $=1\times 0.16$) with a total of 20 FFD control points ($N_x\times N_y=10\times2$, refer to Figure~\ref{fig:FFD-box}). The geometric design variable $\bm{D}$ is set as the $y$-coordinates of the FFD control points,
\begin{equation}
 \bm{D} = \big( P_{0,0,y},\;P_{0,1,y},...,\;P_{9,0,y},\;P_{9,1,y} \big).
 \label{eq:design-variable-FFD}
\end{equation}
The initial design variable $\bm{D}^{(0)}$, which represents the shape of the RAE-2822 airfoil, is set to
\begin{equation}
 \bm{D}^{(0)} = \big( -0.08,\;0.08,...,-0.08,\;0.08 \big).
 \label{eq:D0}
\end{equation}

The idea of the RBF method for mesh deformation is to deform the inner volume grid $\bm{X}_{\text{vol}}$ according to the movement of the surface grid $\Delta \bm{X}_{\text{surf}}$ controlled by the design variable $\bm{D}$, and the relation between $\Delta \bm{X}_{\text{vol}}$ and $\Delta \bm{X}_{\text{surf}}$ is interpolated by the RBF.

The displacement $\Delta \bm{X}_{\text{vol}}$ is a vector collecting the displacement of each inner volume node $\Delta \bm{X}_{\text{vol}}=\big(\Delta \bm{x}_v \big)$, with $\Delta \bm{x}_v=\big(\Delta x_{v}(\bm{x}_v),\Delta y_{v}(\bm{x}_v)\big)$, $\Delta x_{v},\Delta y_{v}$ are movement of $\bm{x}_v$ in the $x,y$ direction, respectively,
\begin{equation}
\left\{
\begin{aligned}
 & \Delta x_{v}(\bm{x}_v)=\sum_{b=1}^{N_b}\alpha_{x,b} \phi\left( \| \bm{x}_v-\bm{x}_b \| \right),\\
 & \Delta y_{v}(\bm{x}_v)=\sum_{b=1}^{N_b}\alpha_{y,b} \phi\left( \| \bm{x}_v-\bm{x}_b \| \right).
\end{aligned}
\right.
\label{eq:RBFs}
\end{equation}
Here, $\bm{x}_b$ denotes one of all $N_b$ boundary nodes, $\| \bm{x}_v-\bm{x}_b \|$ is the distance between the inner node $\bm{x}_v$ and boundary node $\bm{x}_b$, and $\bm{\alpha}=\left(\alpha_{x,b},\alpha_{y,b}\right)_b$ are undetermined weights, $\phi(\|\bm{x}\|)$ is a RBF kernel with the form of
\begin{equation}
 \phi\left(\|\bm{x}\|\right) = \left\{
\begin{aligned}
 &\left( 1-\frac{\|\bm{x}\|}{r} \right)^2  \quad && 0\leq \|\bm{x}\|\leq r \\
 &\quad 0 &&\quad \|\bm{x}\|>r
\end{aligned}
\right.
\label{eq:RBF-function}
\end{equation}
where $r>0$ is the compact support radius.

To determine the undetermined weights $\bm{\alpha}$ in~\eqref{eq:RBFs}, the following linear system with the size of $2\times N_b$ needs to be solved
\begin{equation}
\Delta \bm{X}_{\text{surf}} = \bm{\Phi}_{b,b}\; \bm{\alpha}.
\label{eq:RBFs-matrix-vector}
\end{equation}
Here, $\bm{\Phi}_{b,b}$ is the universal basis matrix, and its position $(b_i, b_j)$ is expressed as
\begin{equation*}
 \Phi_{b_i,b_j} = \phi\left( \| \bm{x}_{b_i}-\bm{x}_{b_j} \| \right),
\end{equation*}
where $\bm{x}_{b_i},\bm{x}_{b_j}$ are two boundary nodes.

Based on~\eqref{eq:RBFs}, the movement of inner volume gird $\Delta \bm{X}_{\text{vol}}$ is determined by
\begin{equation}
\Delta \bm{X}_{\text{vol}} = \bm{\Phi}_{v,b}\; \bm{\alpha}.
\label{eq:RBFs-deformation}
\end{equation}
$\bm{\Phi}_{v,b}$ with the size of $N_v\times N_b$ is the universal basis matrix whose element in position $(v_i, b_j)$ is expressed as
\begin{equation*}
 \Phi_{v_i,b_j} = \phi\left( \| \bm{x}_{v_i}-\bm{x}_{b_j} \| \right),
\end{equation*}
where $\bm{x}_{v_i},\bm{x}_{b_j}$ denote a inner volume node and a boundary node, respectively.

As a result, relation between $\Delta\bm{X}_{\text{vol}}$ and $\Delta\bm{X}_{\text{surf}}$ can be represented as follows,
\begin{equation}
\Delta \bm{X}_{\text{vol}} = \bm{\Phi}_{v,b}(\bm{X}_{\text{vol}},\bm{X}_{\text{surf}})\; \bm{\Phi}_{b,b}^{-1}(\bm{X}_{\text{surf}})\;\Delta\bm{X}_{\text{surf}}.
\label{eq:RBFs-deformation2}
\end{equation}

\subsection{The lightweight data-driven model}
\label{sec:3.2}
\qquad The motivation of this work is to leverage a model driven by sparse data to capture the coarse structural features of the flow-field, and then utilize CFD to refine these coarse features to resolve more intricate flow details. The offline data-driven model is therefore expected to be (a) lightweight to ensure efficiency, and (b) easily updatable to accommodate continuously changing data set.

\begin{figure}[htbp]
  \centering
  \includegraphics[width=13cm]{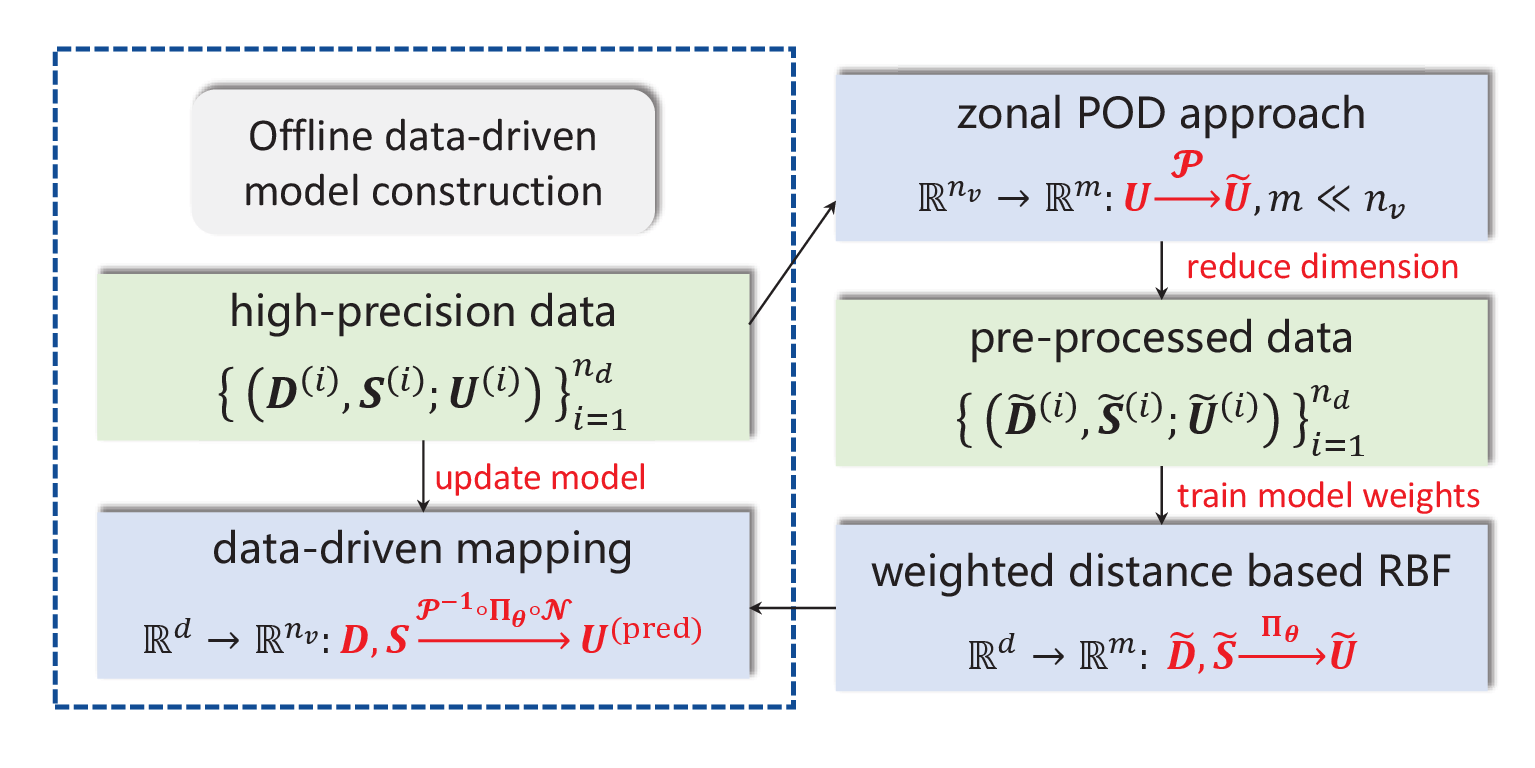}
  \caption{The workflow for the offline constructed data-driven model.}
  \label{fig:AImodel}
\end{figure}

To achieve the two properties mentioned above for the data-driven model, the following two efforts are made: (a) utilizing the zonal proper orthogonal decomposition (POD) method to reduce the dimension of the flow-field $\bm{U}$ (model output) while preserving more near-wall features; (b) adopting the weighted-distance radial basis function (RBF) interpolation to reduce the DoFs within the model and endow the model with the property of easy updatability. The entire workflow for constructing and updating the offline data-driven model is shown in Figure~\ref{fig:AImodel}, and the core of the data-driven model consists of three operators:
\begin{itemize}
  \item Embedding operator $\bm{\mathcal{P}}:\mathbb{R}^{n_v}\to\mathbb{R}^m$, which is the embedding of full-order flow-field variable $\bm{U}$ into the reduced-order $\widetilde{\bm{U}}$ through zonal POD.
  \item Approximation operator $\bm{\Pi}_{\bm{\theta}}:\mathbb{R}^{d}\to\mathbb{R}^m$, which establishes an approximation map between the normalized $\widetilde{\bm{D}}$ and $\widetilde{\bm{S}}$ to the reduced-order flow-field $\widetilde{\bm{U}}$.
  \item Reconstruction operator $\bm{\mathcal{P}^{-1}}:\mathbb{R}^{m}\to\mathbb{R}^{n_v}$, which restores the full-order flow-field variable $\bm{U}$ from $\widetilde{\bm{U}}$ through the inverse POD.
\end{itemize}
More details are described in the following three sub-sections.

\subsubsection{Data generation via HODG platform}
\qquad The template of data input and output (I/O) in this work is presented as follows,
\begin{equation}
 \big( \bm{D},\;\bm{S};\;\bm{U} \big).
\label{eq:data-template}
\end{equation}

During the data generation phase, only inviscid flow is considered, thus
$$\bm{S}=\left(\text{Ma}, \text{AoA}\right).$$
The geometric design variable $\bm{D}$ is described as~\eqref{eq:design-variable-FFD}. The design spaces of $\bm{S}$ and $\bm{D}$ are defined by
$$\mathcal{S}=\Big\{(\text{Ma,AoA})\in\mathbb{R}^2\Big|\;\text{Ma}\in[0.7,0.95],\;\text{AoA}\in[-5^\circ,5^\circ]\Big\}.$$
$$\mathcal{D}=\left\{\bm{D}\in\mathbb{R}^{20}\Big|\;\Delta D_i\leq0.25|D_i^{(0)}|,\;i=1,2..,20\right\}.$$

The sampling in $\mathcal{S}$ adopts uniform sampling, with 6 uniformly distributed sample points of Mach number and 11 uniformly distributed sample points of angle of attack. The sampling in $\mathcal{D}$ adopts the symmetric Latin hypercube sampling (SLHS)~\cite{kenny2000algorithmic} with 14 sample points for each freestream state $\bm{S}^{(i)}$.

Once the shape $\bm{D}^{(i)}$ and freestream $\bm{S}^{(i)}$ are given, the flow-field $\bm{U}^{(i)}$ on grid $\bm{X}^{(i)}$ is obtained via the HODG platform \cite{he2023hodg}. In this work, we employ DG(p2) spatial discretization combined with 1st Euler implicit temporal discretization to numerically simulate the transonic inviscid flow around airfoils. More details about the CFD solver will be introduced in subsequent Section~\ref{sec:3.3}.

As a result, a total of $6\times11\times14=924$ sample points are generated with various geometry shapes, freestream conditions and the corresponding flow-field. The generated data is locally stored and continuously updating.

\begin{remark}
Compared to the conservative variables $(\rho, \rho u, \rho v, E)$, selecting the primitive variables $(\rho,u,v,p)$ for the flow-field $\bm{U}$ during data generation exhibits superior robustness, which can better avoid the occurrence of non-physical solutions such negative pressure when constructing the data-driven model.
\end{remark}

\subsubsection{Dimensionality reduction of flow-field via zonal POD method}
\qquad The POD, also known as the principal component analysis (PCA), is an efficient method to extract main features of the high-dimensional systems. According to Section 3.2.1, $n_d$ flow solutions (snapshots) are generated $\left\{ \bm{U}^{(i)} \right\}_{i=1}^{n_d}$ with $\bm{U}^{(i)}\in \mathbb{R}^{n_v}$, where $n_v=n_e\times n_{\text{cv}}$, and $n_e=1536,\; n_{\text{cv}}=4$ are the numbers of elements and conservative variables, respectively. The key formula of POD is
\begin{equation}
\bm{U}^{(i)}\approx \overline{\bm{U}}+\sum_{k=1}^{m} \widetilde{U}_k^{(i)}\bm{\phi}_k = \overline{\bm{U}}+\bm{\Phi}\cdot \widetilde{\bm{U}}^{(i)},\quad m\ll n_v.
\label{eq:POD}
\end{equation}
In~\eqref{eq:POD}, $\overline{\bm{U}}\in\mathbb{R}^{n_v}$ represents the mean flowfield vector of all samples. $\widetilde{\bm{U}}^{(i)}=\left(\widetilde{U}_1^{(i)},..,\widetilde{U}_m^{(i)}\right)^{\text{T}}$ is the POD coefficient vector of the original flow-field vector $\bm{U}^{(i)}$, and regarded as the dimension-reduced vector of $\bm{U}^{(i)}$ as $m\ll n_v$. $\bm{\Phi}=\left(\bm{\phi}_1,..,\bm{\phi}_{m}\right)$ is the orthogonal POD basis matrix and $\bm{\phi}_k\in\mathbb{R}^{n_v}$ is the $k$-th POD basis vector. Each basis vector $\bm{\phi}_k$ corresponds to a singular value $\lambda_k$ of the snapshot matrix $\bm{A}=\left(\bm{U}^{(1)},..,\bm{U}^{(n_d)}\right)$. The singular values are sorted in the descending order, $\lambda_1\geq\lambda_2\geq\dots\geq\lambda_m$, and each singular value $\lambda_k$ demonstrates the energy weight of $\bm{\phi}_k$. The reduced dimension $m$ depends on the distribution of singular values. A commonly used strategy for selecting $m$ is to retain the energy weights that accounts for more than $\theta$ fraction of the total energy weights, i.e.,
\begin{equation*}
 m=\min_{k} \left\{ \frac{\sum_{i=1}^k\;\lambda_i}{\sum_{i=1}^{n_d}\;\lambda_i}\geq \theta \right\}.
\end{equation*}
Depending on different application scenarios, $\theta\in(0,1)$ can be set to $0.8,\;0.99$, etc. Specific algorithmic steps for solving POD basis matrix $\bm{\Phi}$ and POD coefficients $\{\widetilde{\bm{U}}^{(i)}\}_{i=1}^{n_d}$ can be referred to in \cite{chatterjee2000introduction}.

To improve the efficacy of flow-field feature extraction, this work adopts a zonal POD approach for dimensionality reduction. Based on the prior knowledge, the characteristic structures in the transonic inviscid flows, such as shock waves and expansion waves, are primarily concentrated near the wall, whereas the far-field is largely governed by freestream conditions. Therefore, the flow-field $\bm{U}$ is strategically segmented into three sub-vectors:
\begin{equation}
\bm{U}=(\bm{U}_{\text{wall}}, \bm{U}_{\text{mid}}, \bm{U}_{\text{far}}),
\label{eq:sub-U}
\end{equation}
and then the POD methods with distinct thresholds $\theta$ are individually applied to each sub-vector in~\eqref{eq:sub-U}. The specific strategy is described as follows,
\begin{itemize}
  \item For $\bm{U}_{\text{wall}}$, $\theta_{\text{wall}}$ is set to 0.9 to ensure the extraction of more near-wall flow features.
  \item For $\bm{U}_{\text{mid}}$, $\theta_{\text{mid}}$ is set to 0.6 for transition of the intermedium flow-field.
  \item For $\bm{U}_{\text{far}}$, $\theta_{\text{far}}$ is set to 0.5, indicating that a moderate amount of far-field information is discarded to maintain the overall efficiency of the POD process.
\end{itemize}
We compare the performance of the zonal POD method mentioned above with the original POD method ($\theta=0.8$ and $\theta=0.9$) in terms of dimensionality reduction effects. The results are shown in the Table~\ref{tab:POD},
\begin{table}[htbp]
\centering \caption{Comparison of dimensionality reduction effects between the zonal POD and original POD with different threshold $\theta$.}
\begin{center}
\begin{tabular}{m{3cm}<{\centering}|m{3.5cm}<{\centering}|m{3.5cm}<{\centering}}
\toprule
  & Original dimension & Reduced dimension \\ \hline
  \hspace{0.3em} zonal POD \hspace{0.3em} &6,144 & 46  \\ \hline
  POD, $\theta=0.8$ &6,144 & 42  \\ \hline
  POD, $\theta=0.9$ &6,144 & 85 \\
\bottomrule
\end{tabular}
\end{center}
\label{tab:POD}
\end{table}

As can be seen from the Table~\ref{tab:POD}, the dimensionality reduction effect of the original POD with $\theta=0.8$ is slightly better than that of the zonal POD, and both are significantly superior to the POD with $\theta=0.9$. The potential strengths of zonal POD will be further manifested in its impact on the predictive performance of the data-driven model, which will be comprehensively analyzed in the Section~\ref{sec:4}.

POD-based dimensionality reduction is easily updatable. With the continuous expansion of dataset, if the representation of new data samples $\bm{U}^{(q)}$ under the current POD bases fails to reach the required accuracy, new orthogonal basis $\bm{\phi}_q$ can be added to the POD basis matrix $\bm{\Phi}$ using the following formula to enhance the representation capability,
\begin{equation}
 \bm{\phi}_q = \frac{\bm{U}^{(q)}-\overline{\bm{U}}-\sum_{k=1}^m \widetilde{U}_k^{(q)}\bm{\phi}_k}{\left\| \bm{U}^{(q)}-\overline{\bm{U}}-\sum_{k=1}^m \widetilde{U}_k^{(q)}\bm{\phi}_k \right\|},
\label{eq:POD-update}
\end{equation}
where $\widetilde{U}_k^{(q)}$ is the projection coefficient of $\bm{U}^{(q)}$ onto the $k$-th orthogonal basis $\bm{\phi}_k$.

At the end of this sub-section, the dimensionality reduction of (zonal) POD and the reconstruction of inverse (zonal) POD for flow-field $\bm{U}$ are summarized as follows,
\begin{equation}
 \bm{U} \xrightarrow[\text{dimension reduction}]{\text{(zonal) POD}} \widetilde{\bm{U}} \xrightarrow[\;\;\;\;\;\text{ reconstruction }\;\;\;\;\;]{\text{inverse POD}} \overline{\bm{U}}+\bm{\Phi}\cdot\widetilde{\bm{U}}.
 \label{eq:zonal-POD}
\end{equation}
The mean flowfield vector $\overline{\bm{U}}$ and the POD basis matrix $\bm{\Phi}$ are locally stored and continuously updating. After reducing the dimensionality of the flow-field $\bm{U}$ to $\widetilde{\bm{U}}$ using the zonal POD approach, the input and output dimensions of the prediction problem~\eqref{eq:problem} are successfully maintained at a medium-scale.

\subsubsection{The weighted-distance RBF interpolation}
\qquad The RBF interpolation mapping is selected as the data-driven model in this work to further reduce the DoFs within the model, thereby decreasing the runtime memory. In addition, the RBF model is easily updatable, with minimal costs for updating its DoFs when the dataset is updated.

The RBF interpolation model~\cite{powell1992theory, huang2016optimization} uses a weighted sum of basis functions to represent complicated input-output mappings. Denoting
\begin{equation}
\widetilde{\bm{X}} = (\widetilde{\bm{D}}, \widetilde{\bm{S}}),\quad \widetilde{\bm{Y}} = \widetilde{\bm{U}},
\label{eq:RBF-x}
\end{equation}
where $\widetilde{\bm{D}}, \widetilde{\bm{S}}$ are normalized geometric and freestream design variables, respectively. The form of the RBF mapping $\bm{\Pi}_{\bm{\theta}}:\mathbb{R}^{d}\to\mathbb{R}^{m}$ is presented as follows,
\begin{equation}
 \widetilde{\bm{Y}} = \bm{\Pi}_{\bm{\theta}}(\widetilde{\bm{X}}) = \bm{w} \bm{\psi} =\sum_{i=1}^{n_d} \bm{w}_i \cdot \psi\left( \big\| \widetilde{\bm{X}}-\widetilde{\bm{X}}^{(i)} \big\|_s \right), \quad \widetilde{\bm{X}}\in\mathbb{R}^{d}.
 \label{eq:RBFs}
\end{equation}
Here, where $n_d$ denotes the number of sample points, $\widetilde{\bm{X}}^{(i)}$ denotes the input of the $i$-th sample point, and $\bm{w}=(\bm{w}_1,\bm{w}_2,..,\bm{w}_{n_d})\in\mathbb{R}^{m\times n_d}$ is the weight matrix. The parameter $\{\bm{w}_i\}_{i=1}^{n_d}$ and $\{\widetilde{\bm{X}}^{(i)}\}_{i=1}^{n_d}$ in~\eqref{eq:RBFs} make up of the DoFs $\bm{\theta}$ within the RBF. The Euclidean norm $\|\cdot\|_s=\|\cdot\|_2$ is utilized, and $\psi\left( \cdot \right)$ is an RBF kernel or RBF basis function. This work employs the following cubic RBF basis function with a linear tail,
\begin{equation}
 \psi^{\text{cubic}}(r)=r^3.
 \label{eq:RBF-cubic}
\end{equation}

The undetermined augmented weight matrix $\bm{w}$ are estimated according to the interpolation condition
\begin{equation}
\bm{\Pi}_{\bm{\theta}}(\widetilde{\bm{X}}^{(i)})=\widetilde{\bm{Y}}^{(i)}, \quad i=1,\cdots,n_d.
\label{eq:interpolation-condition}
\end{equation}
The DoFs $\bm{w}$ can be determined by solving the linear system~\eqref{eq:interpolation-condition}.

To enhance the generalization capability of the prediction model, this work adopts a weighted-distance RBF interpolation mapping. We firstly train a fully connected neural network using Pytorch to fit the normalized and dimensionality-reduced data, and then utilize its powerful autograd functionality to obtain the sensitivity of the first few POD coefficients (modes) concerning various design variables. Based on the prior knowledge and sensitivity analysis, the Mach number (Ma) and the angle of attack (AoA) exhibit a greater influence on the flow-field, whereas individual components of geometric design variables have a relatively minor impact on the flow-field. Therefore, the distance $\|\cdot\|_s$ used in~\eqref{eq:RBFs} is modified into a weighted distance according to the sensitivity analysis,
\begin{equation}
\| \widetilde{\bm{X}}-\widetilde{\bm{X}}^{(i)} \|_s^2=d\big\|\widetilde{\bm{S}}-\widetilde{\bm{S}}^{(i)}\big\|^2+\big\|\widetilde{\bm{D}}-\widetilde{\bm{D}}^{(i)}\big\|^2=:\| \widetilde{\bm{X}}-\widetilde{\bm{X}}^{(i)} \|_w^2
\label{eq:weighted-distance}
\end{equation}
The advantage of the weighted distance defined in~\eqref{eq:weighted-distance} is that it allows variables with greater influence on the flow-field to have larger contribution weights, thereby enhancing the generalization capability of the RBF interpolation.

In the Section~\ref{sec:4}, detailed test and analysis are conducted to access the generalization capability of the RBF models using the classical Euclidean distance $\|\cdot\|_2$ and the weighted distance $\|\cdot\|_w$.

At the end of this sub-section, workflow of the entire data-driven prediction model is summarized below,
\begin{equation}
 \left(\bm{D},\bm{S}\right) \xrightarrow[]{\text{\tiny{normalization}}} (\widetilde{\bm{D}},\widetilde{\bm{S}}) \xrightarrow[]{\text{\tiny{RBF prediction}}} \widetilde{\bm{U}}\xrightarrow[]{\;\;\;\;\text{\tiny{inverse POD}}\;\;\;\;} \bm{U}.
 \label{eq:AI-model}
\end{equation}
The model weight matrix $\bm{w}$ are locally stored and continuously updating.

\subsection{CFD solver with initial DoFs assignment}
\label{sec:3.3}
\qquad This work employs high-order discontinuous Galerkin methods (DGMs) as the CFD solver to generate data and correct the steady flow-field predicted by the data-driven model. For the 2D compressible Euler equations,
\begin{equation}
\left\{
\begin{aligned}
 &\nabla \cdot \bm{\mathcal{F}}_I(\bm{u})= \bm{0}  &&\bm{x}\in\Omega \\
 &\mathcal{B}\bm{u}(\bm{x})=\bm{0}      &&\bm{x}\in \partial \Omega=:\Gamma.
\end{aligned}
\right.
\label{eq:2deuler}
\end{equation}
The conservative variable vector $\bm{u}$ and the inviscid flux matrix $\bm{\mathcal{F}}_I(\bm{u})$ are defined by
\begin{equation}
\bm{u}=\begin{pmatrix}
\rho \\
\rho u \\
\rho v \\
\rho E
\end{pmatrix},\quad
\bm{\mathcal{F}}_I(\bm{u})= \begin{pmatrix}  \rho u  &\rho v   \\ \rho u^2+p  &\rho uv \\ \rho uv  &\rho v^2+p\\  u(E+p)  &v(E+p)  \end{pmatrix}.
\label{eq:euler-flux}
\end{equation}
Here, $\rho,\;p$ and $E$ denote the density, pressure and total energy of the fluid, respectively. $(u,v)$ is the velocity vector of the flow. The pressure variable $p$ can be computed from the equation of state
\begin{equation*}
 p=(\gamma-1)\left(E-\frac{1}{2}\rho (u^2+v^2)\right),
\end{equation*}
where $\gamma$ is the ratio of the specific heats.

DGMs are used to discretize the weak form of the governing equations~\eqref{eq:2deuler} over the domain $\Omega$
 \begin{equation}
  \int_{\Gamma}\bm{\mathcal{F}}_I(\bm{u})\cdot \bm{n} \; \phi\;d\Gamma -
  \int_{\Omega}\bm{\mathcal{F}}_I(\bm{u})\cdot \nabla \phi \; d\Omega =\bm{0},
  \label{eq:weak}
 \end{equation}
where $\Gamma(=\partial \Omega)$ denotes the boundary of $\Omega$ and $\bm{n}$ is the unit outward normal vector to the boundary.

Substituting variable $\bm{u}$ and test function $\phi$ in~\eqref{eq:weak} by DG solution $\bm{U}_h^{(e)}$ and its elemental basis function $\phi_i^{(e)}$, we can obtain the spatial DG discretization of~\eqref{eq:weak},
\begin{equation}
 \oint_{\Gamma_e}\widehat{\bm{F}}_{I}(\bm{U}_h^-,\;\bm{U}_h^+,\;\bm{n}_e)\;\phi_i^{(e)}d\Gamma - \int_{\Omega_e}\bm{\mathcal{F}}_{I}(\bm{U}_h^{(e)})\cdot \nabla\phi_i^{(e)}\;d\Omega=\bm{0}, \quad e\leq n_e,\;i\leq n_k.
 \label{eq:semi-DG}
\end{equation}
Here, $n_e,n_k$ denote the number of elements and basis functions, respectively. $\bm{U}_h^{(e)}=\sum_{i=0}^{n_k} \bm{U}^{(e)}_i\cdot \phi_i^{(e)}(\bm{x})$ is the piecewise polynomial DG solution inside $\Omega_e$, and $\phi_i^{(e)}$ is the $i$-th basis function. Taylor basis functions~\cite{luo2008discontinuous} are used in this work. $\bm{U}_h^-,\;\bm{U}_h^+$ respectively represent the internal and external values of DG solution at cell interfaces $\Gamma_e$, and $\widehat{\bm{F}}_I(\bm{U}_h^-,\;\bm{U}_h^+,\;\bm{n}_e)$ is an approximate Riemann solver for the inviscid numerical flux vector $\bm{\mathcal{F}}_I(\bm{u})\cdot \bm{n}_e$, and $\bm{n}_e=(n_x,n_y)$ denotes the unit outward normal vector at interfaces $\Gamma_e$.

Appropriate shock-processing techniques are needed to avoid high-order solution suffering from spurious oscillations. This work employs a characteristic-compression embedded shock indicator~\cite{feng2021characteristic2} + strong-residual based artificial viscosity (AV)~\cite{he2022ddg} + bound and positivity-preserving limiter~\cite{cheng2014positivity} to handle shock waves and near-vacuum regions in transonic flows, greatly enhancing the robustness of DGMs in simulating high-speed flows.

The AV-modified spatial DG discretization is expressed by
\begin{equation}
 \oint_{\Gamma_e}\widehat{\bm{F}}_{I}(\bm{U}_h^-,\;\bm{U}_h^+,\;\bm{n}_e)\;\phi_i^{(e)}d\Gamma - \int_{\Omega_e}\bm{\mathcal{F}}_{I}(\bm{U}_h^{(e)})\cdot \nabla\phi_i^{(e)}\;d\Omega + \epsilon_e\;\int_{\Omega_e}\nabla\bm{U}_h^{(e)}\cdot \nabla\phi_i^{(e)}\;d\Omega=\bm{0}.
 \label{eq:semi-DG-artificial}
\end{equation}
$\epsilon_e$ is a Hartmann-type artificial viscosity coefficient with the form of
\begin{equation*}
\begin{aligned}
\epsilon_e=&C_{\epsilon}h_e^{2-\beta}\frac{\int_{\Omega_e}\left|\nabla \cdot \bm{\mathcal{F}}_I\right|d\Omega}{|\Omega_e|},\quad or \;\;
\epsilon_e=&C_{\epsilon}h_e^{2-\beta}\frac{\oint_{\Gamma_e}\left|\bm{\mathcal{F}}_I\cdot \bm{n}_e\right|d\Gamma}{|\Gamma_e|},
\end{aligned}
\end{equation*}
where $C_{\epsilon}$ and $\beta$ are usually set to $0.001\sim0.1$, $0.01\sim0.5$, respectively.

Collecting the DG DoFs as the global solution vector $\bm{U}_a = \big( \bm{U}_i^{(e)} \big)$, \eqref{eq:semi-DG-artificial} is actually a nonlinear algebraic equation system $\bm{R}(\bm{U}_a)=\bm{0}$ about $\bm{U}_a$. A pseudo-time term is introduced to accelerate the process of solving~\eqref{eq:semi-DG-artificial}, and~\eqref{eq:semi-DG-artificial} is then rewritten as the following time-dependent ordinary differential equation system,
\begin{equation}
\bm{M}\frac{d\bm{U}_a}{dt}+\bm{R}(\bm{U}_a)=\bm{0},
\label{eq:semi-DG-matrix}
\end{equation}
where $\bm{M}$ is the mass matrix.

In this work, $1^{\text{st}}$ order backward Euler implicit temporal discretization is used to treat~\eqref{eq:semi-DG-matrix},
\begin{equation}
  \bm{M}\frac{\Delta\bm{U}_a^n}{\Delta t}+\bm{R}(\bm{U}_a^{n+1})=\bm{0},
  \label{eq:time}
\end{equation}
using $1^{\text{st}}$ order approximation $\bm{R}(\bm{U}_a^{n+1})\approx\bm{R}(\bm{U}_a^{n})+(\frac{\partial \bm{R}}{\partial \bm{U}_a})^n\Delta \bm{U}_a^n$, and nonlinear system~\eqref{eq:time} is now linearized as
\begin{equation}
  \left(\frac{\bm{M}}{\Delta t}\bm{I}+(\frac{\partial \bm{R}}{\partial \bm{U}_a})^n\right)\Delta \bm{U}_a^n=-\bm{R}(\bm{U}_a^n),
  \label{eq:time2}
\end{equation}
where $\Delta \bm{U}_a^n=\bm{U}_a^{n+1}-\bm{U}_a^{n}$. The linear equation system~\eqref{eq:time2} at each time step is solved by GMRES method with the LU-SGS preconditioner \cite{Hong1998A}. Specially, growth strategy of local time step size \cite{feng2024adjoint} are utilized in this work to speed up convergence in the implicit scheme.

In order to further accelerate convergence process of the steady solution, the initial DG DoFs are assigned jointly through the data-driven model (for cell-averages) and reconstruction (for high-order moments). Specifically, the data-driven model outlined in Section 3.2 is employed to predict the cell-averages of the DG solution on the deformed mesh. Subsequently, the weighted essentially non-oscillatory (WENO) \cite{zhu2008runge} reconstruction is utilized to obtain high-order moments of the DG solution. The positivity-preserving limiter \cite{cheng2014positivity} is applied to high-order moments to maintain the positivity of pressure within the DG solutions.

In subsequent Section~\ref{sec:4} and \ref{sec:5}, we will design numerical experiments to show the predictive performance of the data-driven models in Section \ref{sec:3.2}, and demonstrate the accelaration performance of the proposed initial DoFs assignment strategy in Section \ref{sec:3.3}.

\section{Predictive performance test for data-driven models}
\label{sec:4}
\begin{figure}[htbp]
  \centering
  \includegraphics[width=10cm]{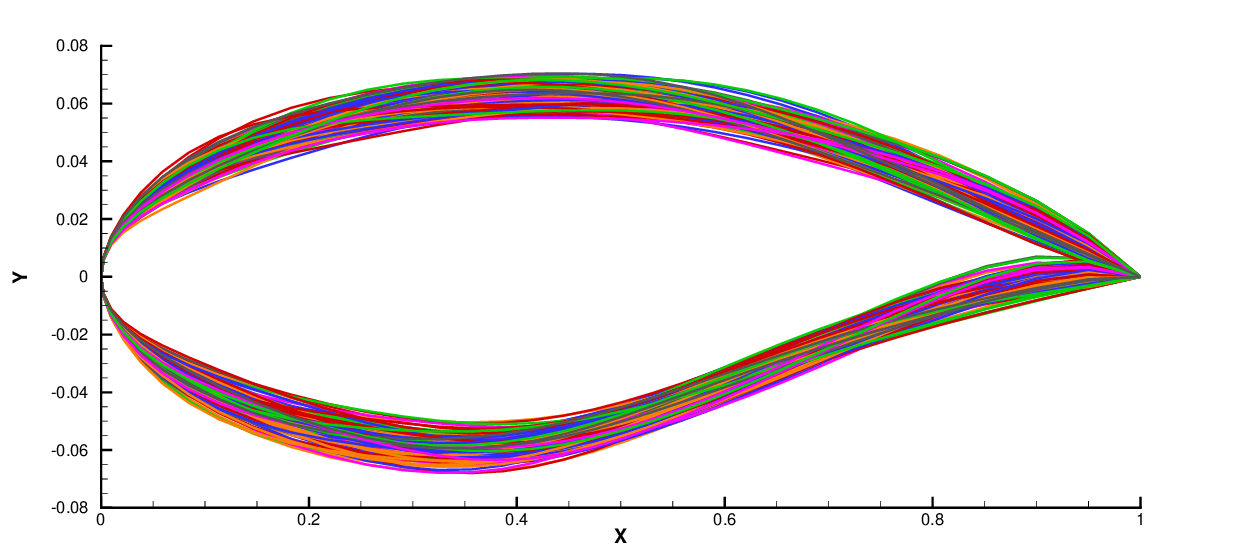}
  \caption{The aerodynamic shapes of airfoils.}
  \label{fig:shapes}
\end{figure}

\qquad In this section, the ML-enhanced DGM proposed in Section~\ref{sec:3} is implemented using the HODG platform, and the predictive performance of the data-driven model on dataset is tested as well. The dataset contains a total of 924 sample points with 900 allocated for model construction and the remaining 24 reserved for validation. A portion of aerodynamic shapes contained within the dataset is depicted in Figure~\ref{fig:shapes}, encompassing a Mach number ranging from 0.7 to 0.95 and an angle of attack ranging from -5 to 5 degrees.

The notation "Zonal POD", "Orig POD", "Mod RBF" and "Orig RBF" are used to refer to the zonal POD, original POD with $\theta=0.8$, weighted-distance RBF and original RBF as mentioned in Section \ref{sec:3.2}, respectively.

\subsection{Predictive performance on training dataset}
\qquad This part investigates the predictive performance of the data-driven model on training set. We randomly select 5 sample points from different Mach number ranges in the training set, and present the results (including the flow-field and the $C_p$ distribution) produced by different strategies and CFD at these sample points. Selecting either Mod RBF or Orig RBF will not affect the predictive performance of the model on the training set.

\begin{figure}[htbp]
 \centering
 \subfigure[model prediction of No.1]{
 \includegraphics[width=4.8cm]{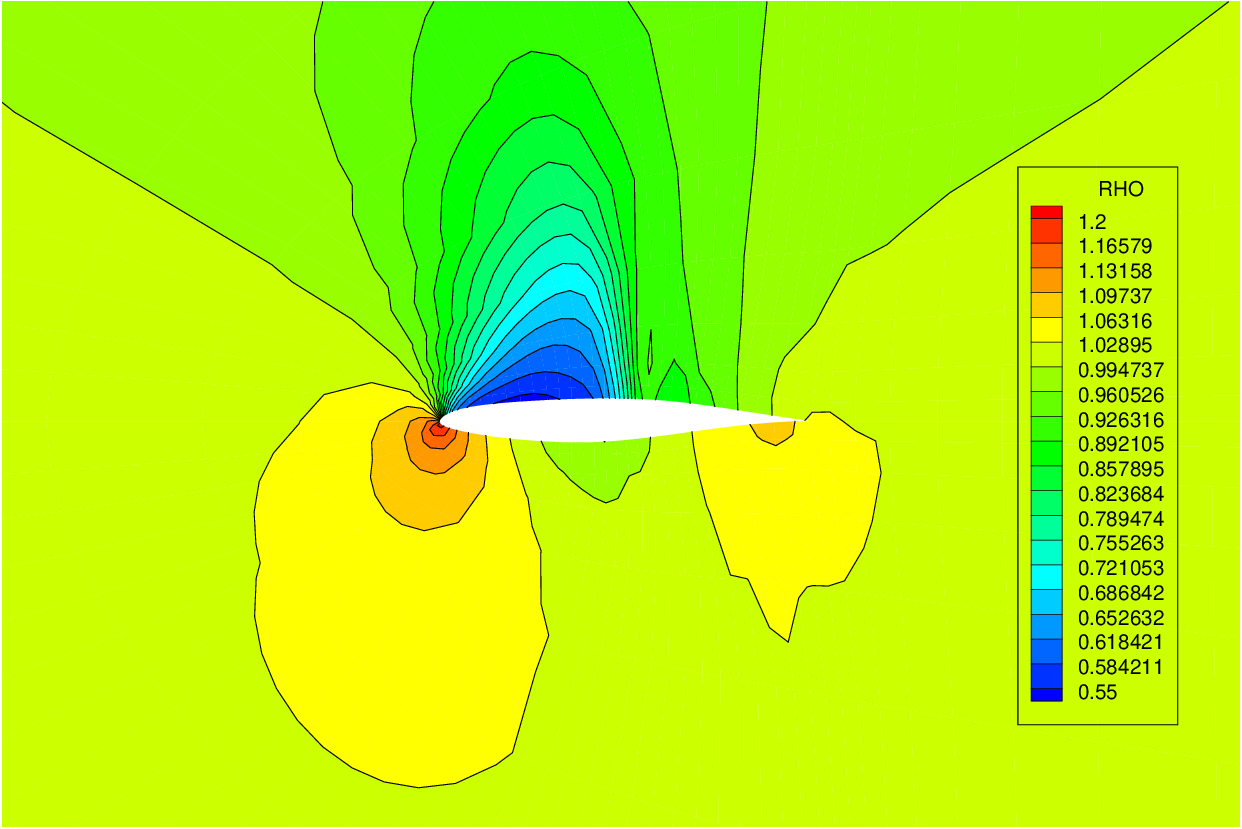}}
 \subfigure[CFD correction of No.1]{
 \includegraphics[width=4.8cm]{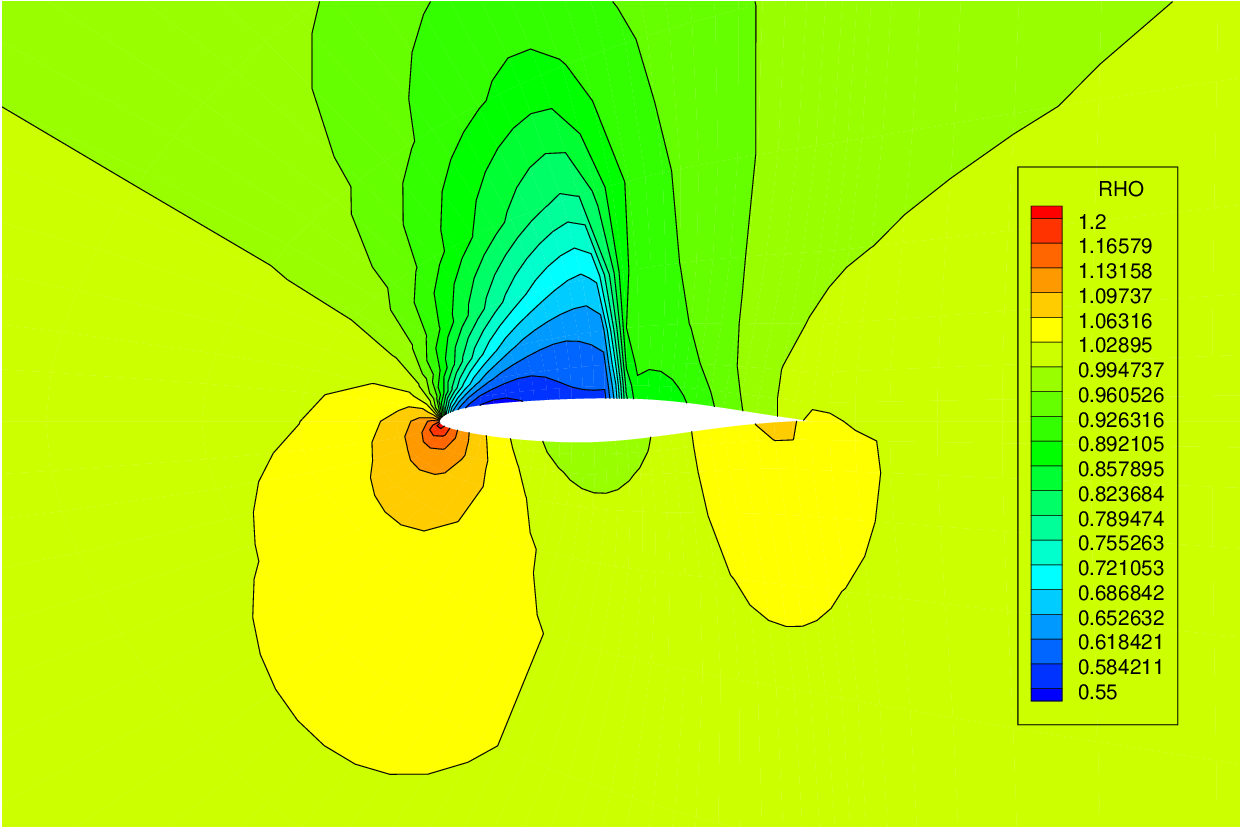}}
 \subfigure[relative error of No.1]{
 \includegraphics[width=4.8cm]{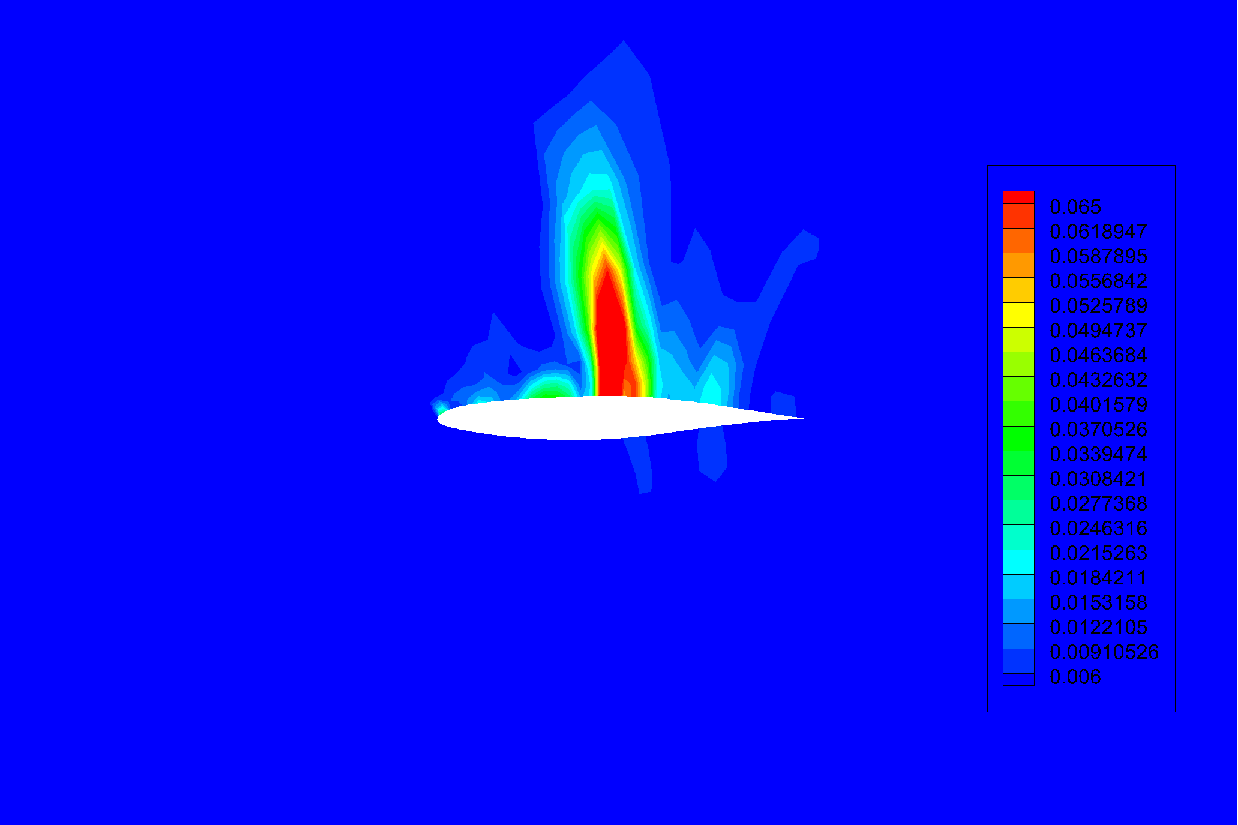}}
 \subfigure[model prediction of No.2]{
 \includegraphics[width=4.8cm]{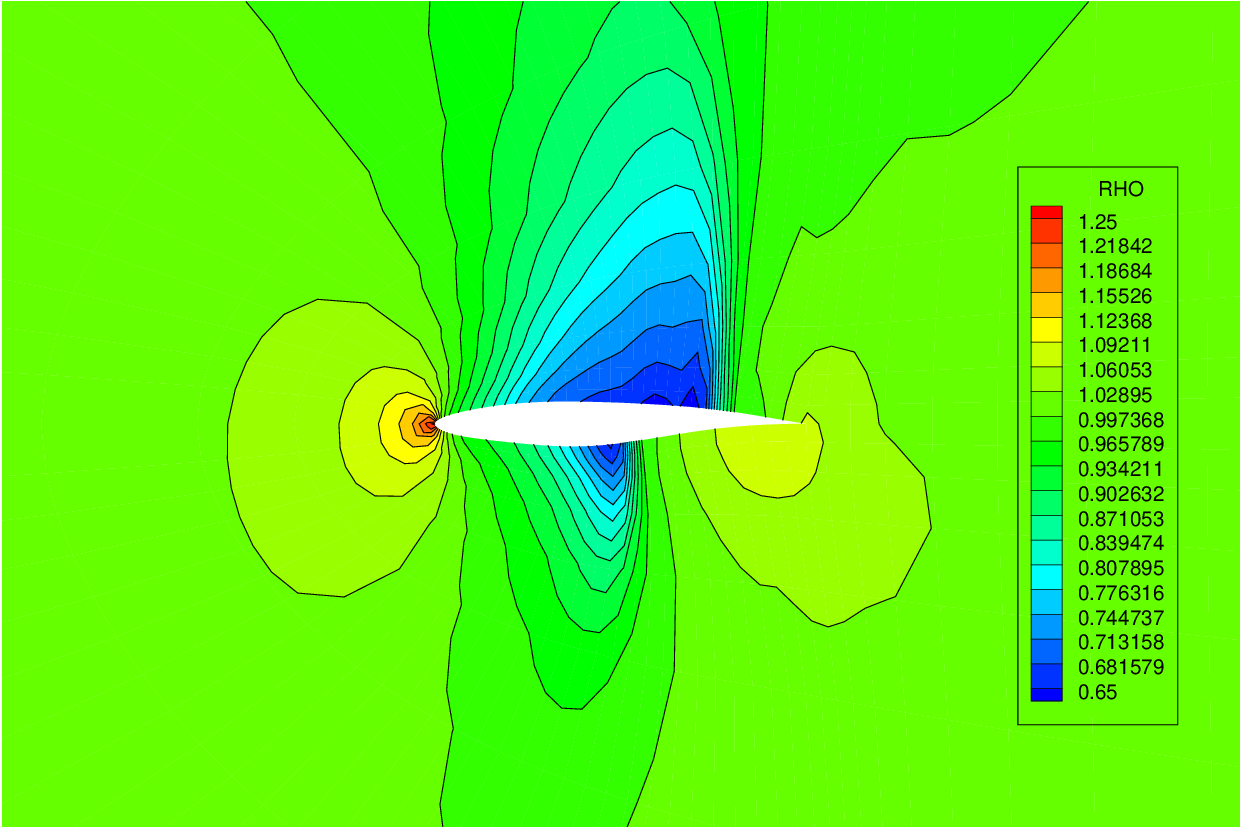}}
 \subfigure[CFD correction of No.2]{
 \includegraphics[width=4.8cm]{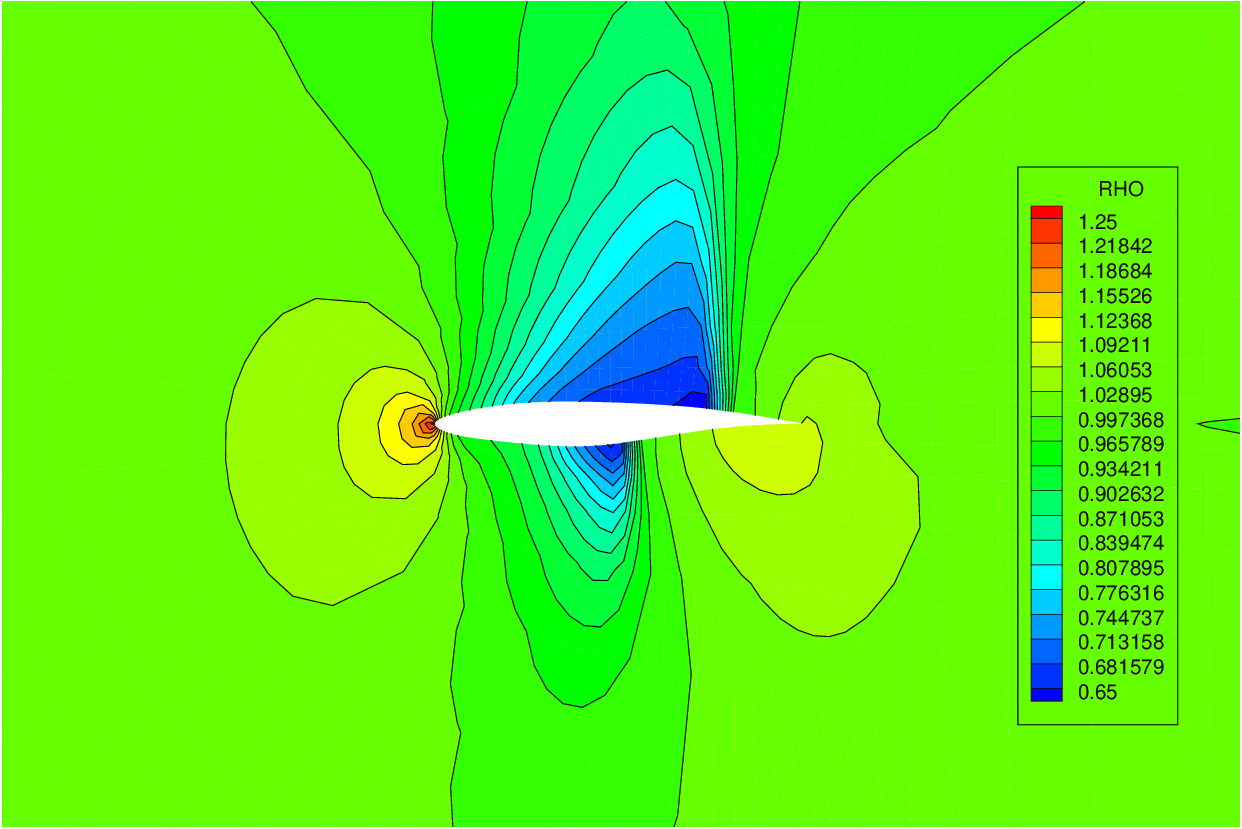}}
 \subfigure[relative error of No.2]{
 \includegraphics[width=4.8cm]{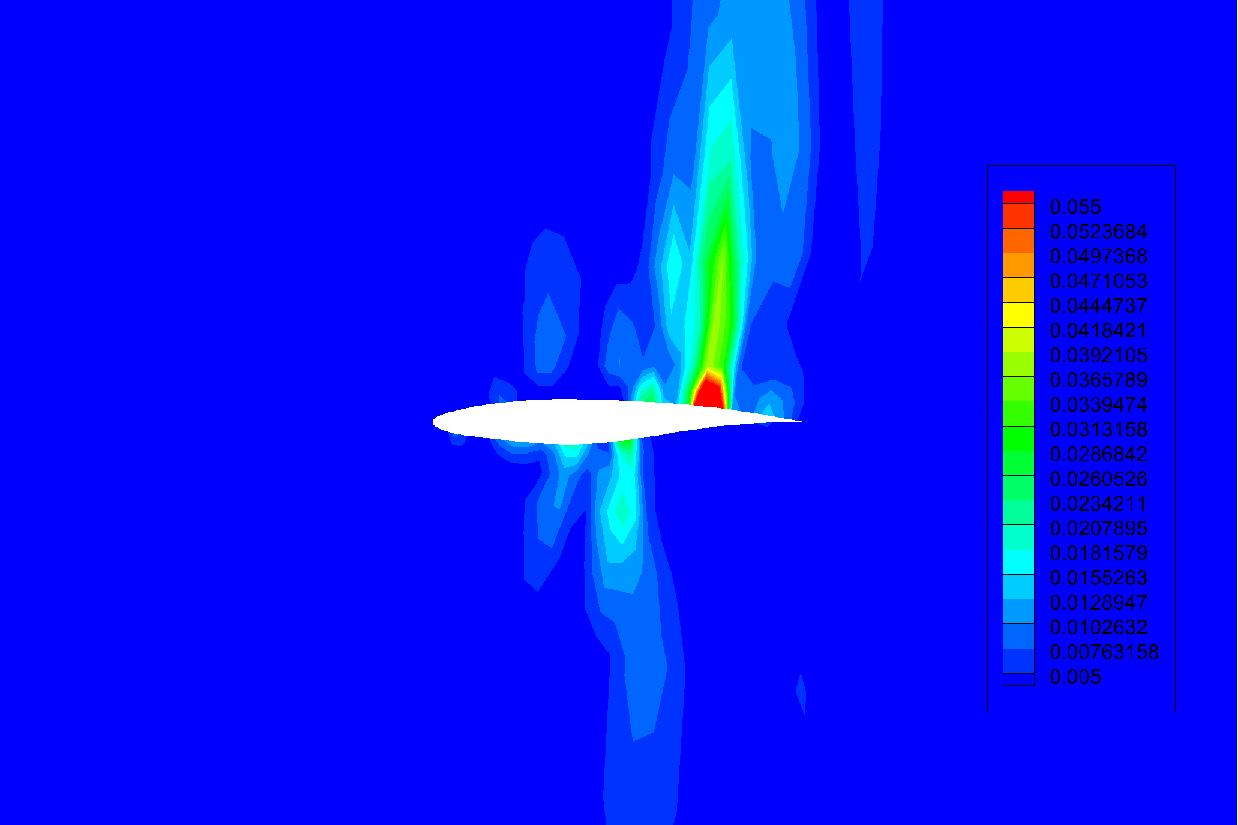}}
 \subfigure[model prediction of No.3]{
 \includegraphics[width=4.8cm]{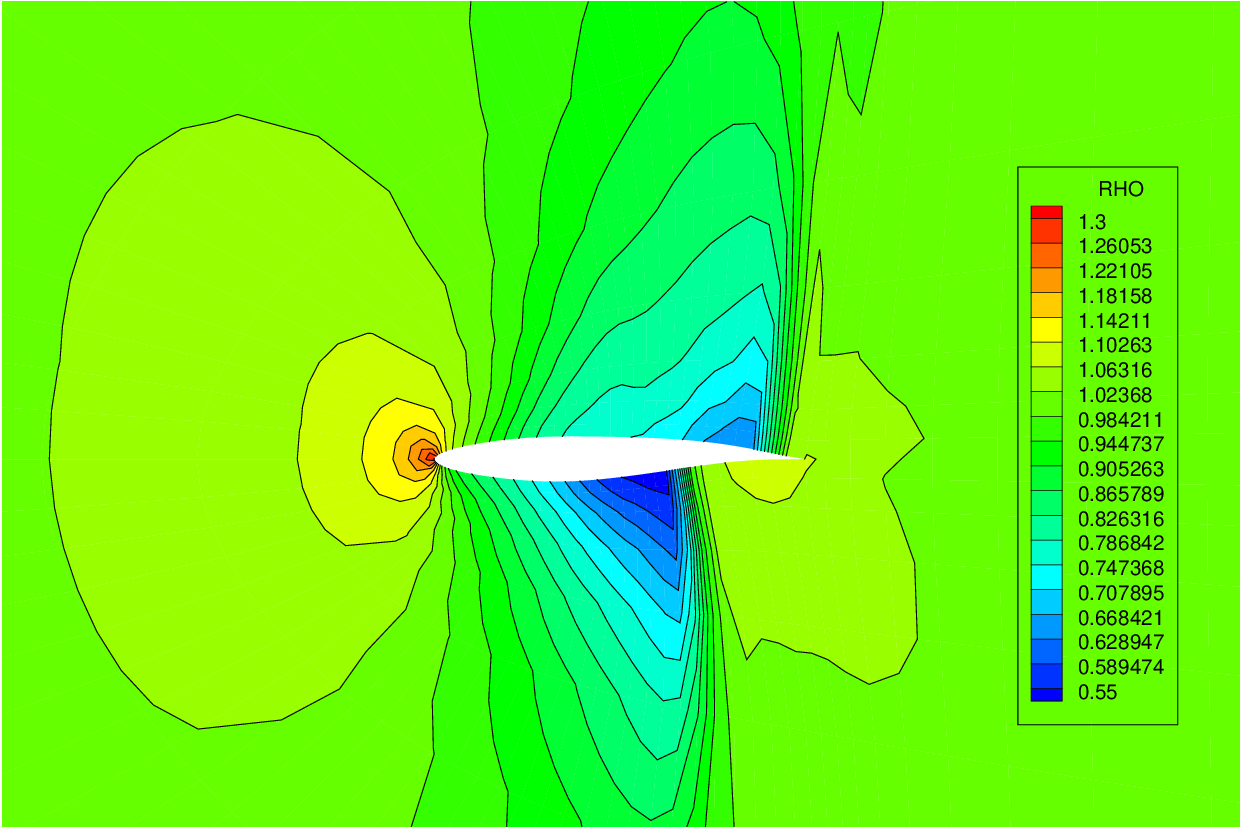}}
 \subfigure[CFD correction of No.3]{
 \includegraphics[width=4.8cm]{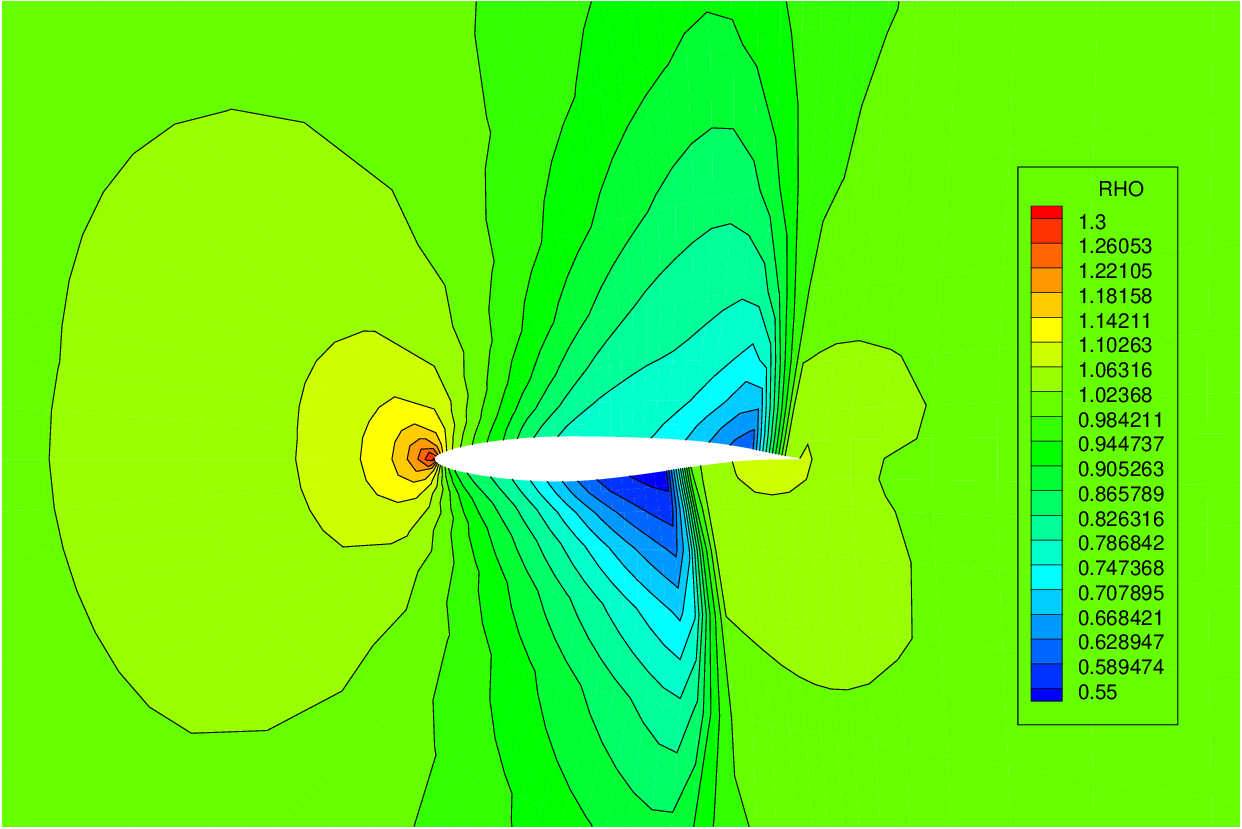}}
 \subfigure[relative error of No.3]{
 \includegraphics[width=4.8cm]{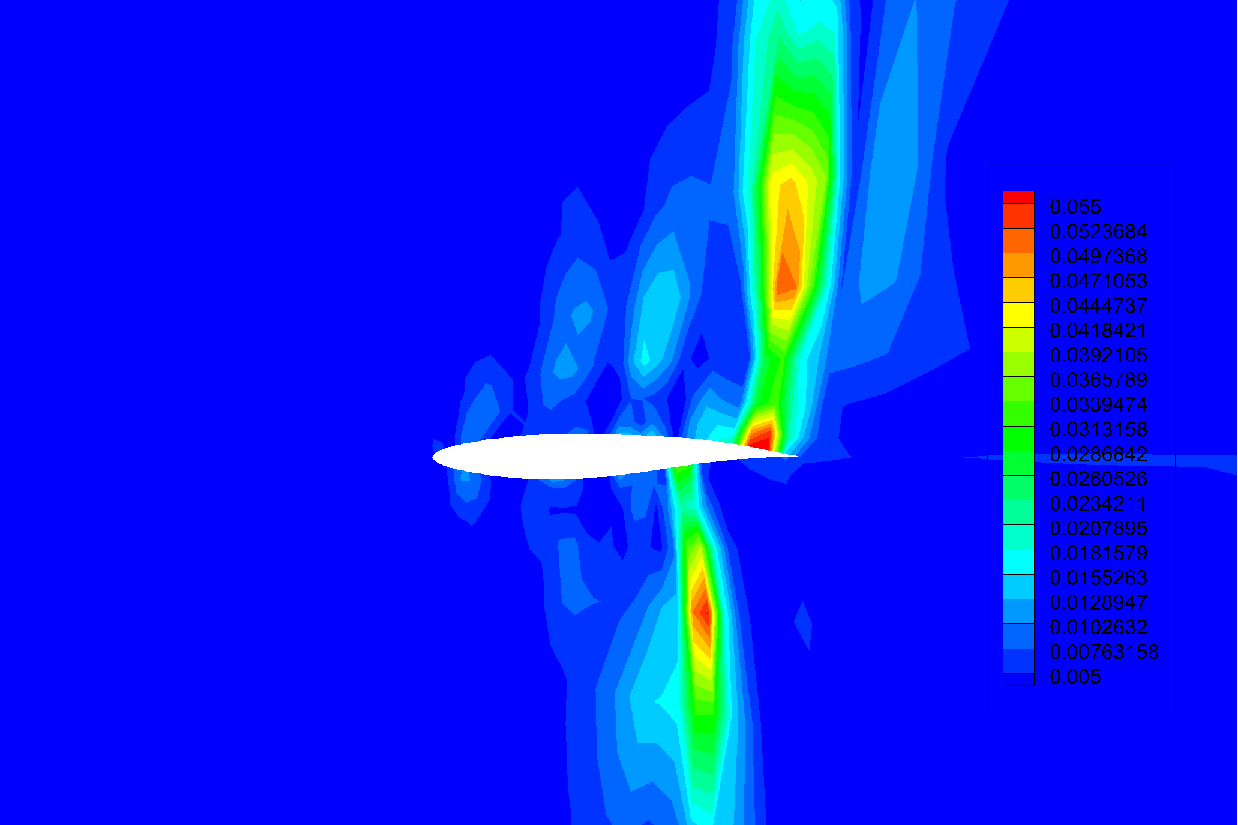}}
 \subfigure[model prediction of No.4]{
 \includegraphics[width=4.8cm]{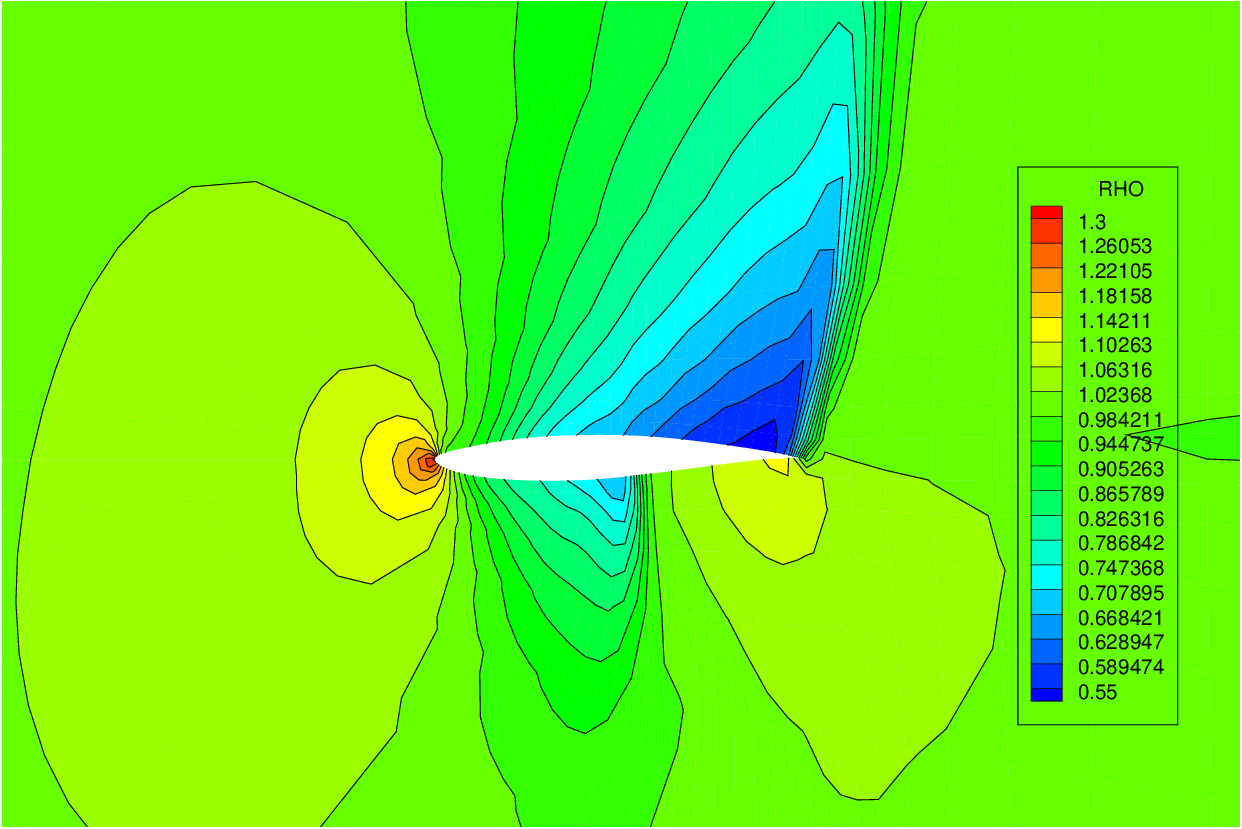}}
 \subfigure[CFD correction of No.4]{
 \includegraphics[width=4.8cm]{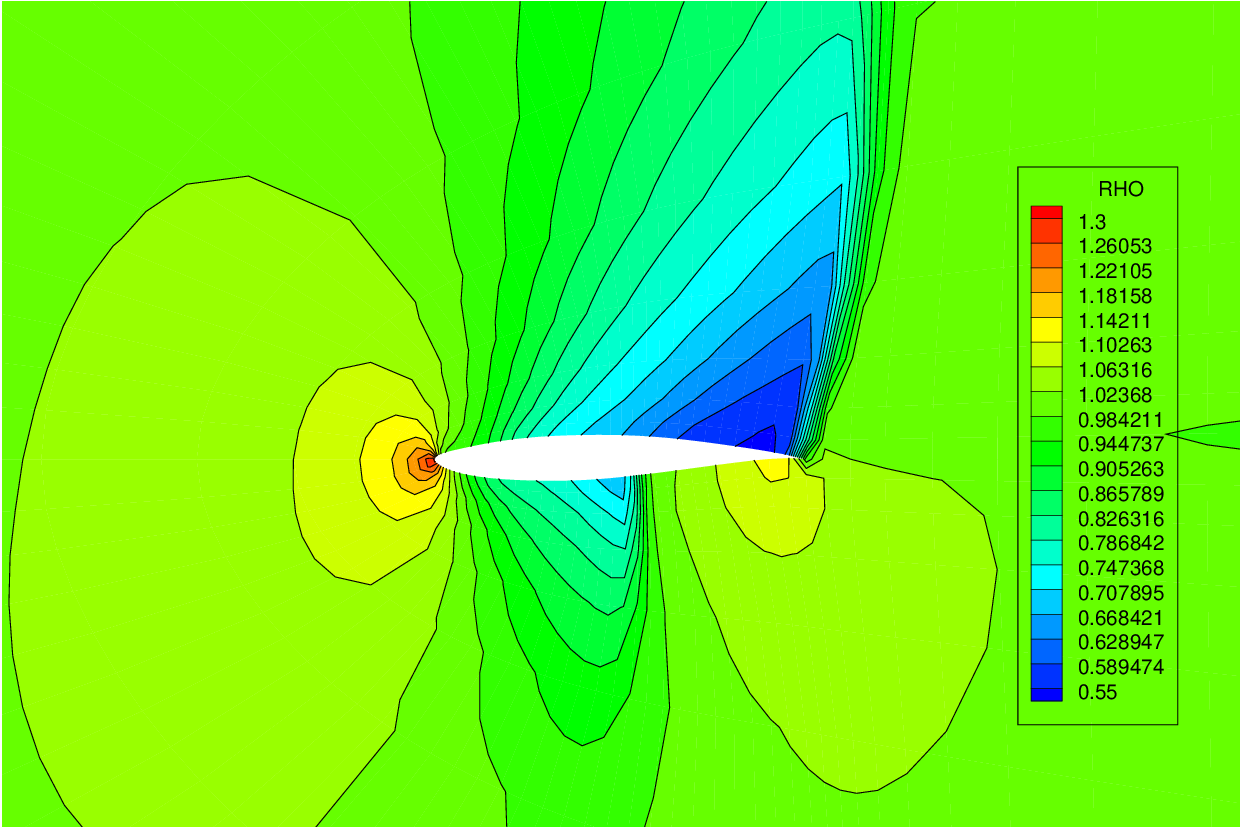}}
 \subfigure[relative error of No.4]{
 \includegraphics[width=4.8cm]{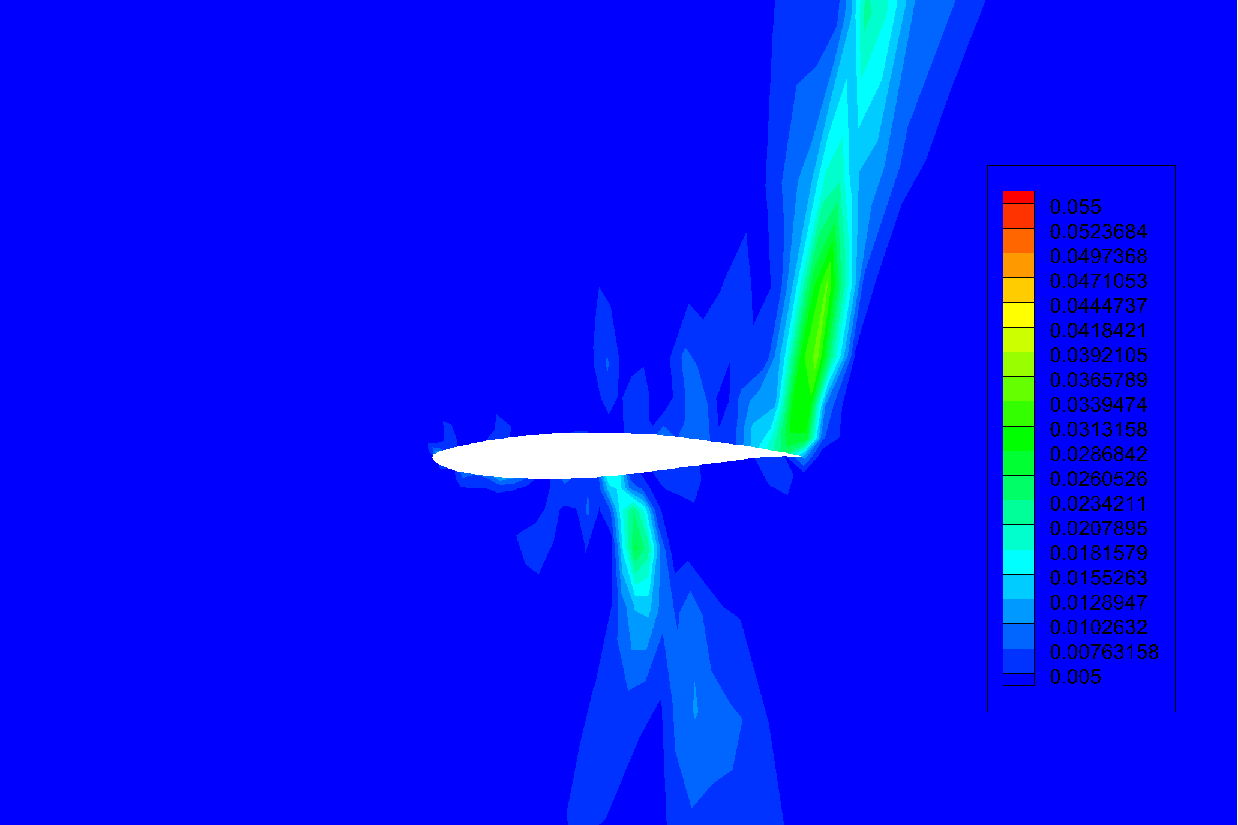}}
 \subfigure[model prediction of No.5]{
 \includegraphics[width=4.8cm]{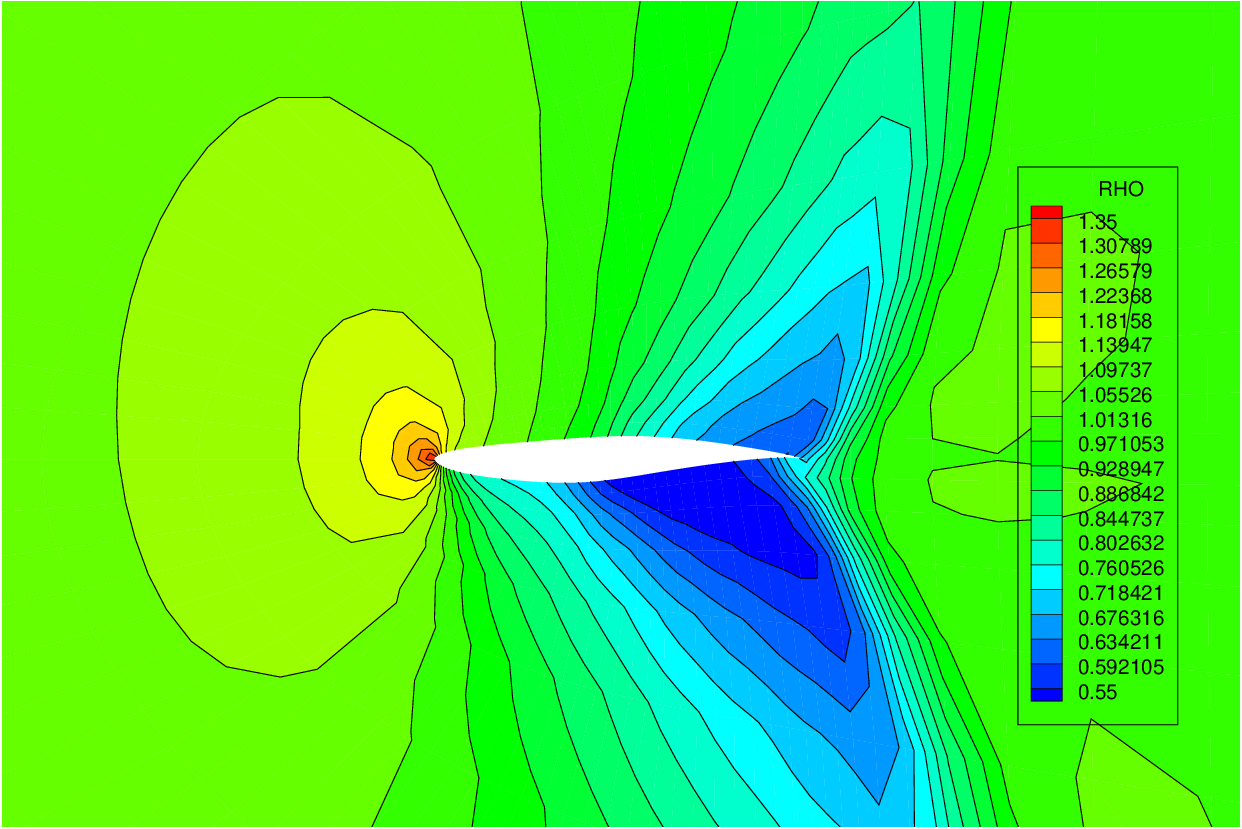}}
 \subfigure[CFD correction of No.5]{
 \includegraphics[width=4.8cm]{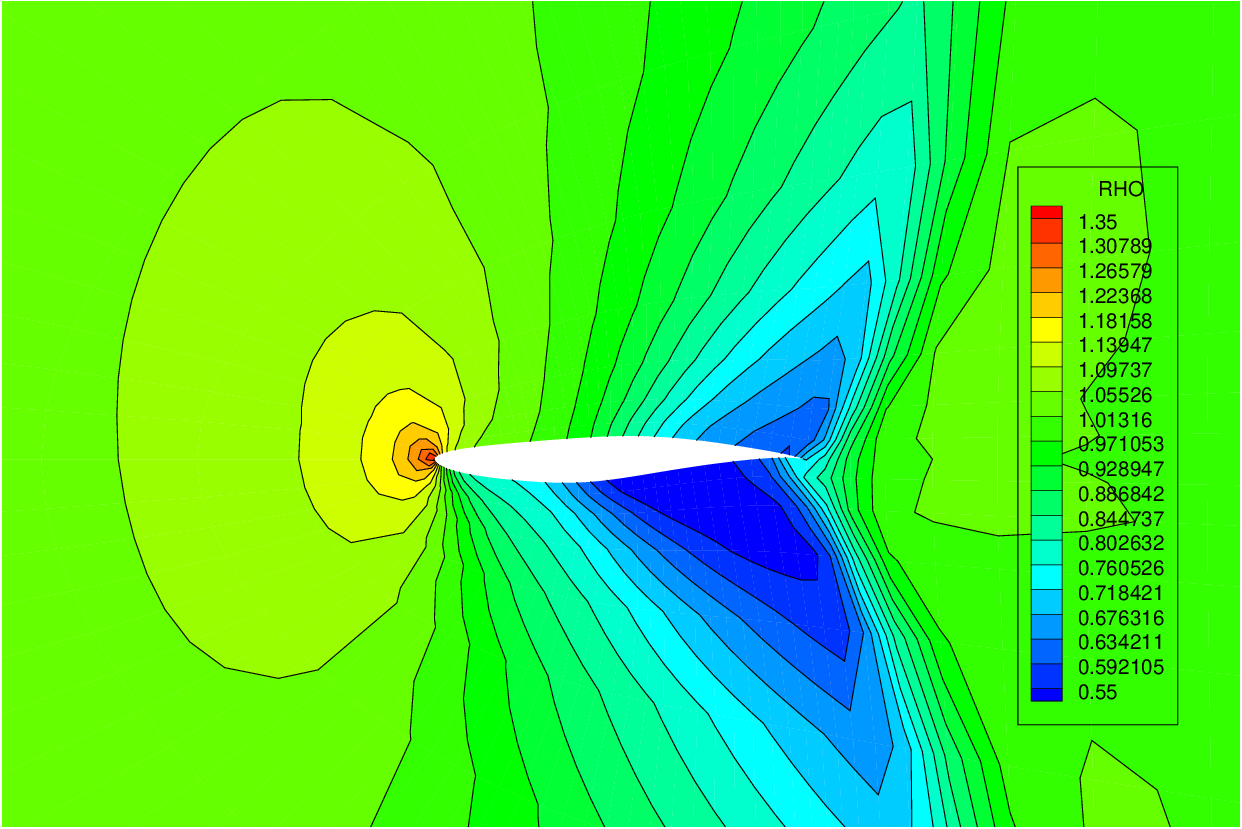}}
 \subfigure[relative error of No.5]{
 \includegraphics[width=4.8cm]{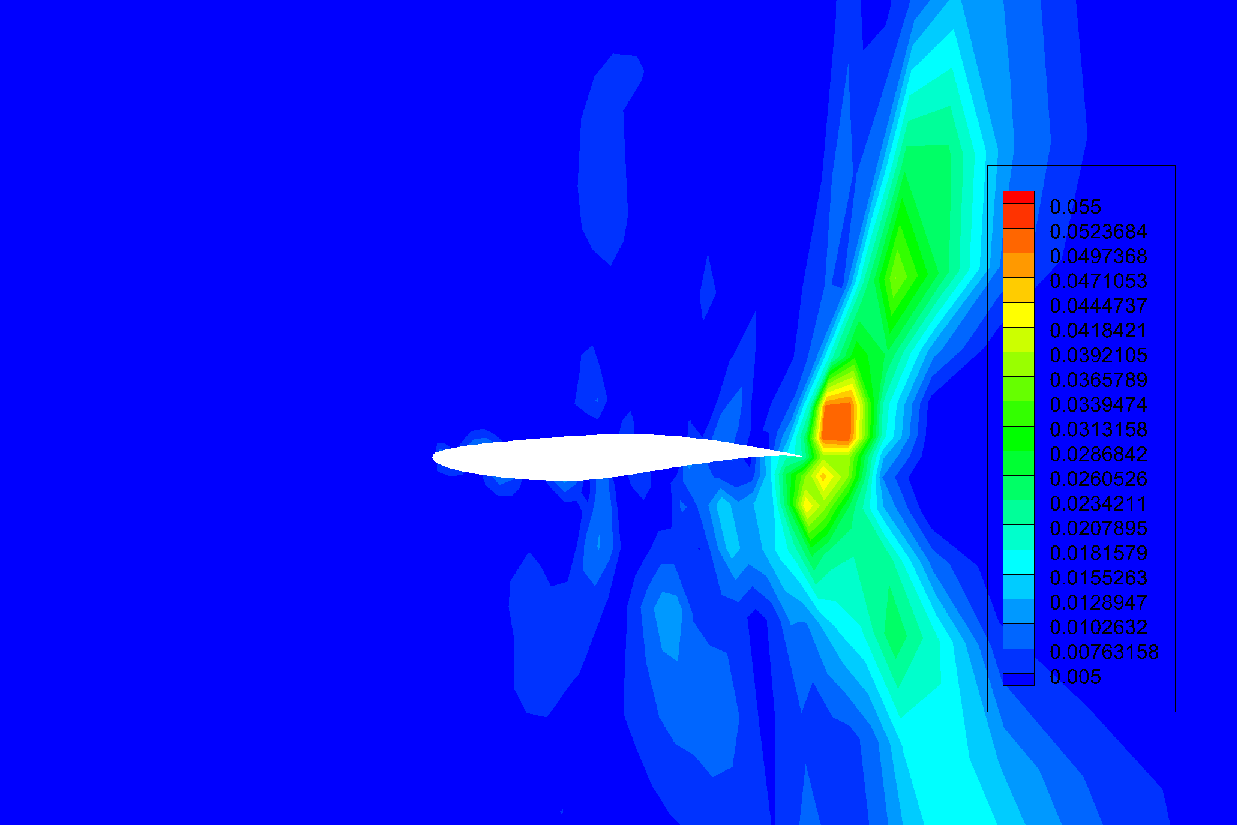}}
 \caption{Flow-field produced by Orig POD + Orig RBF and CFD in training set.}
 \label{fig:rae2822flow1}
\end{figure}

\begin{figure}[htbp]
 \centering
 \subfigure[model prediction of No.1]{
 \includegraphics[width=4.8cm]{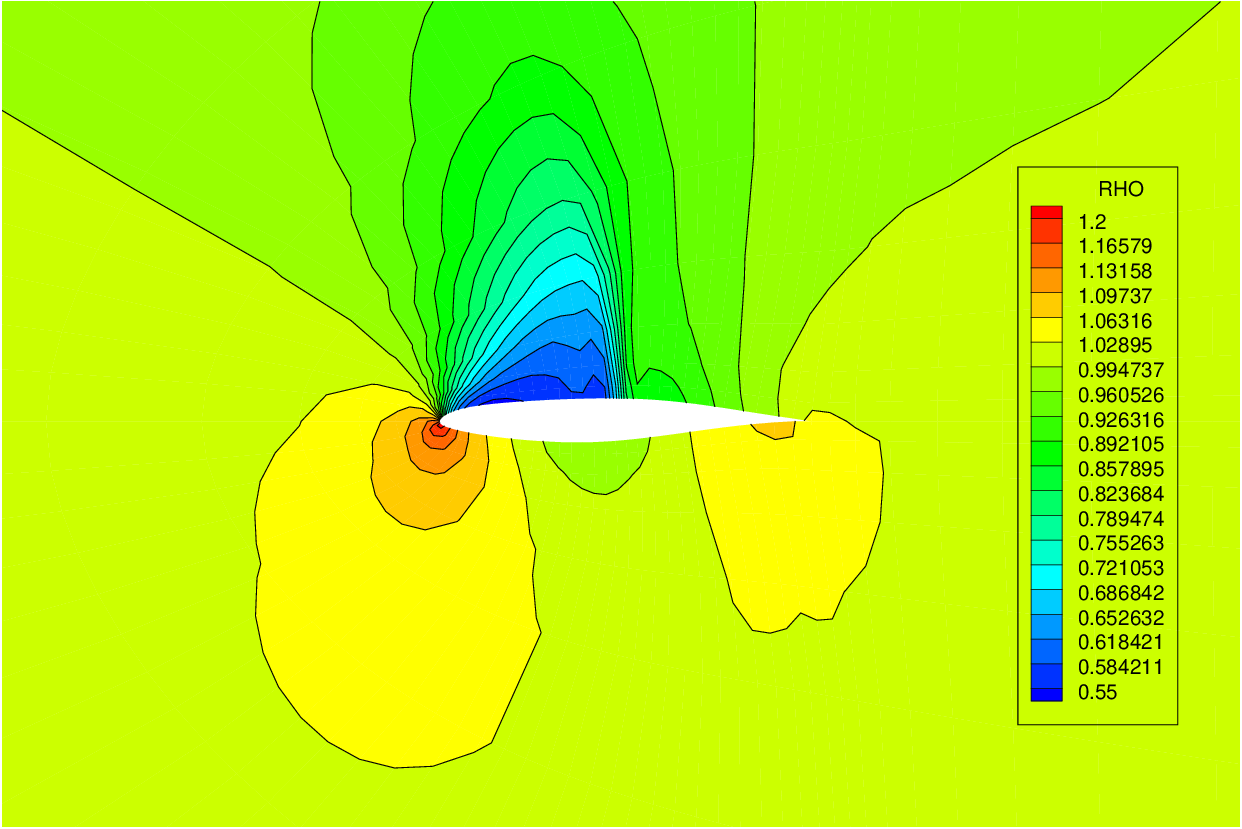}}
 \subfigure[CFD correction of No.1]{
 \includegraphics[width=4.8cm]{figure/flow-CFD-1.eps}}
 \subfigure[relative error of No.1]{
 \includegraphics[width=4.8cm]{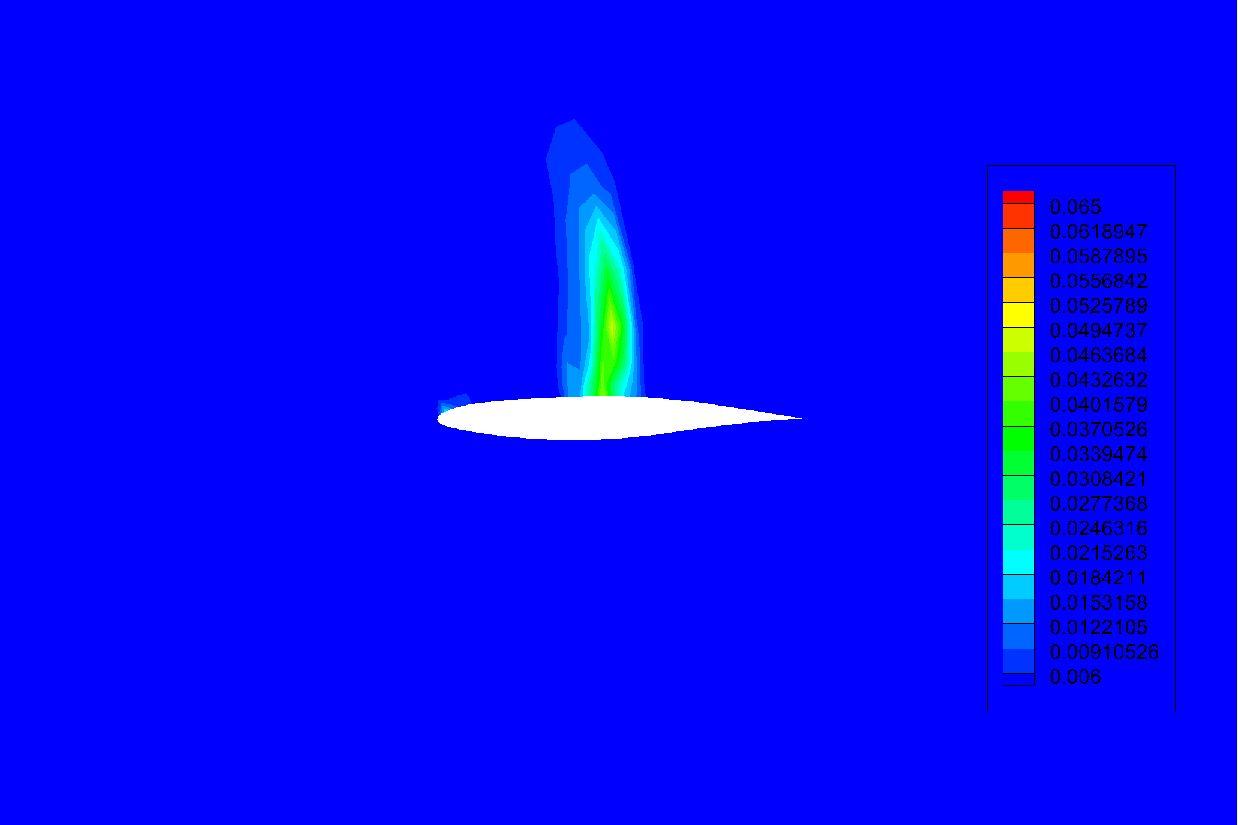}}
 \subfigure[model prediction of No.2]{
 \includegraphics[width=4.8cm]{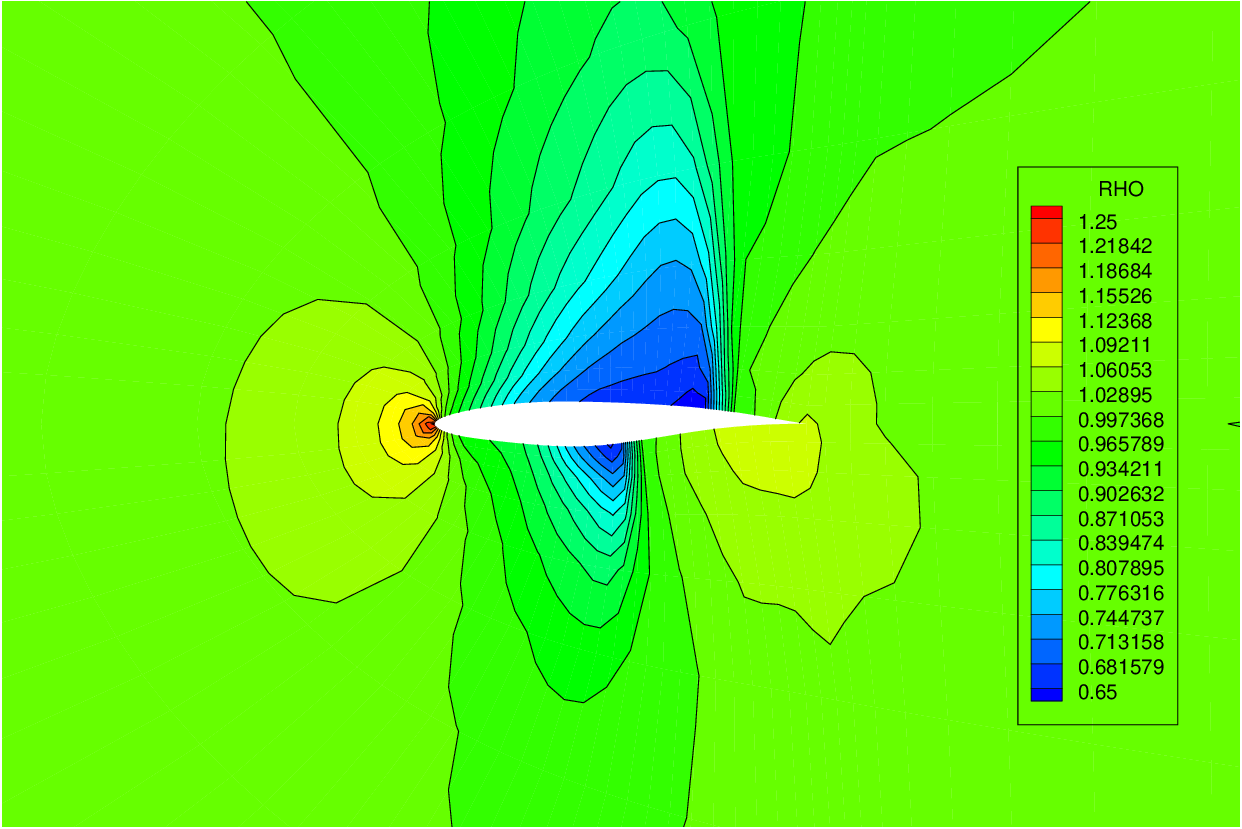}}
 \subfigure[CFD correction of No.2]{
 \includegraphics[width=4.8cm]{figure/flow-CFD-2.eps}}
 \subfigure[relative error of No.2]{
 \includegraphics[width=4.8cm]{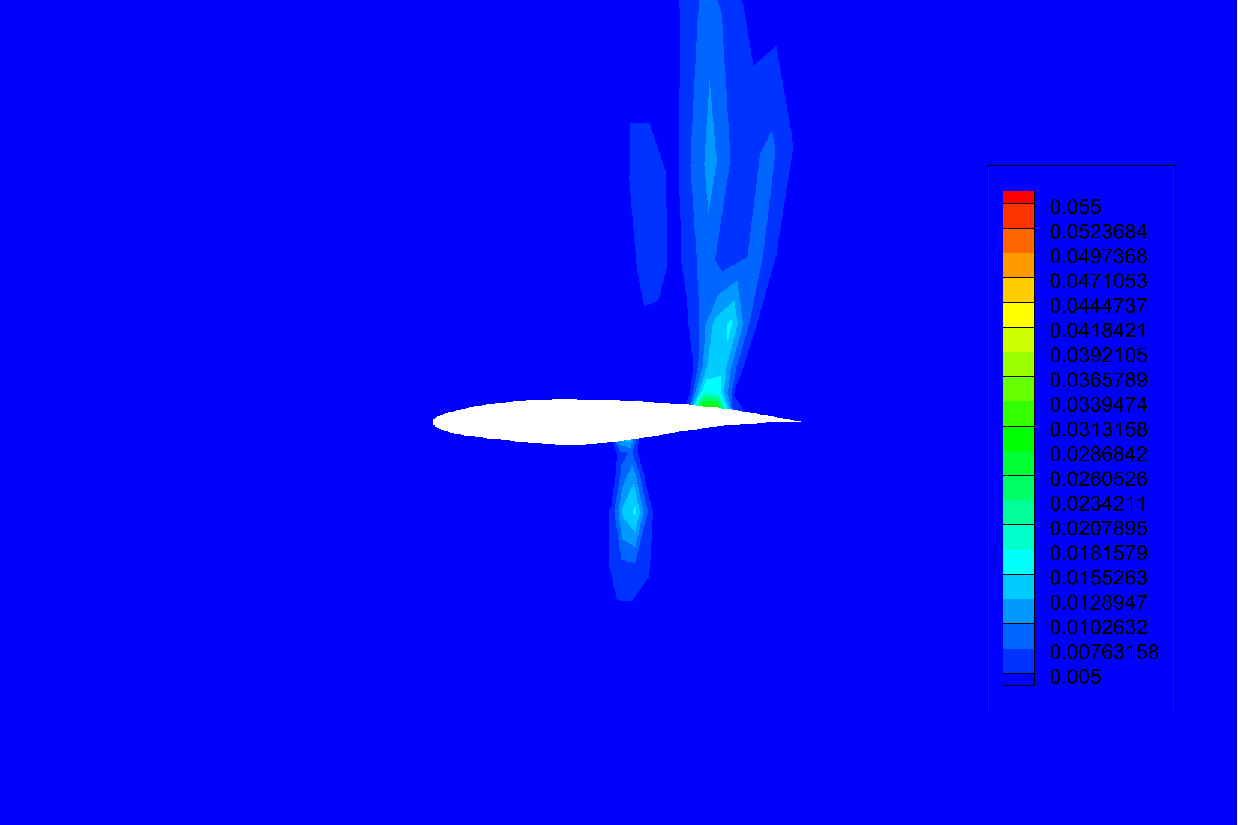}}
 \subfigure[model prediction of No.3]{
 \includegraphics[width=4.8cm]{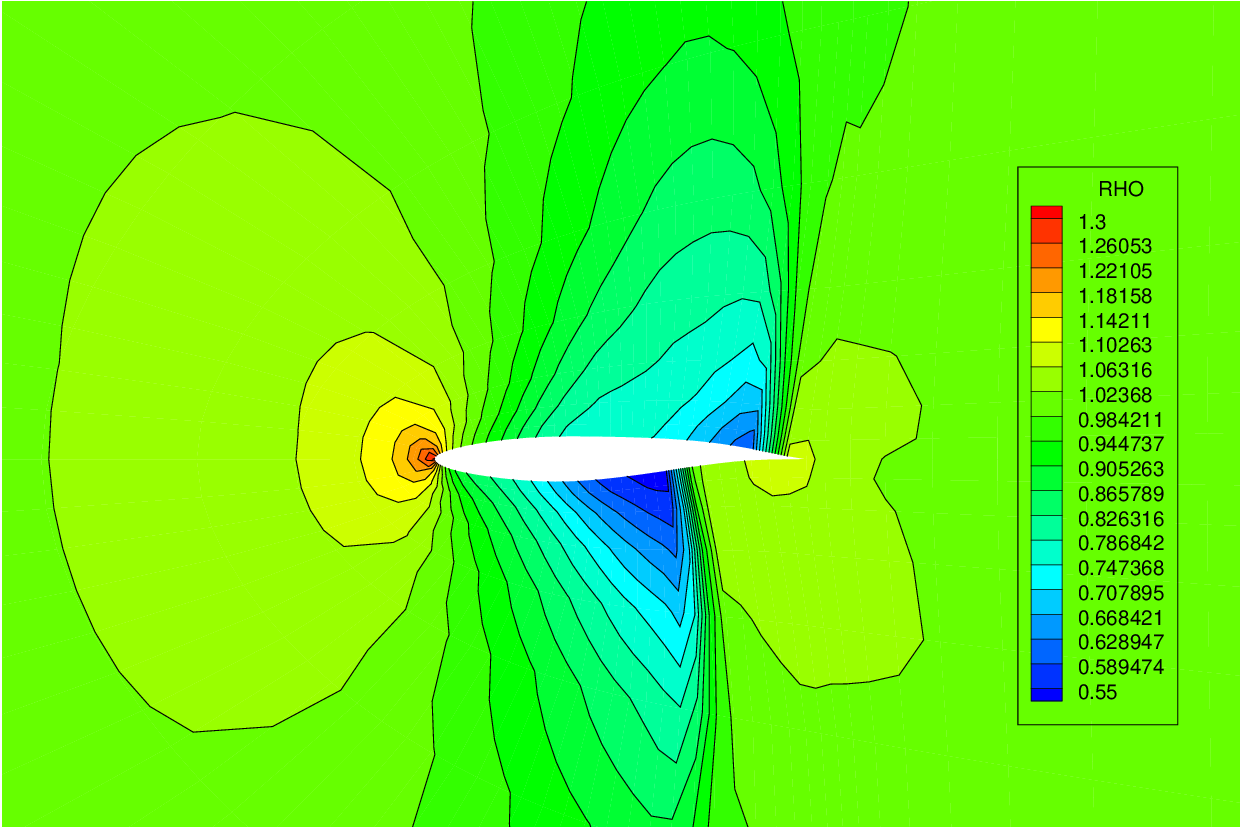}}
 \subfigure[CFD correction of No.3]{
 \includegraphics[width=4.8cm]{figure/flow-CFD-3.eps}}
 \subfigure[relative error of No.3]{
 \includegraphics[width=4.8cm]{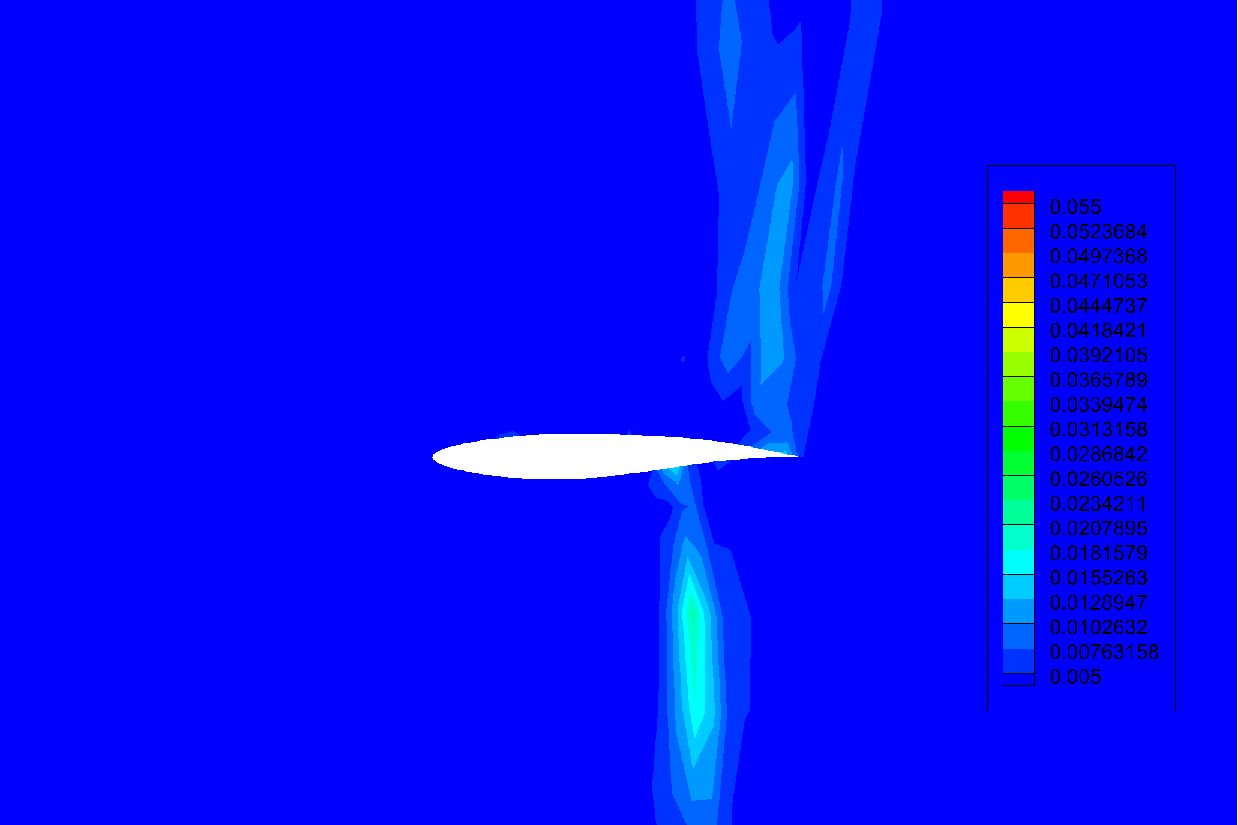}}
 \subfigure[model prediction of No.4]{
 \includegraphics[width=4.8cm]{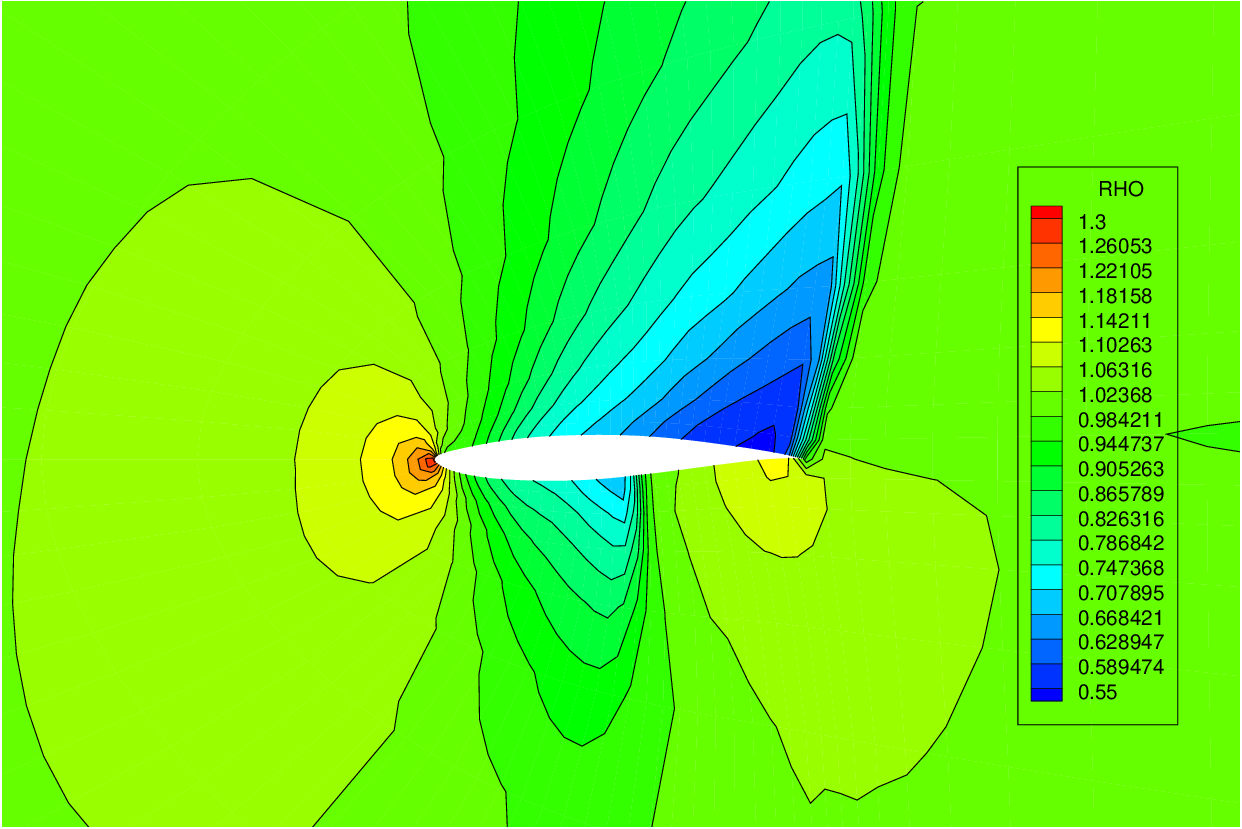}}
 \subfigure[CFD correction of No.4]{
 \includegraphics[width=4.8cm]{figure/flow-CFD-4.eps}}
 \subfigure[relative error of No.4]{
 \includegraphics[width=4.8cm]{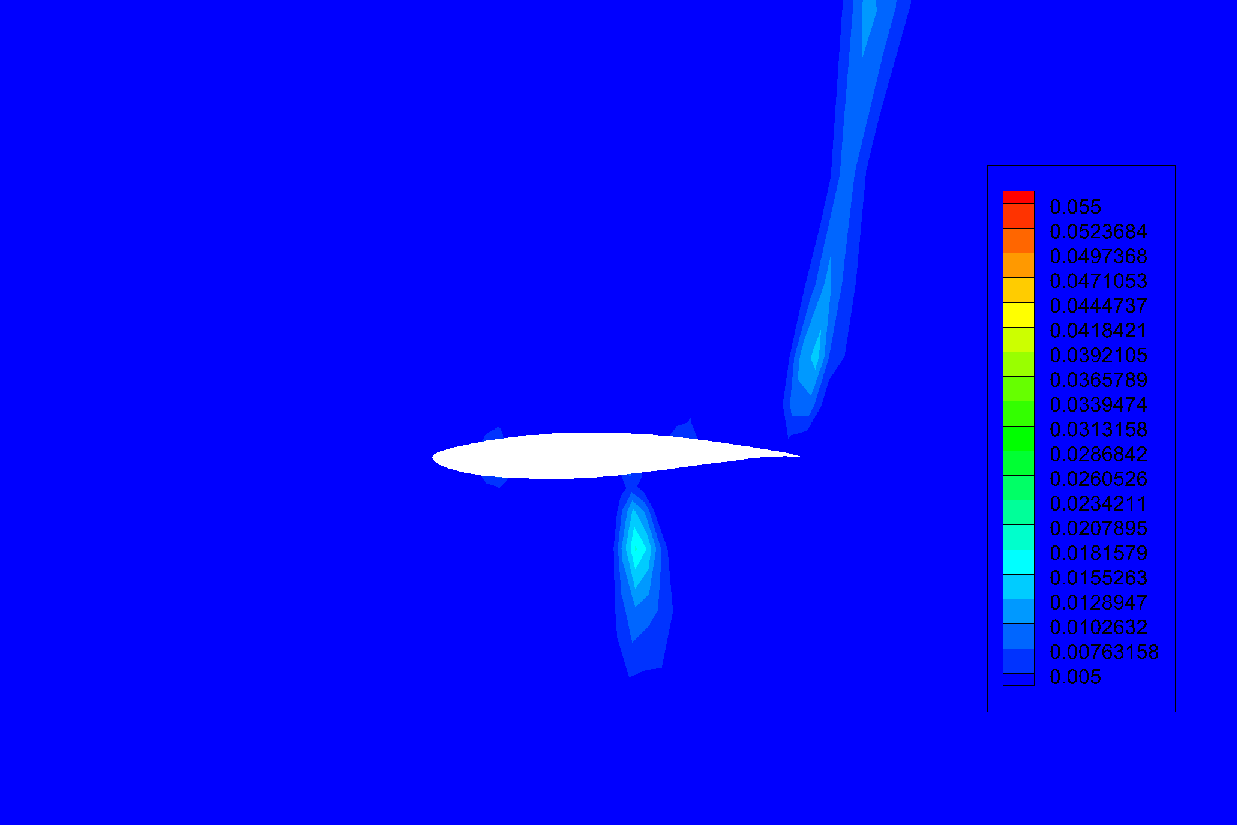}}
 \subfigure[model prediction of No.5]{
 \includegraphics[width=4.8cm]{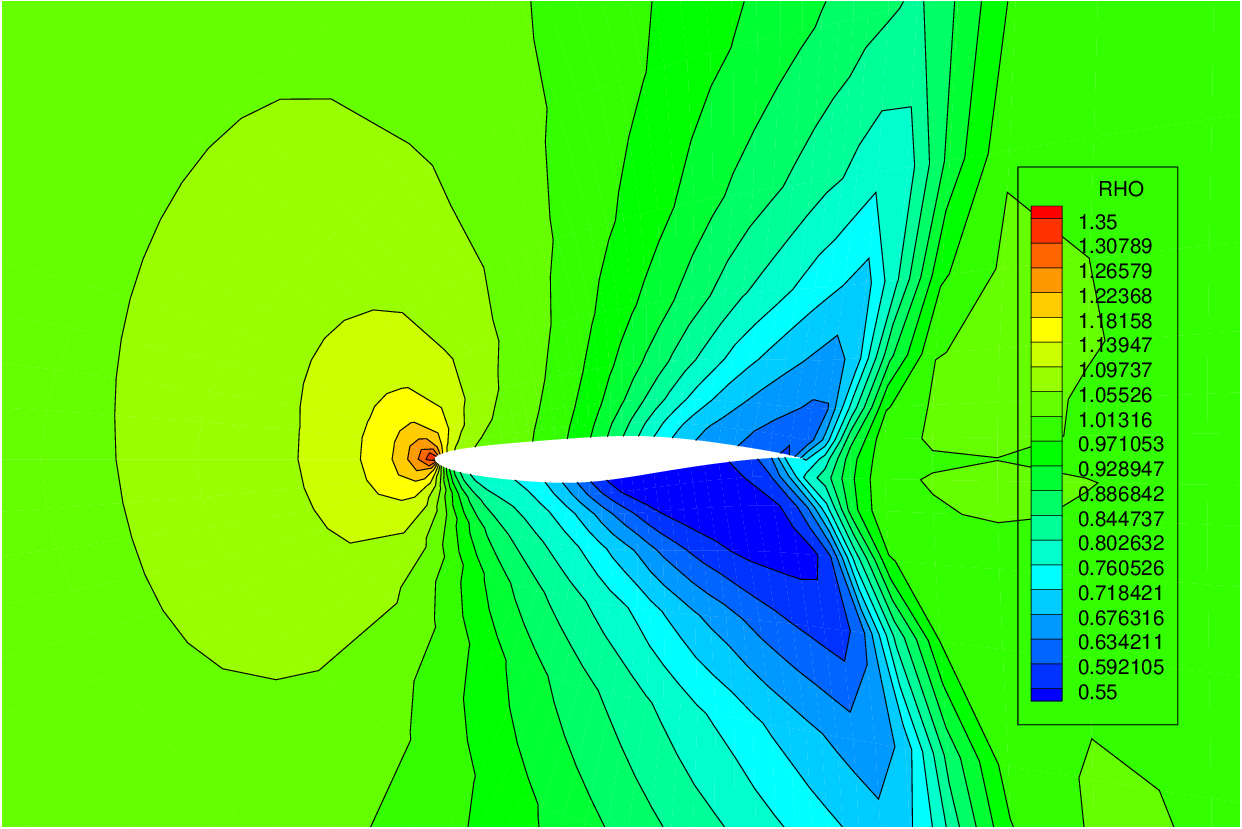}}
 \subfigure[CFD correction of No.5]{
 \includegraphics[width=4.8cm]{figure/flow-CFD-5.eps}}
 \subfigure[relative error of No.5]{
 \includegraphics[width=4.8cm]{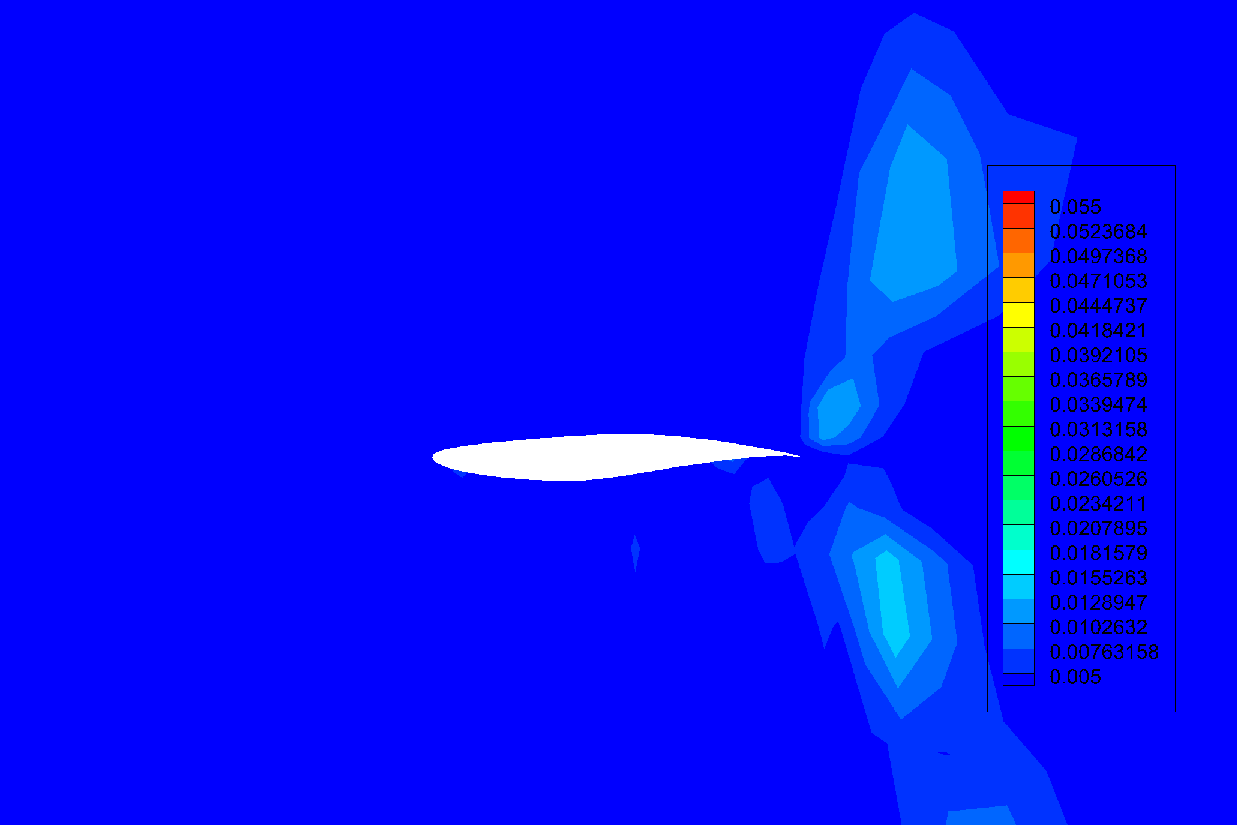}}
 \caption{Flow-field produced by Zonal POD + Orig RBF and CFD in training set.}
 \label{fig:rae2822flow2}
\end{figure}

\begin{figure}[htbp]
 \centering
 \subfigure[$C_p$ curves of No.1]{
 \includegraphics[width=7cm]{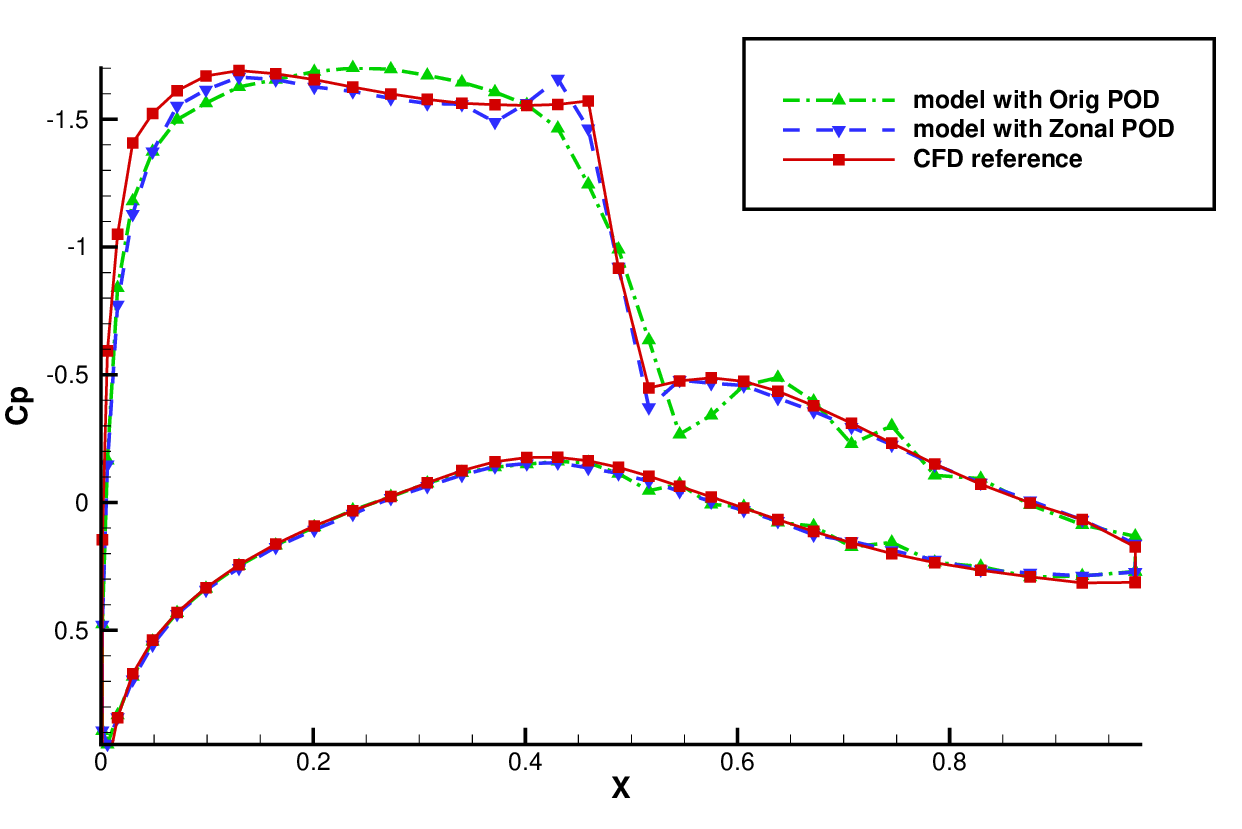}}
 \subfigure[$C_p$ curves of No.2]{
 \includegraphics[width=7cm]{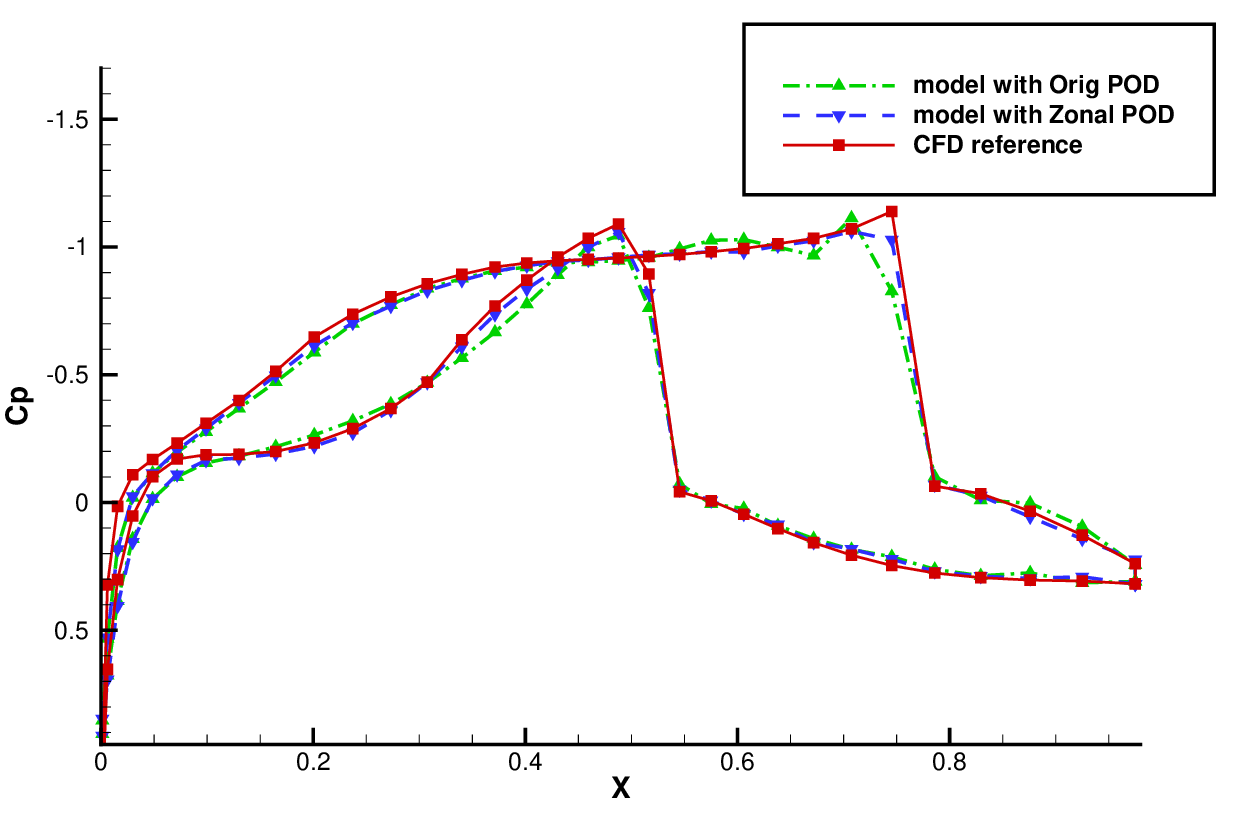}}
 \subfigure[$C_p$ curves of No.3]{
 \includegraphics[width=7cm]{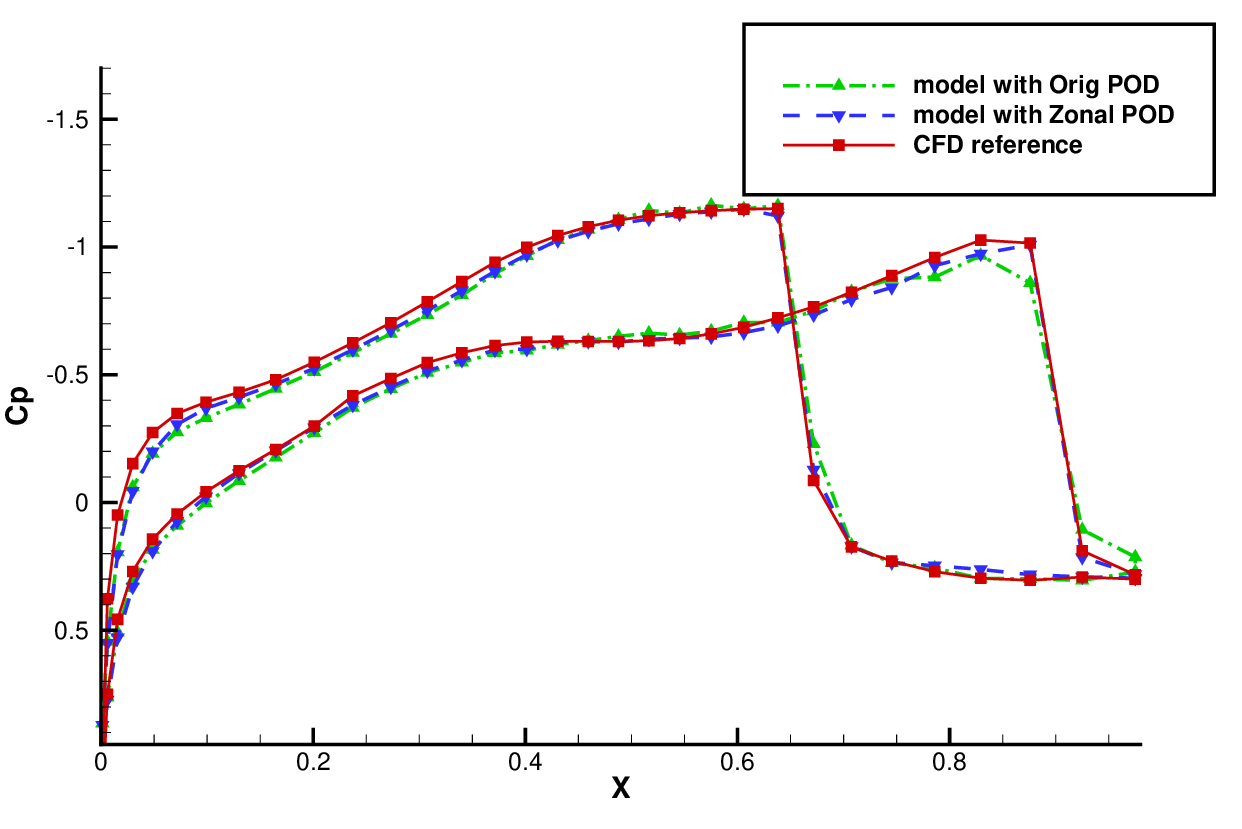}}
 \subfigure[$C_p$ curves of No.4]{
 \includegraphics[width=7cm]{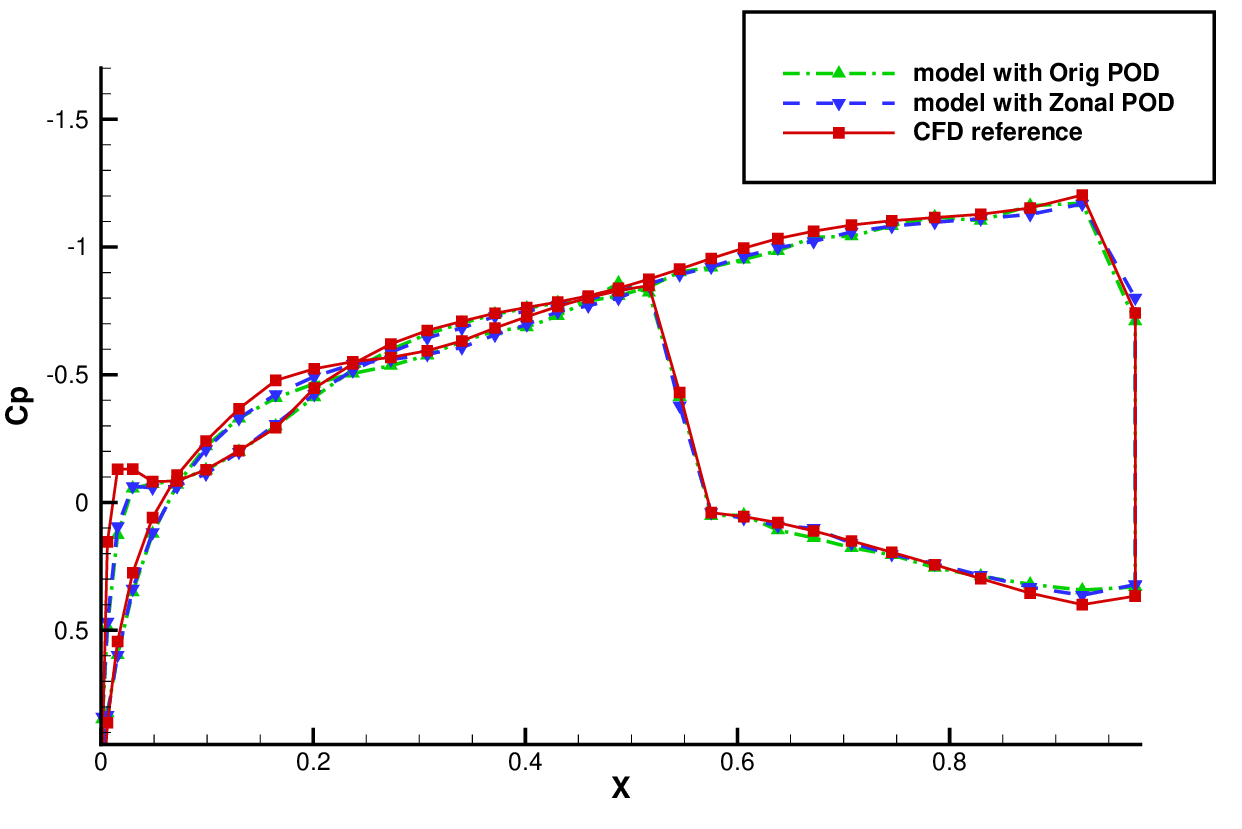}}
 \subfigure[$C_p$ curves of No.5]{
 \includegraphics[width=7cm]{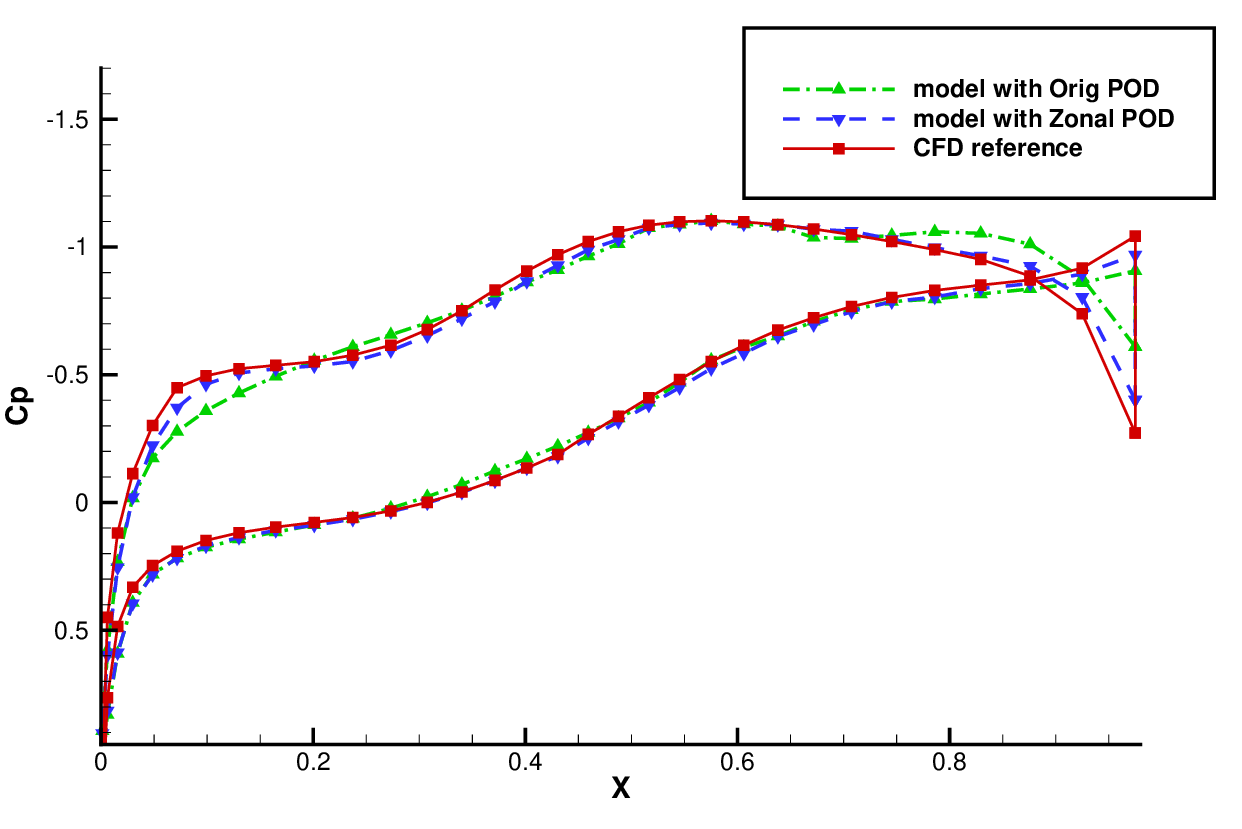}}
 \caption{$C_p$ distribution produced by data-driven models and CFD in training set.}
 \label{fig:rae2822Cp1}
\end{figure}

Figure~\ref{fig:rae2822flow1} and \ref{fig:rae2822flow2} show flow-field produced by different data-driven models and CFD, and the relative error between model prediction and CFD correction is provided as well. Figure~\ref{fig:rae2822Cp1} compares the difference of $C_p$ distribution among different models and CFD.

Based on the dimensionality reduction effects compared in Table~\ref{tab:POD}, it can be summarized from Figures~\ref{fig:rae2822flow1} and \ref{fig:rae2822flow2} that the Zonal POD is capable of using comparable reduced dimensions to capture more precise flow structures near the wall as compared with Orig POD. Especially in the vicinity of shock waves, the predicted flow-field via Zonal POD exhibits significantly smaller relative errors (almost all less than $5\%$) compared to that with Orig POD. Similar behaviors can be observed from the $C_p$ curves shown in Figure~\ref{fig:rae2822Cp1}.

\subsection{Prediction effect on validation dataset}
\qquad This part investigates the predictive performance on validation set. Similarly, we randomly select 4 sample points from different Mach number ranges in the validation set, and present the results (including the flow-field and the $C_p$ distribution) produced by different strategies and CFD at these sample points.

\begin{figure}[htbp]
 \centering
 \subfigure[model prediction of No.1]{
 \includegraphics[width=4.8cm]{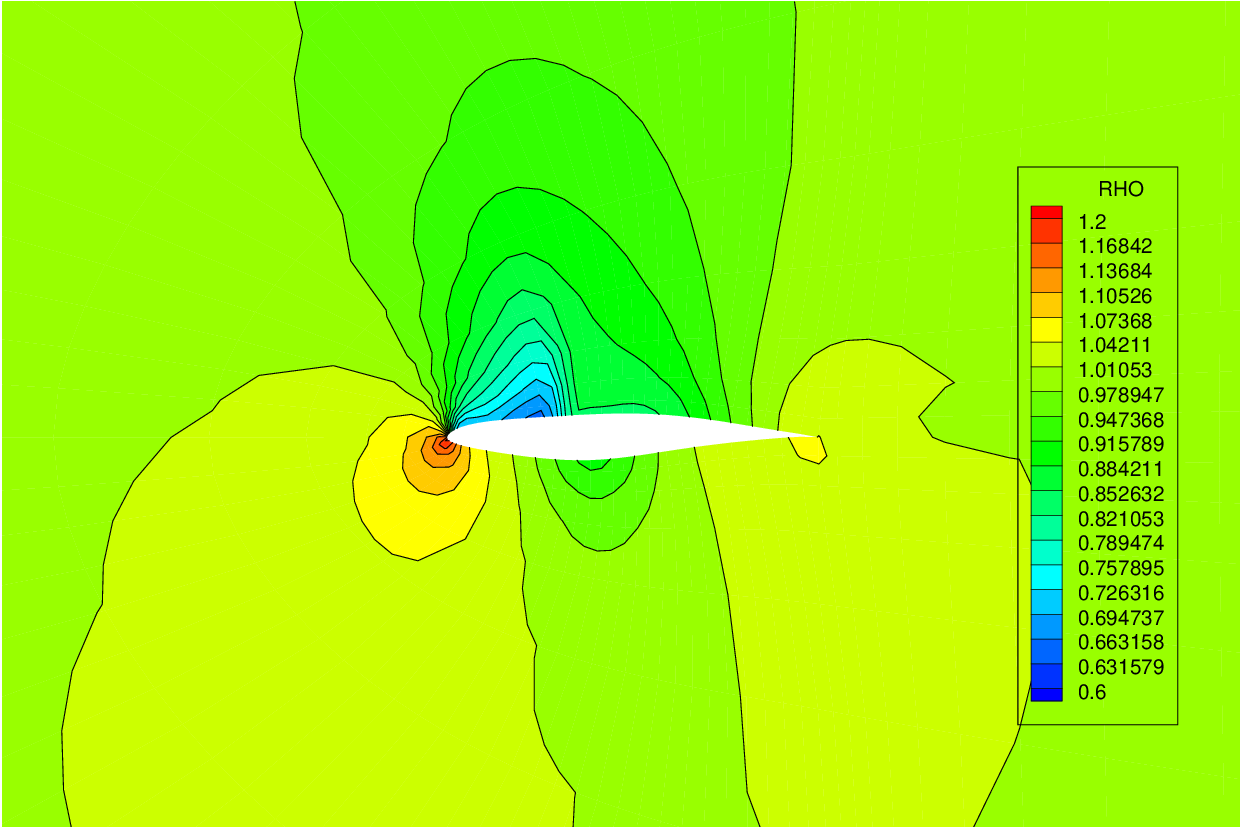}}
 \subfigure[CFD correction of No.1]{
 \includegraphics[width=4.8cm]{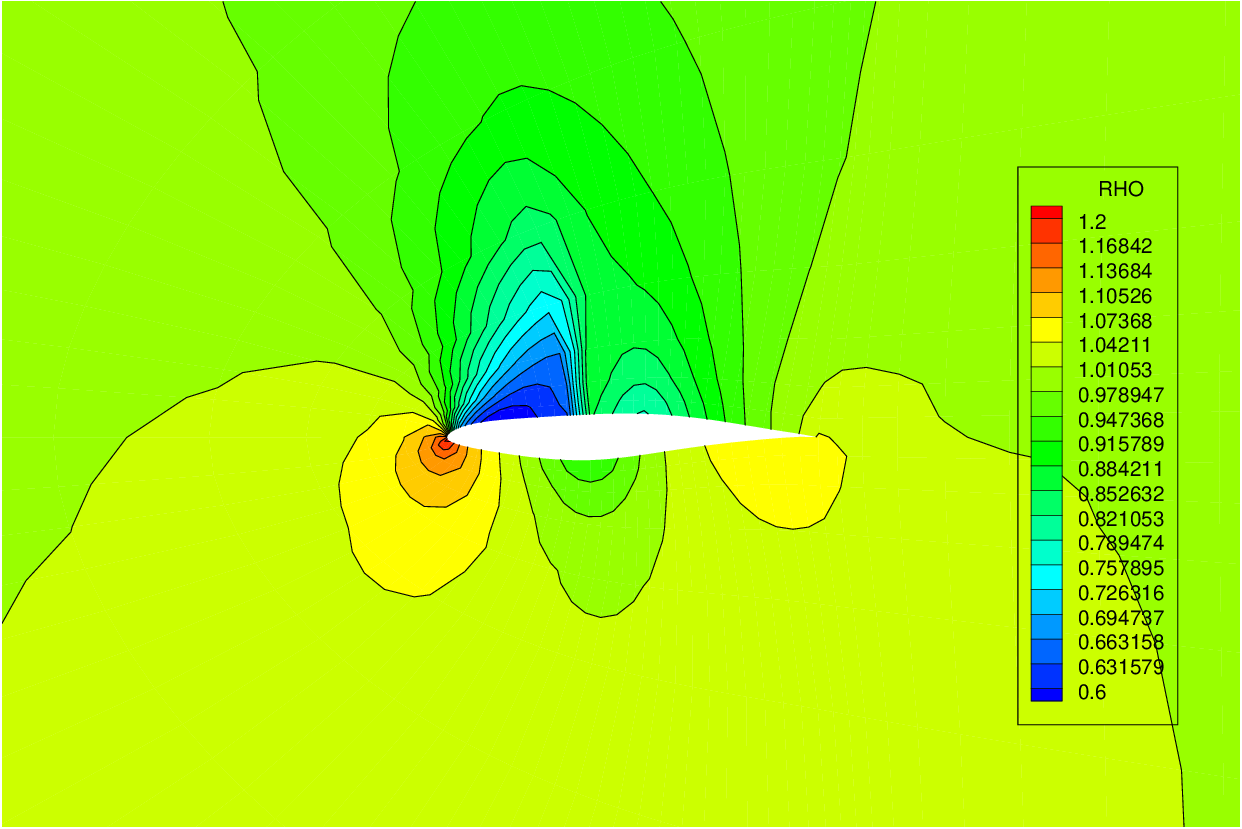}}
 \subfigure[relative error of No.1]{
 \includegraphics[width=4.8cm]{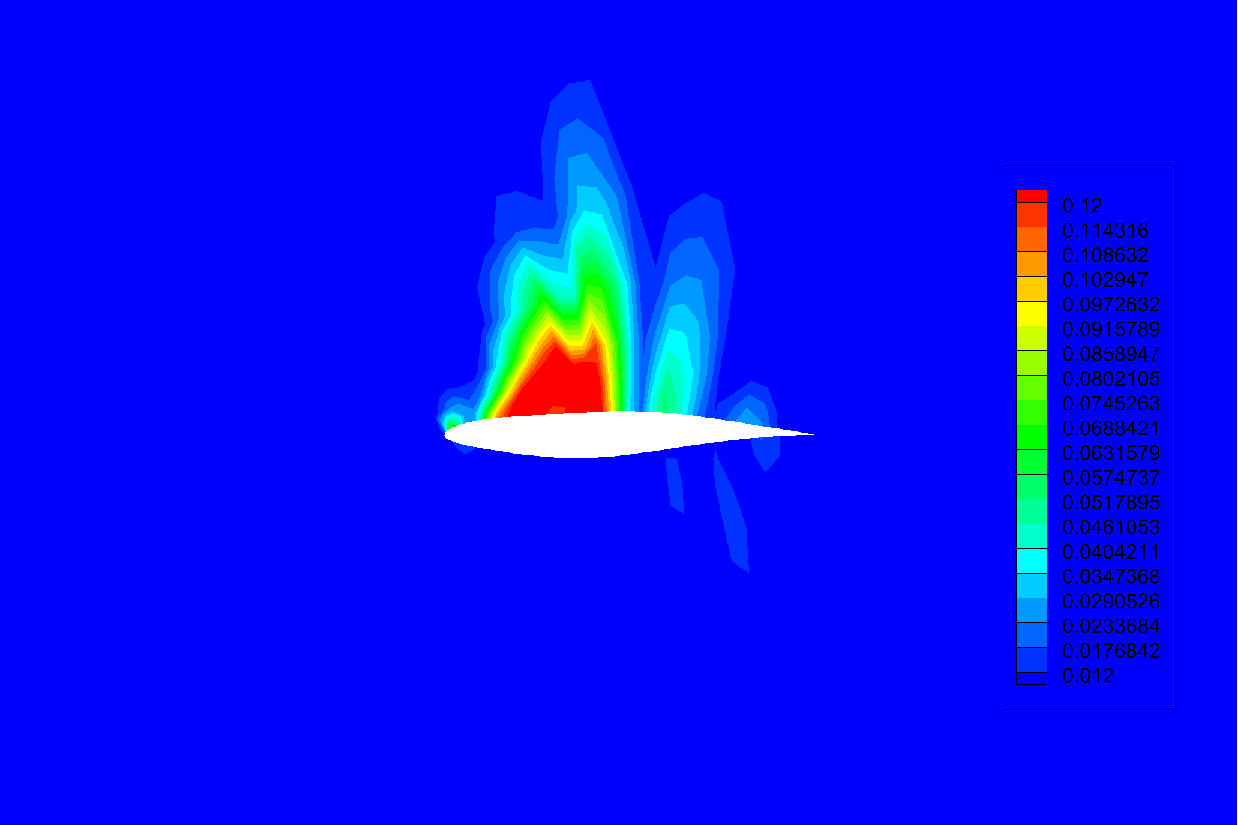}}
 \subfigure[model prediction of No.2]{
 \includegraphics[width=4.8cm]{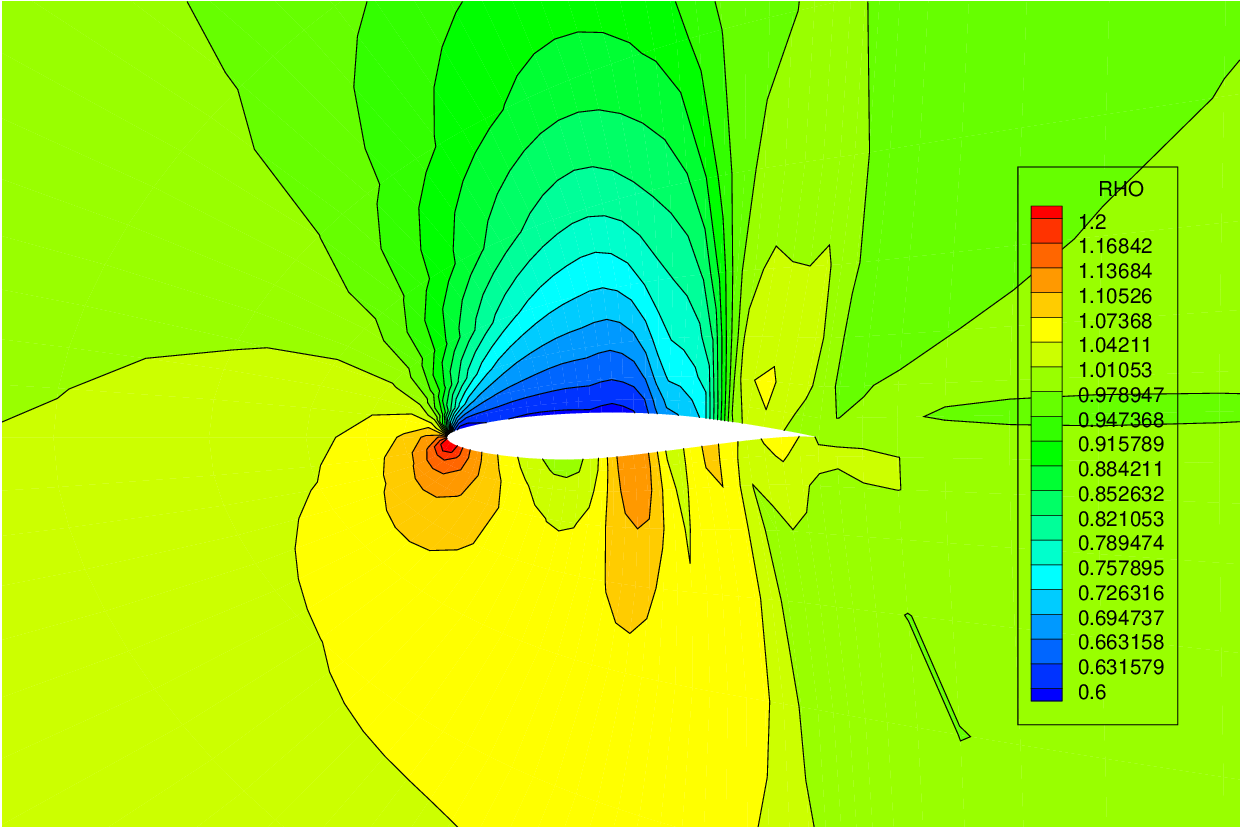}}
 \subfigure[CFD correction of No.2]{
 \includegraphics[width=4.8cm]{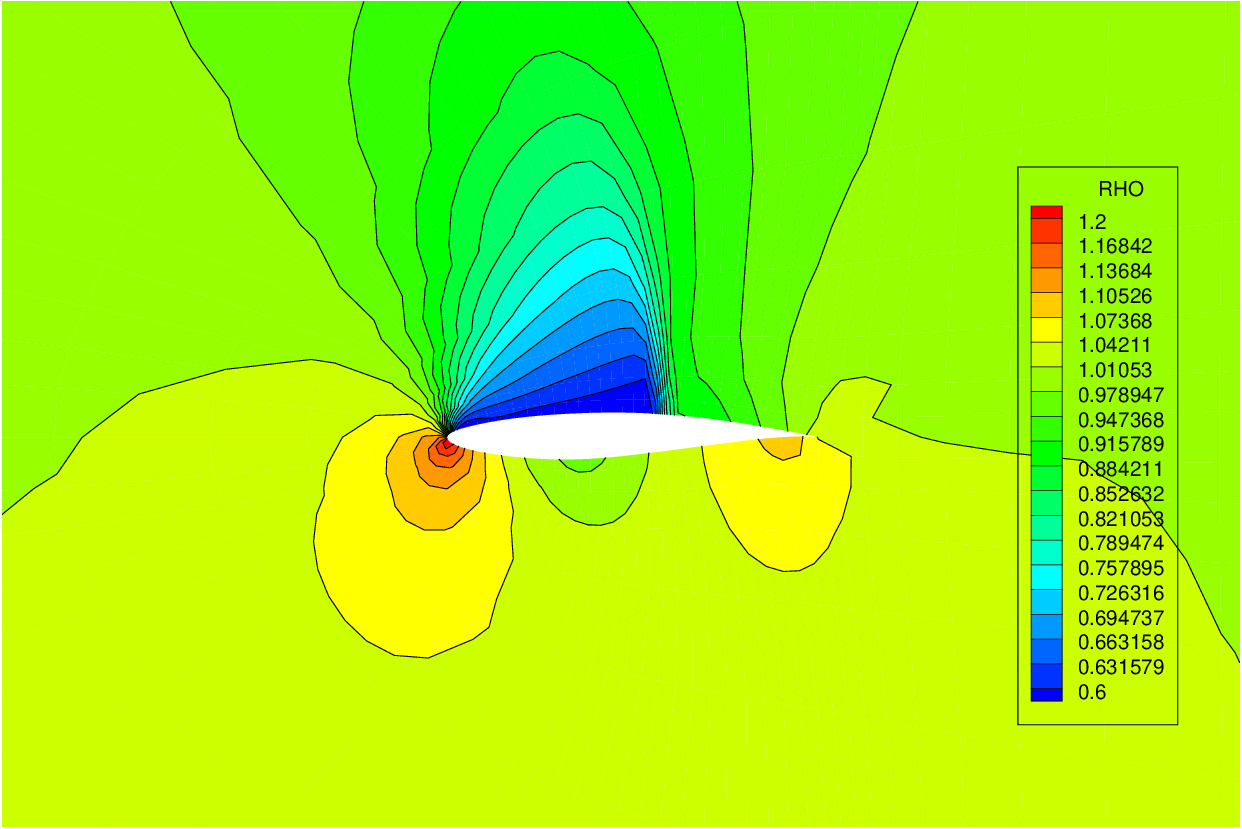}}
 \subfigure[relative error of No.2]{
 \includegraphics[width=4.8cm]{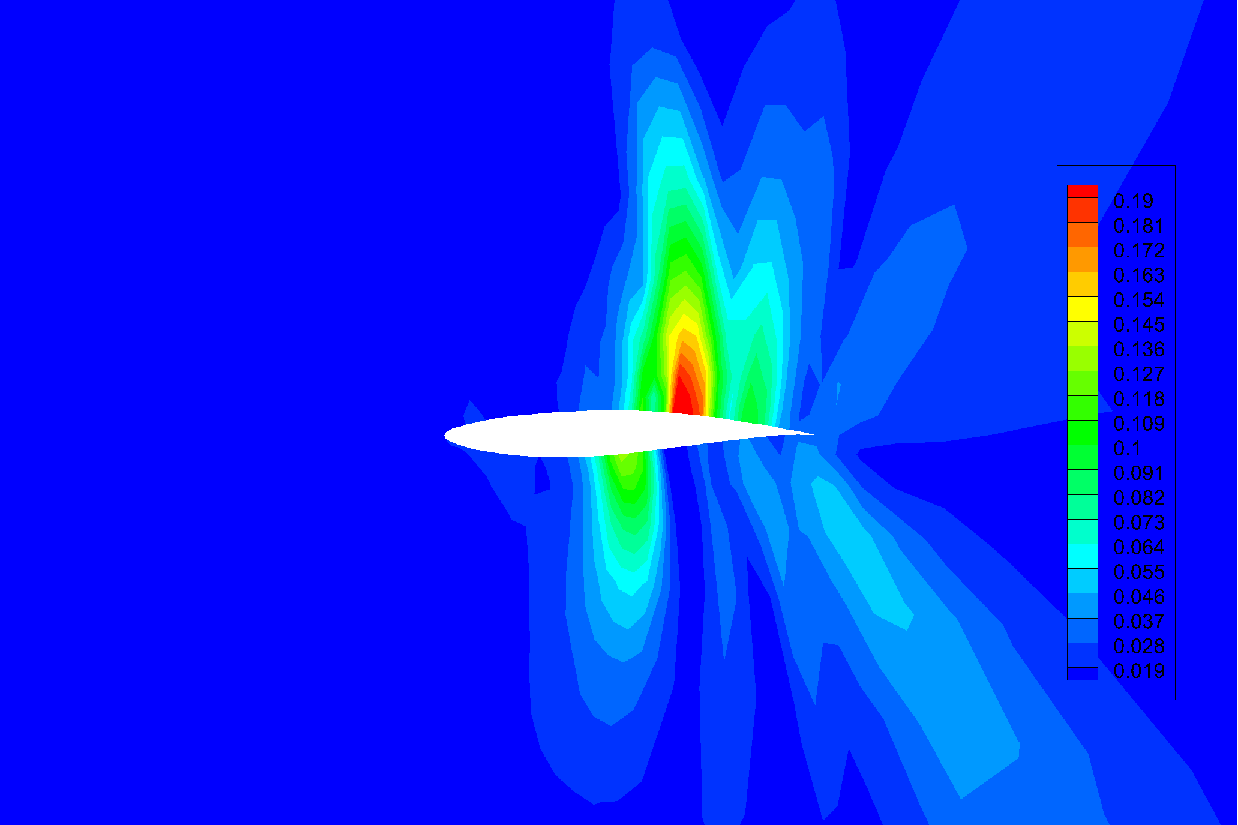}}
 \subfigure[model prediction of No.3]{
 \includegraphics[width=4.8cm]{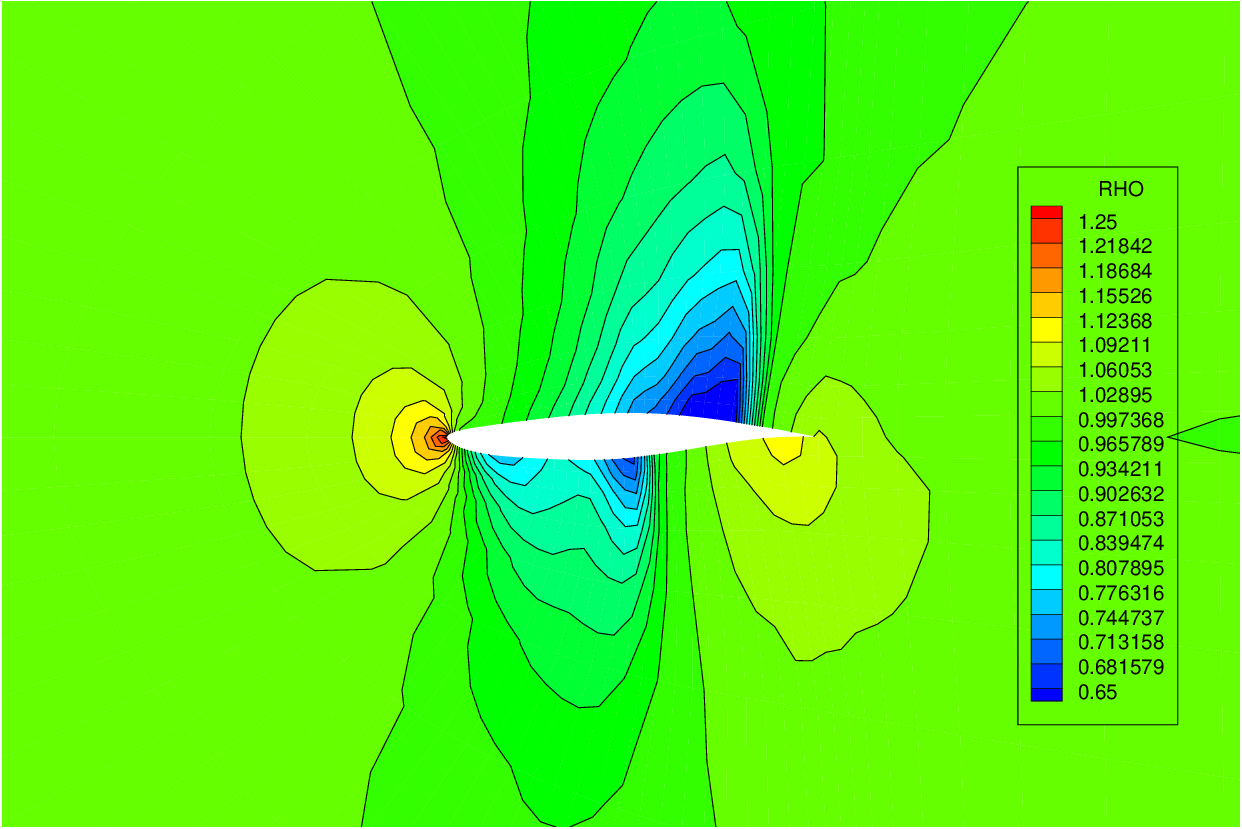}}
 \subfigure[CFD correction of No.3]{
 \includegraphics[width=4.8cm]{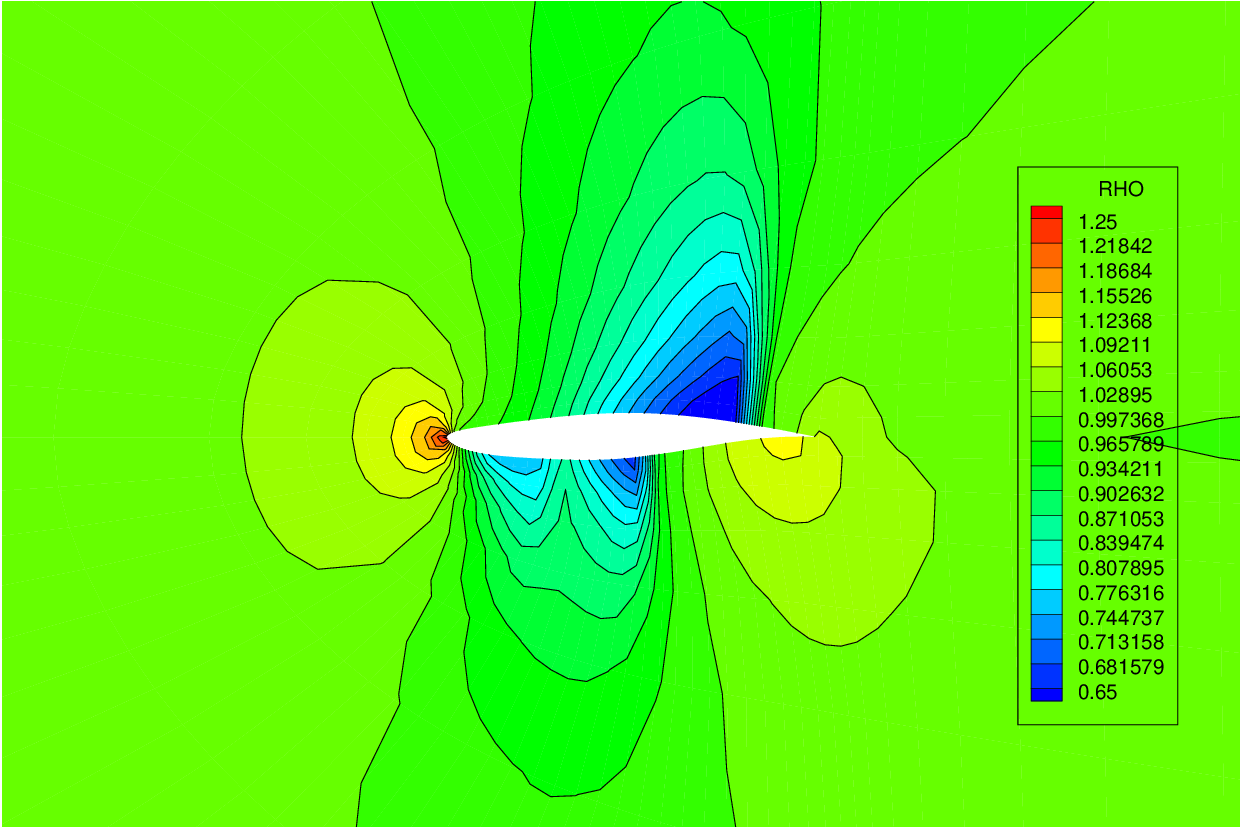}}
 \subfigure[relative error of No.3]{
 \includegraphics[width=4.8cm]{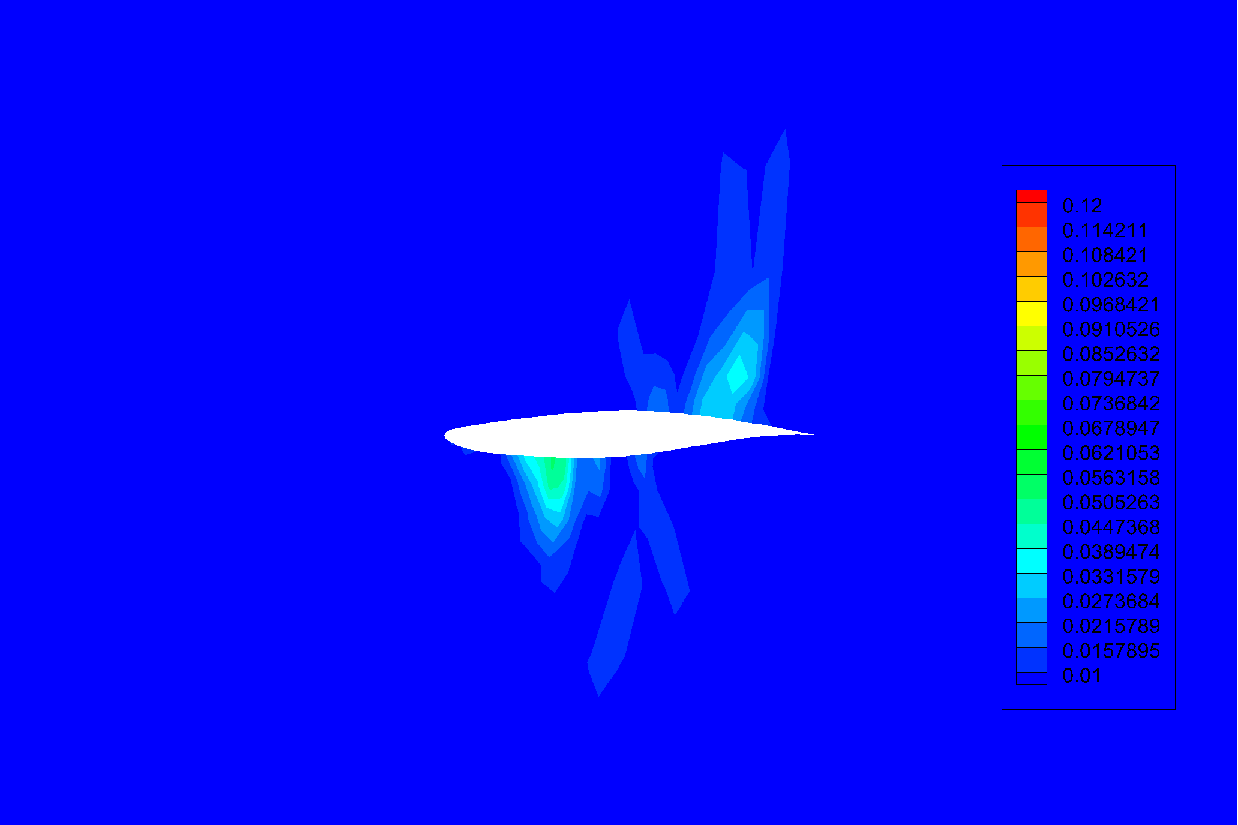}}
 \subfigure[model prediction of No.4]{
 \includegraphics[width=4.8cm]{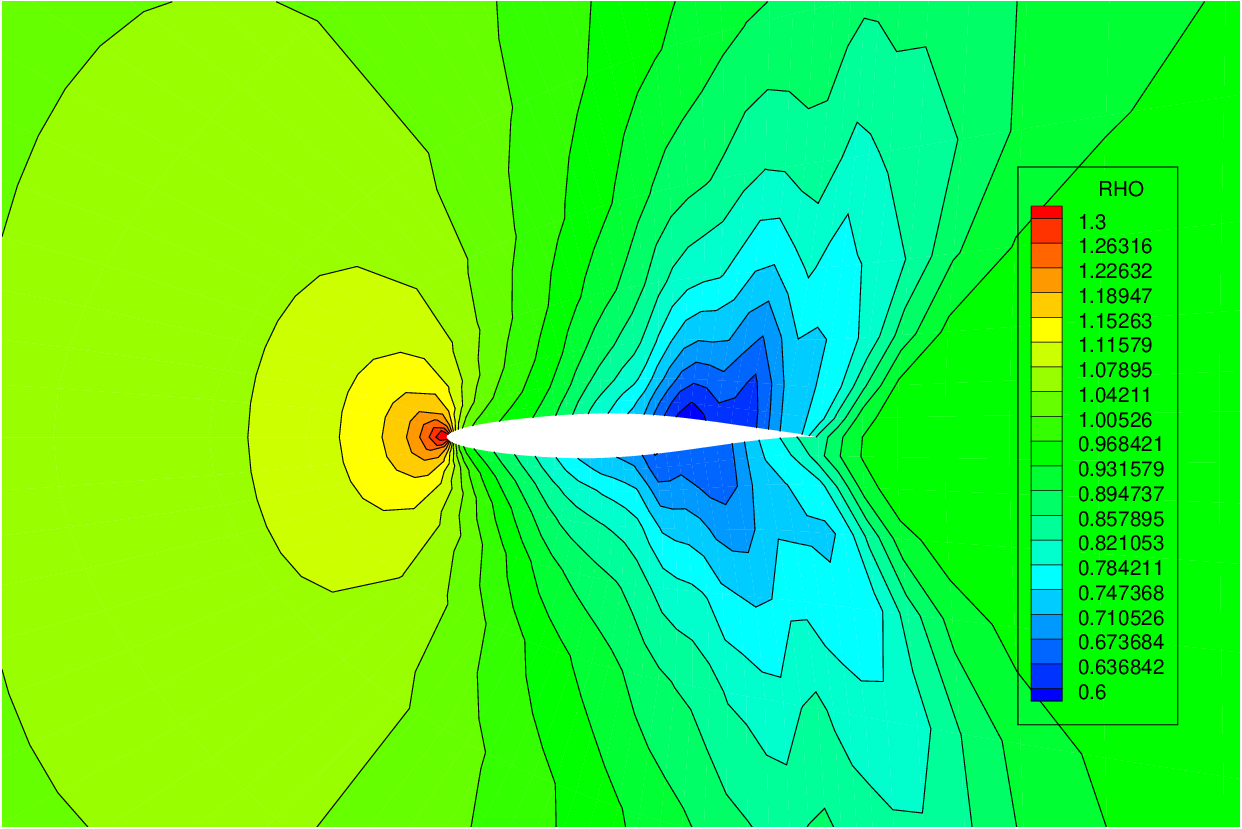}}
 \subfigure[CFD correction of No.4]{
 \includegraphics[width=4.8cm]{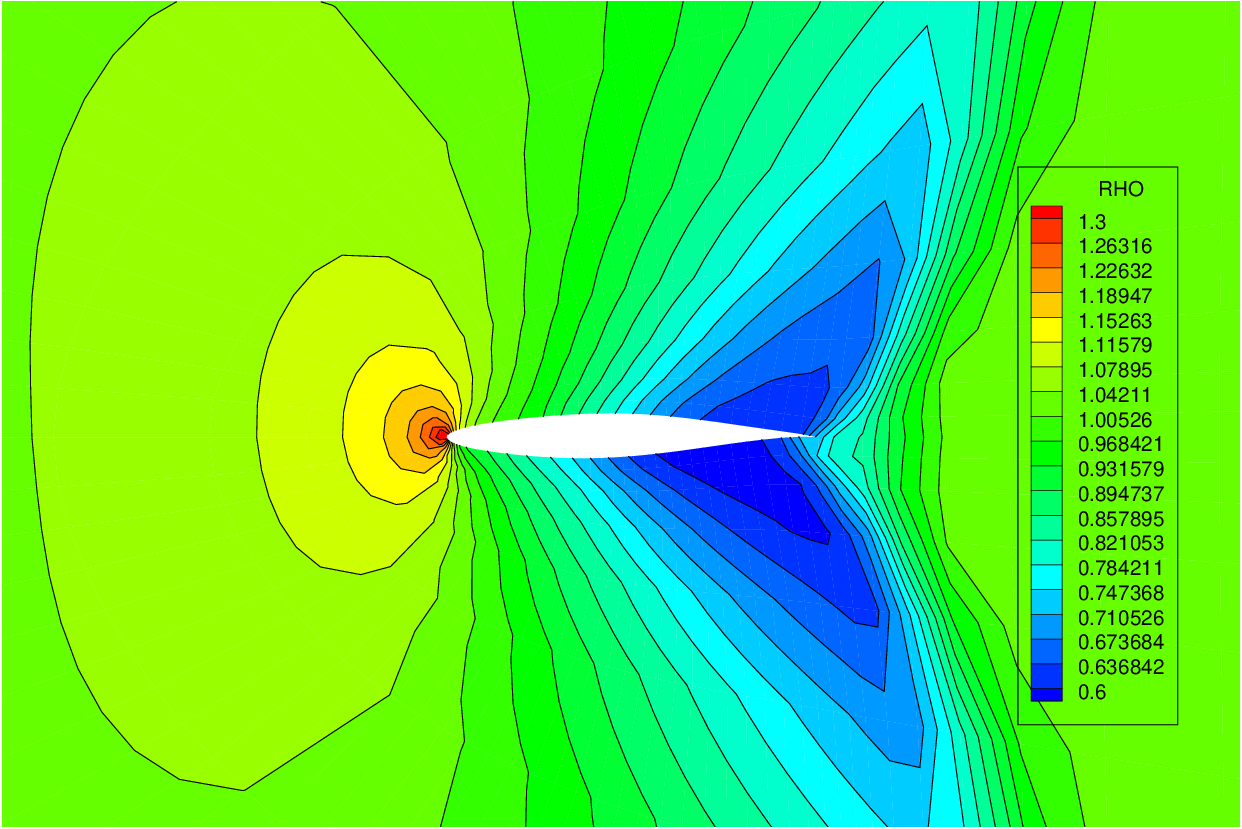}}
 \subfigure[relative error of No.4]{
 \includegraphics[width=4.8cm]{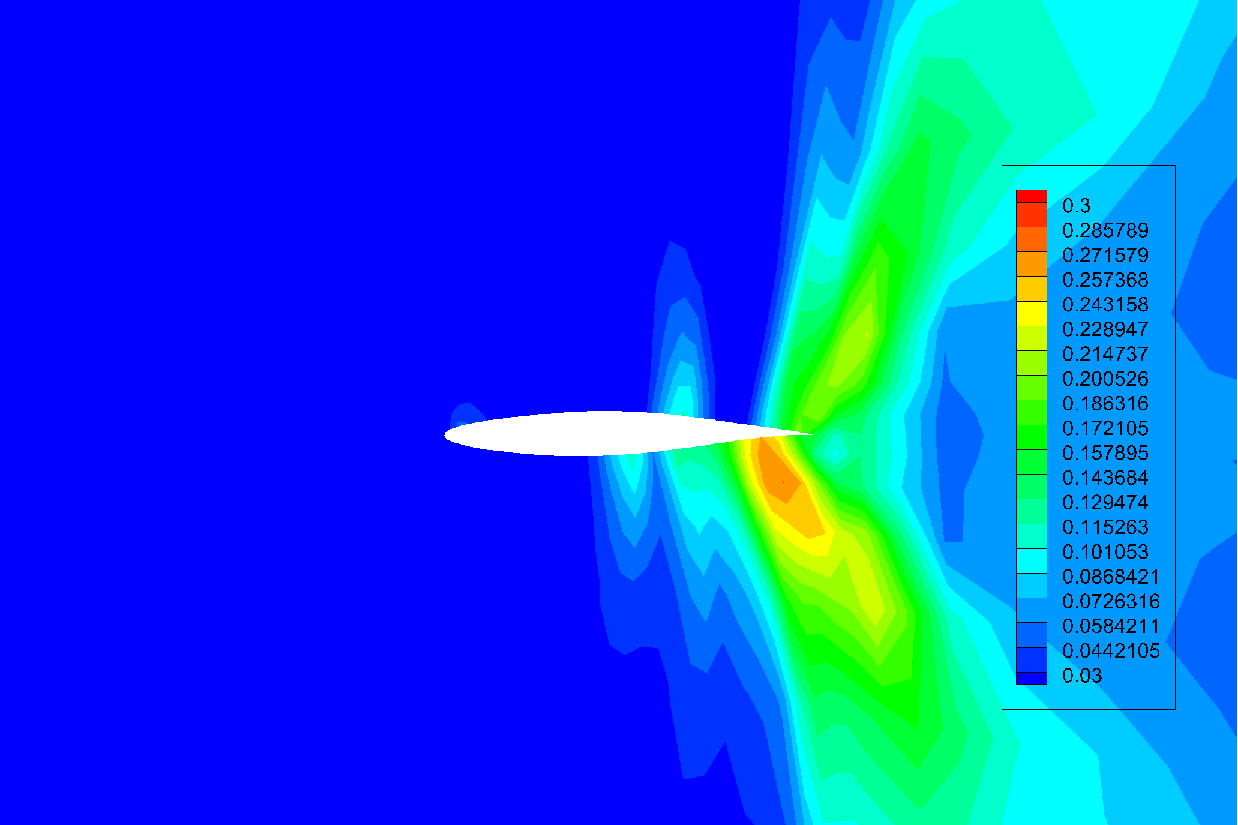}}
 \caption{Flow-field produced by Zonal POD + Orig RBF model and CFD in validation set.}
 \label{fig:rae2822flow3}
\end{figure}

\begin{figure}[htbp]
 \centering
 \subfigure[model prediction of No.1]{
 \includegraphics[width=4.8cm]{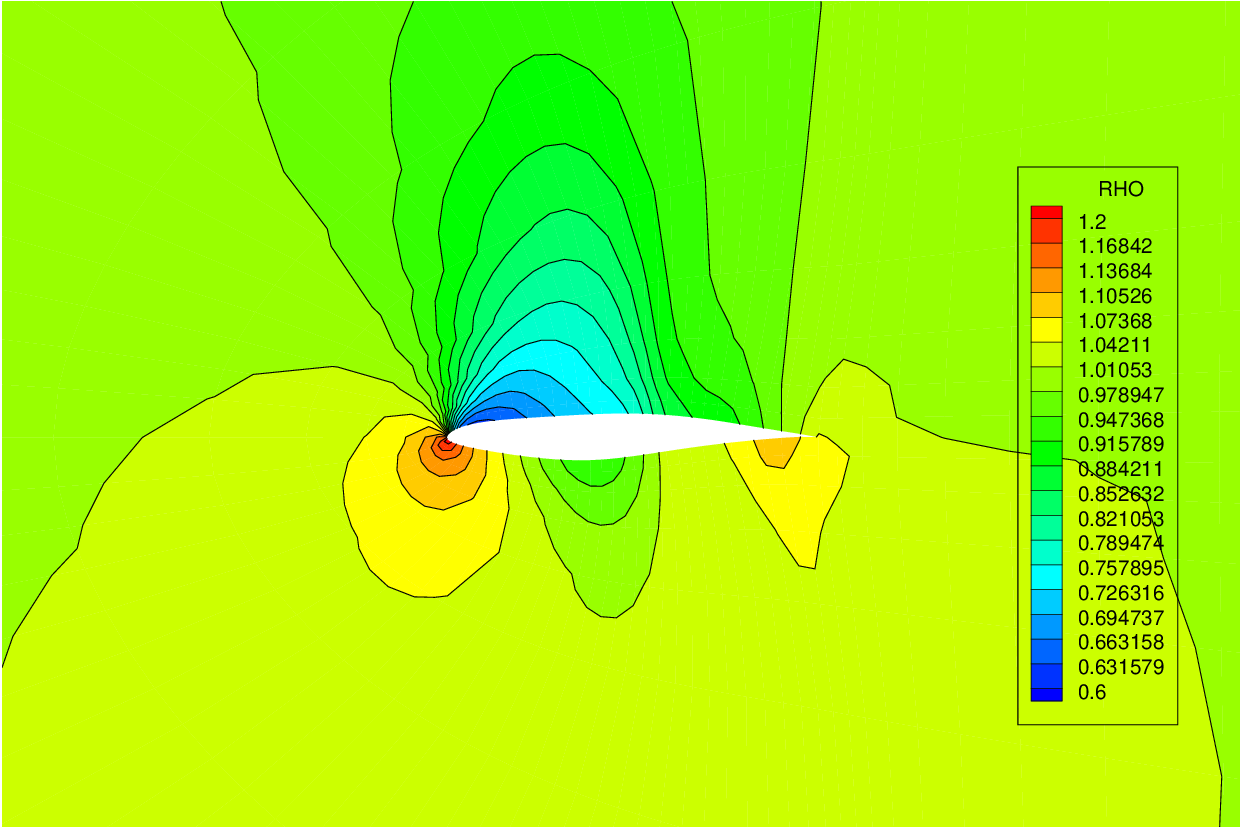}}
 \subfigure[CFD correction of No.1]{
 \includegraphics[width=4.8cm]{figure/flow-CFD-6.eps}}
 \subfigure[relative error of No.1]{
 \includegraphics[width=4.8cm]{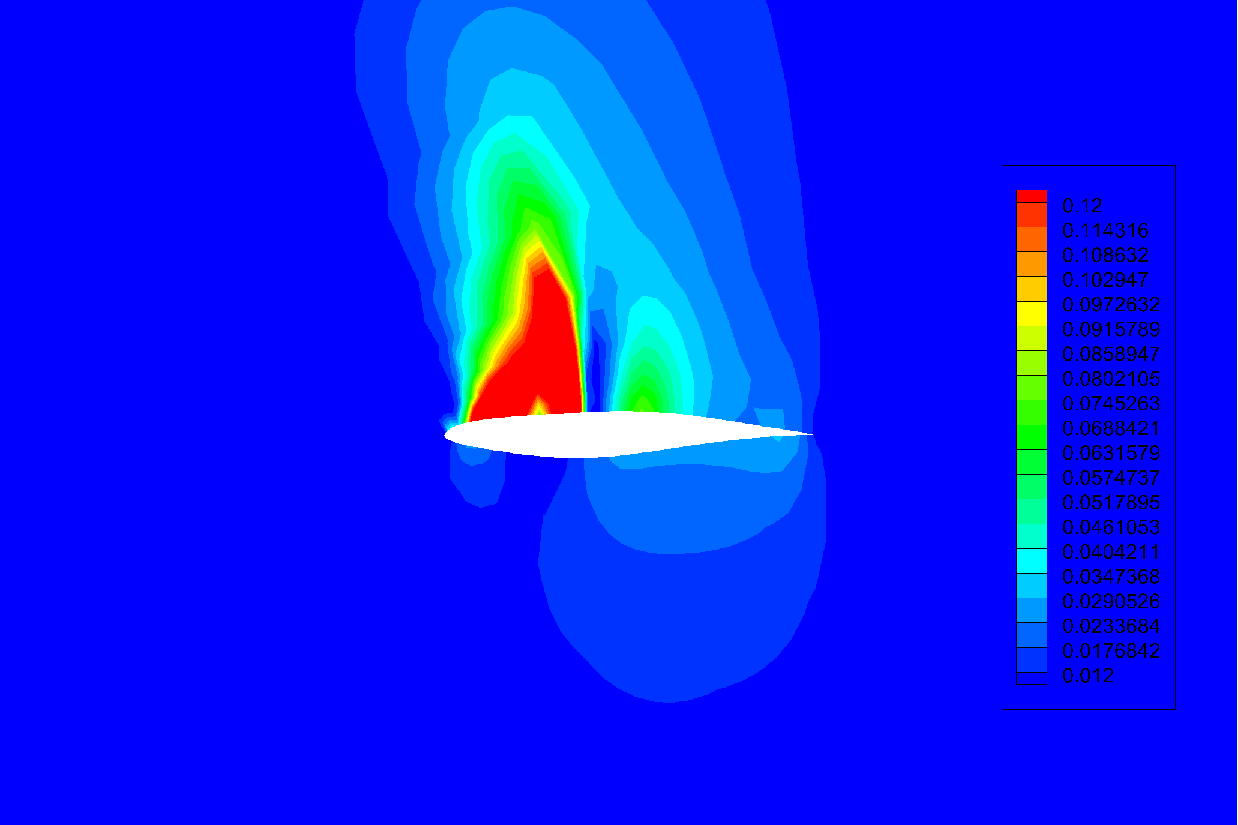}}
 \subfigure[model prediction of No.2]{
 \includegraphics[width=4.8cm]{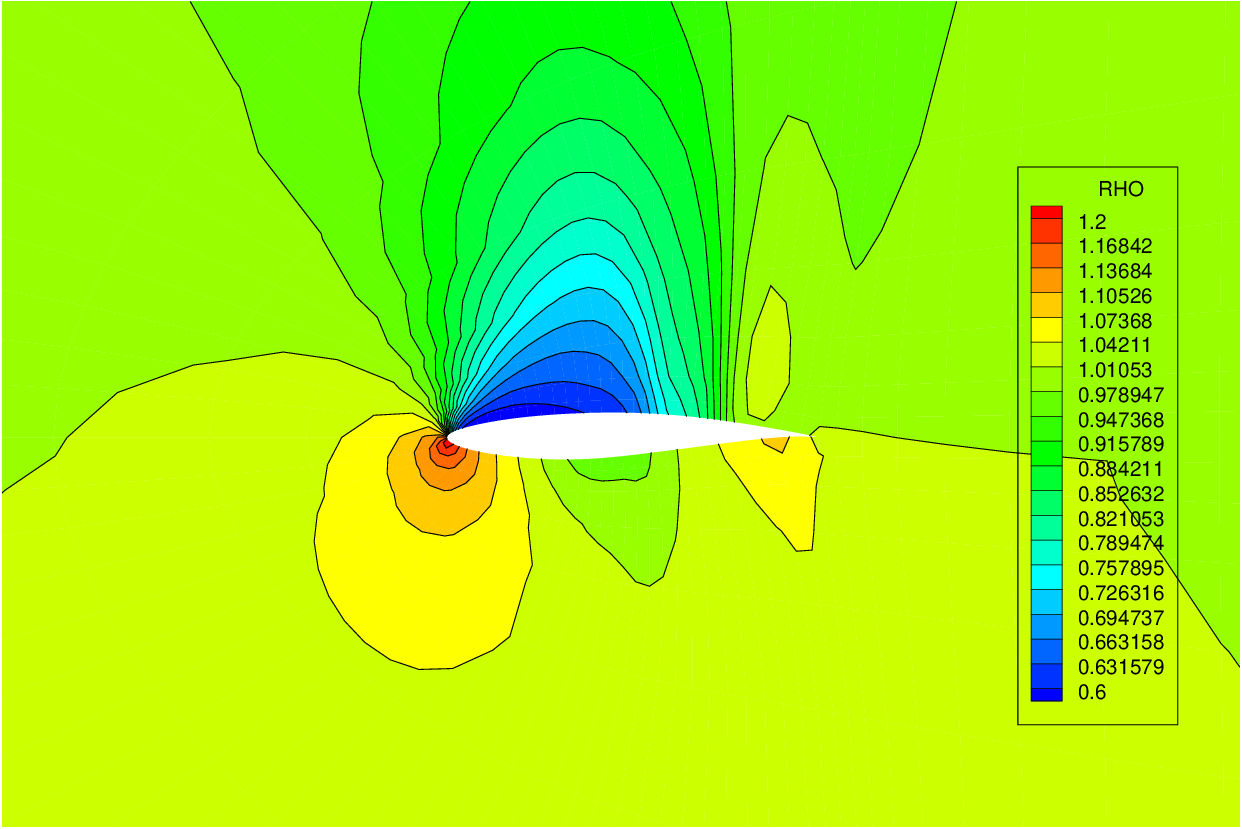}}
 \subfigure[CFD correction of No.2]{
 \includegraphics[width=4.8cm]{figure/flow-CFD-7.eps}}
 \subfigure[relative error of No.2]{
 \includegraphics[width=4.8cm]{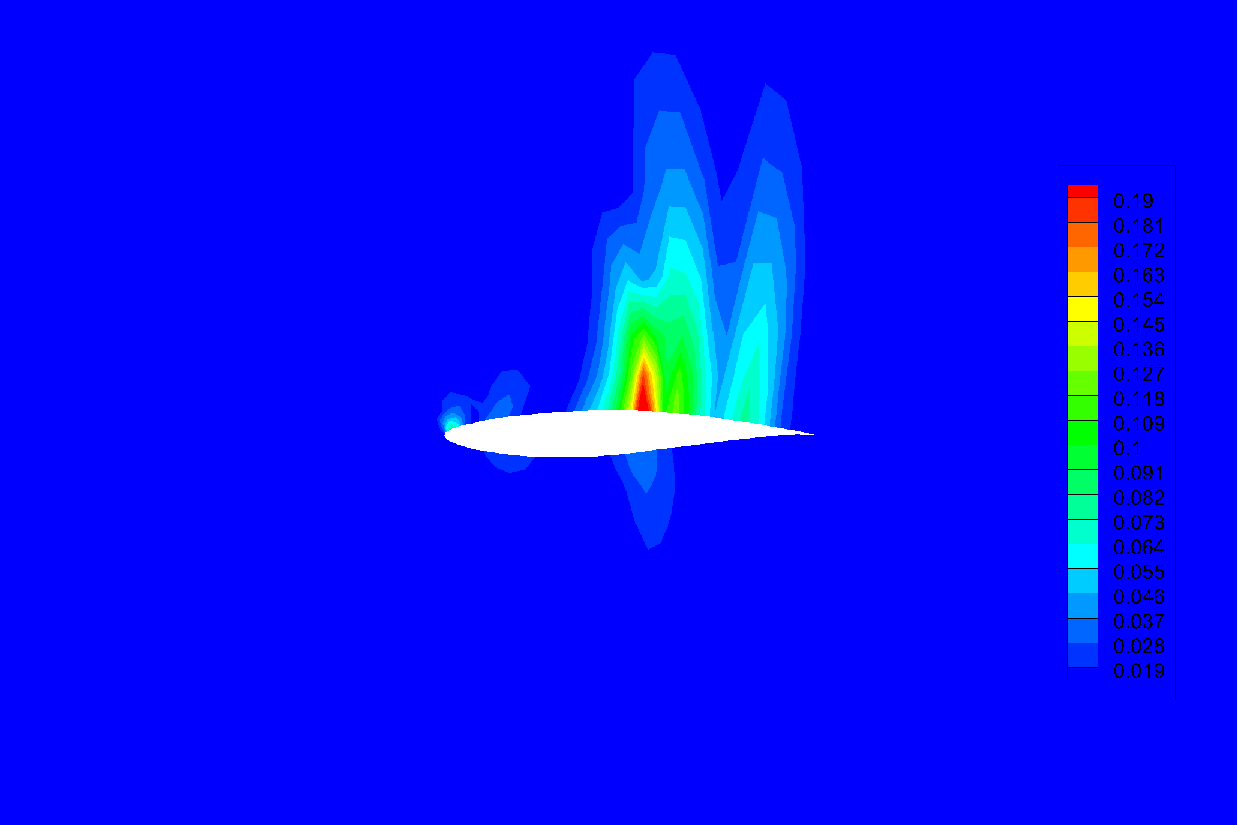}}
 \subfigure[model prediction of No.3]{
 \includegraphics[width=4.8cm]{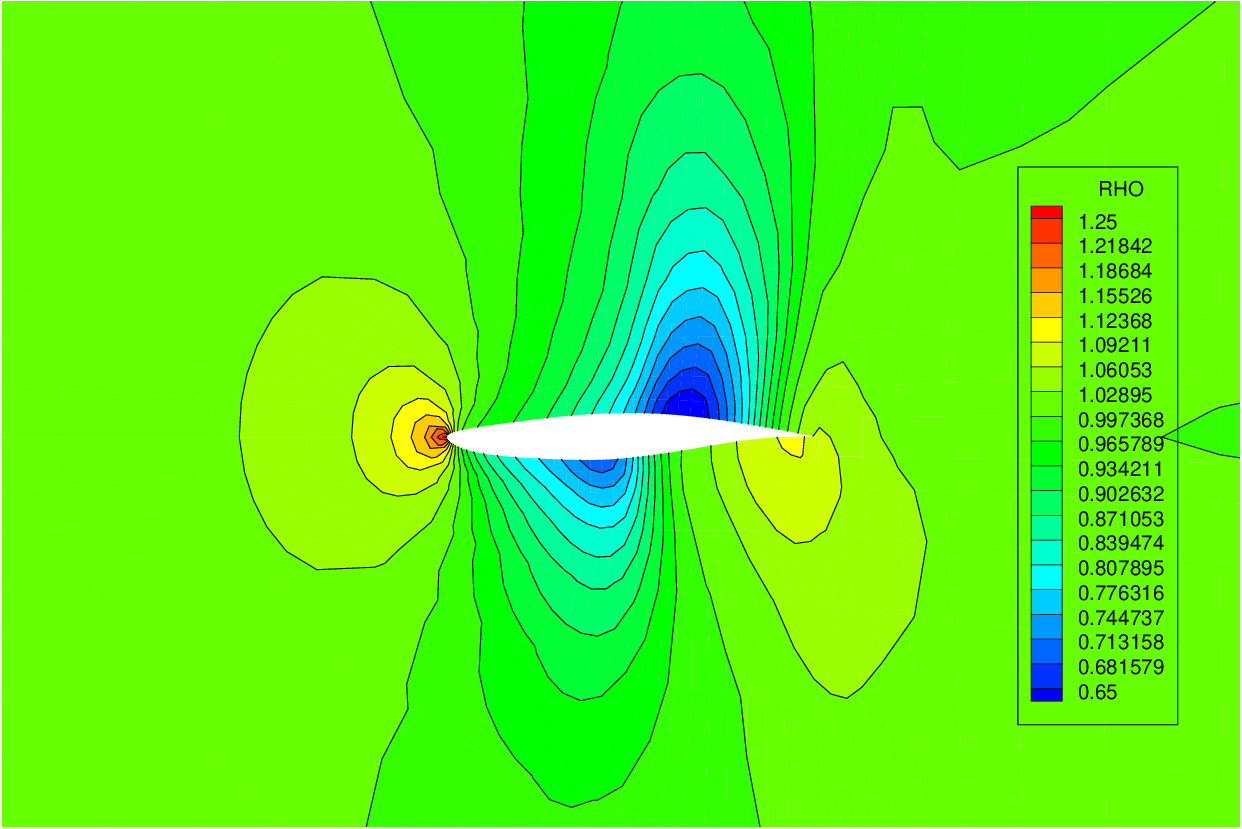}}
 \subfigure[CFD correction of No.3]{
 \includegraphics[width=4.8cm]{figure/flow-CFD-8.eps}}
 \subfigure[relative error of No.3]{
 \includegraphics[width=4.8cm]{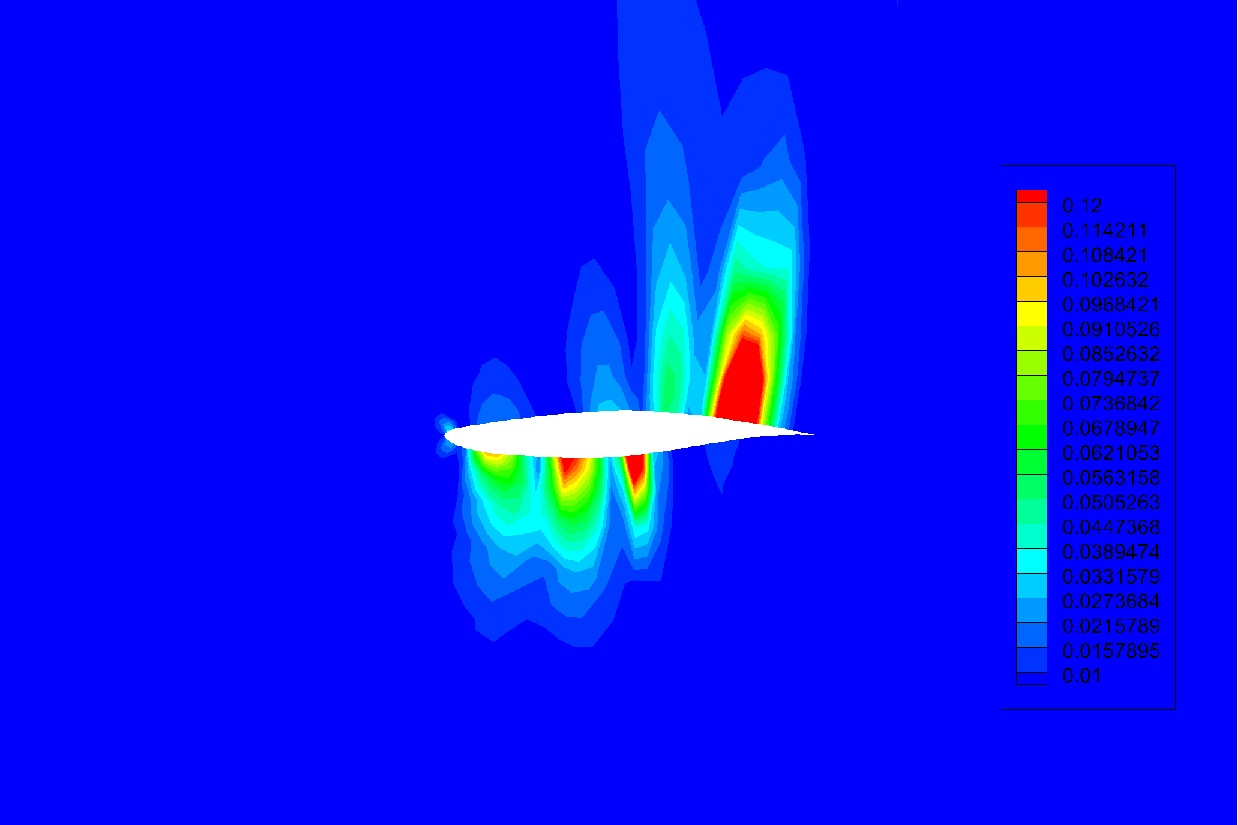}}
 \subfigure[model prediction of No.4]{
 \includegraphics[width=4.8cm]{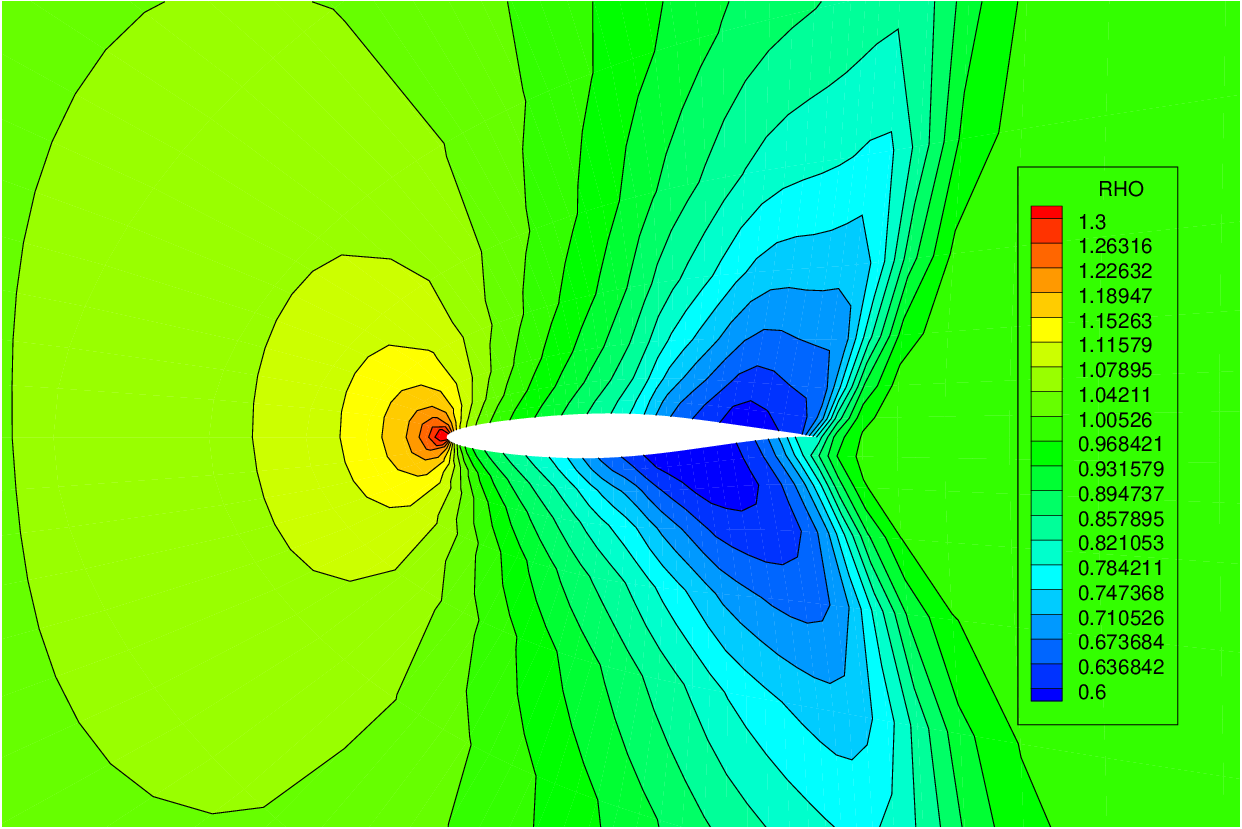}}
 \subfigure[CFD correction of No.4]{
 \includegraphics[width=4.8cm]{figure/flow-CFD-9.eps}}
 \subfigure[relative error of No.4]{
 \includegraphics[width=4.8cm]{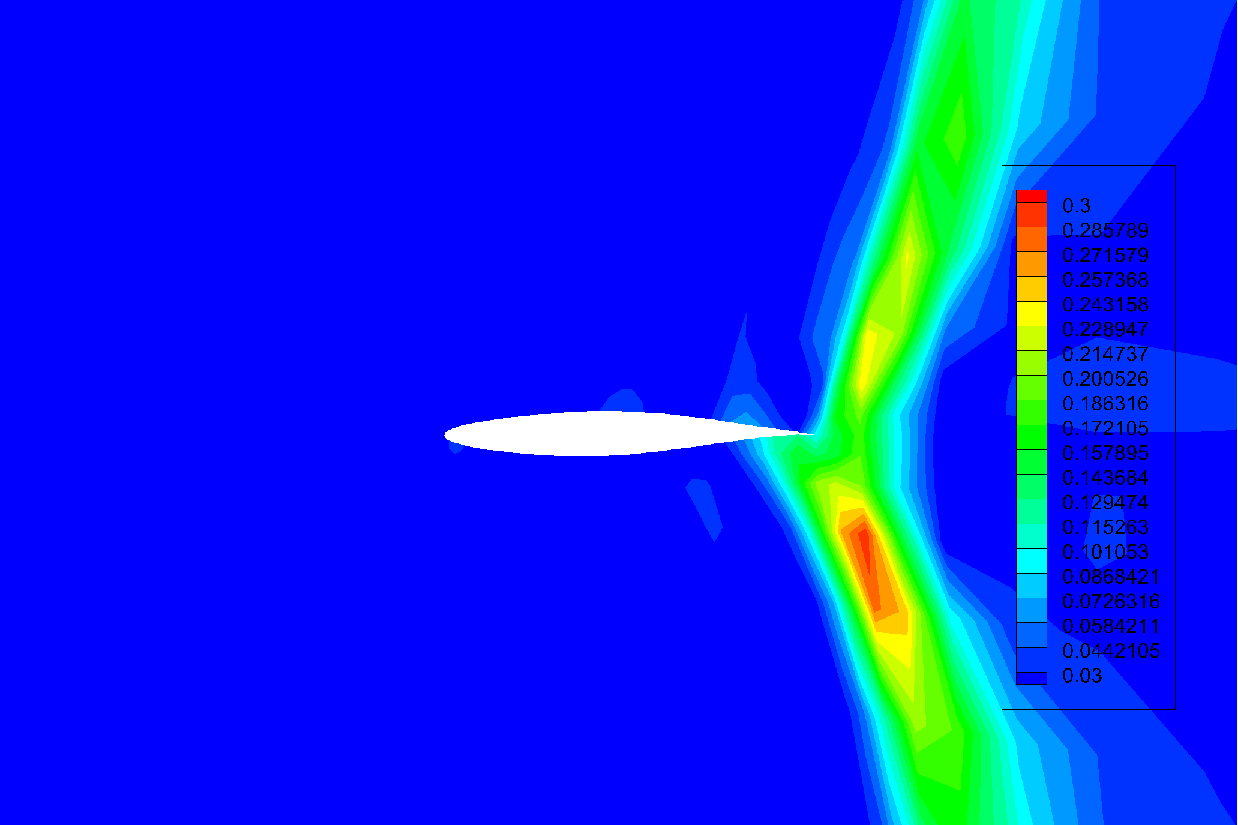}}
 \caption{Flow-field produced by Orig POD + Mod RBF model and CFD in validation set.}
 \label{fig:rae2822flow4}
\end{figure}

\begin{figure}[htbp]
 \centering
 \subfigure[model prediction of No.1]{
 \includegraphics[width=4.8cm]{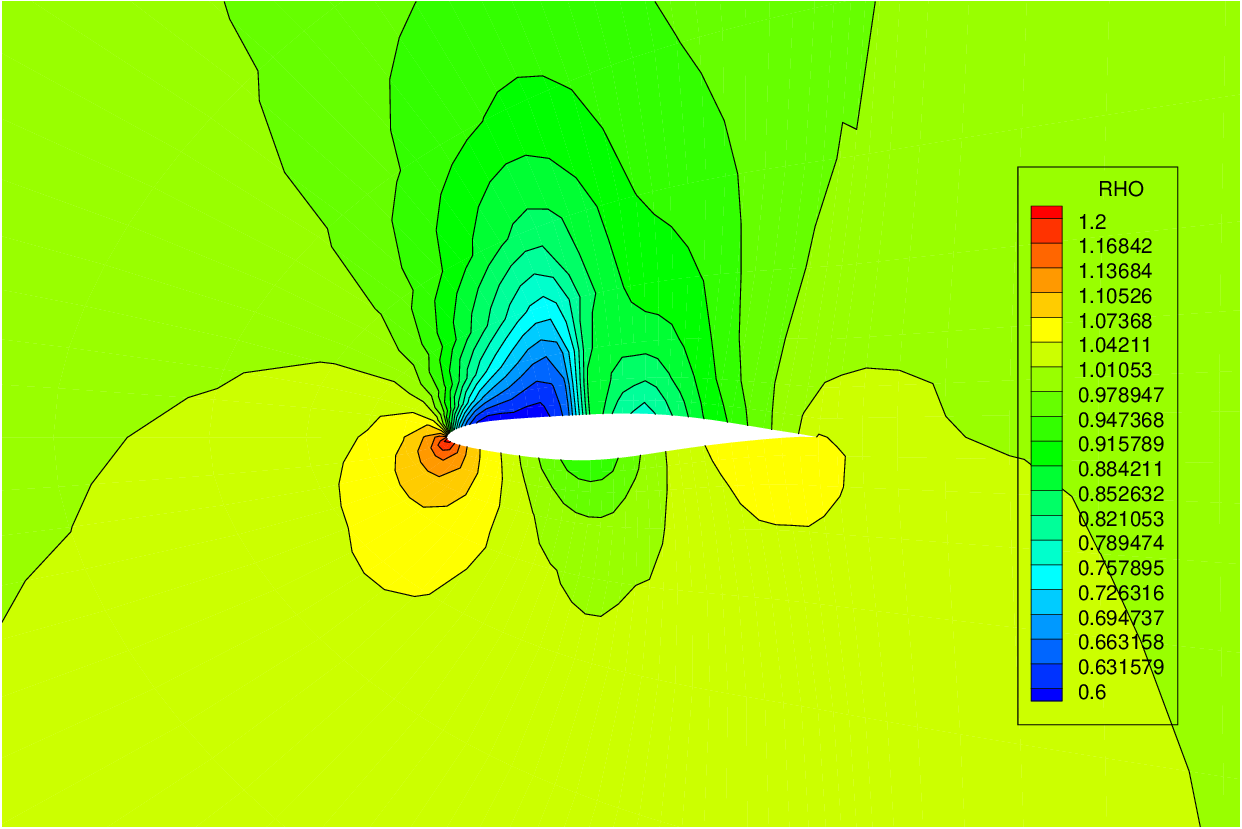}}
 \subfigure[CFD correction of No.1]{
 \includegraphics[width=4.8cm]{figure/flow-CFD-6.eps}}
 \subfigure[relative error of No.1]{
 \includegraphics[width=4.8cm]{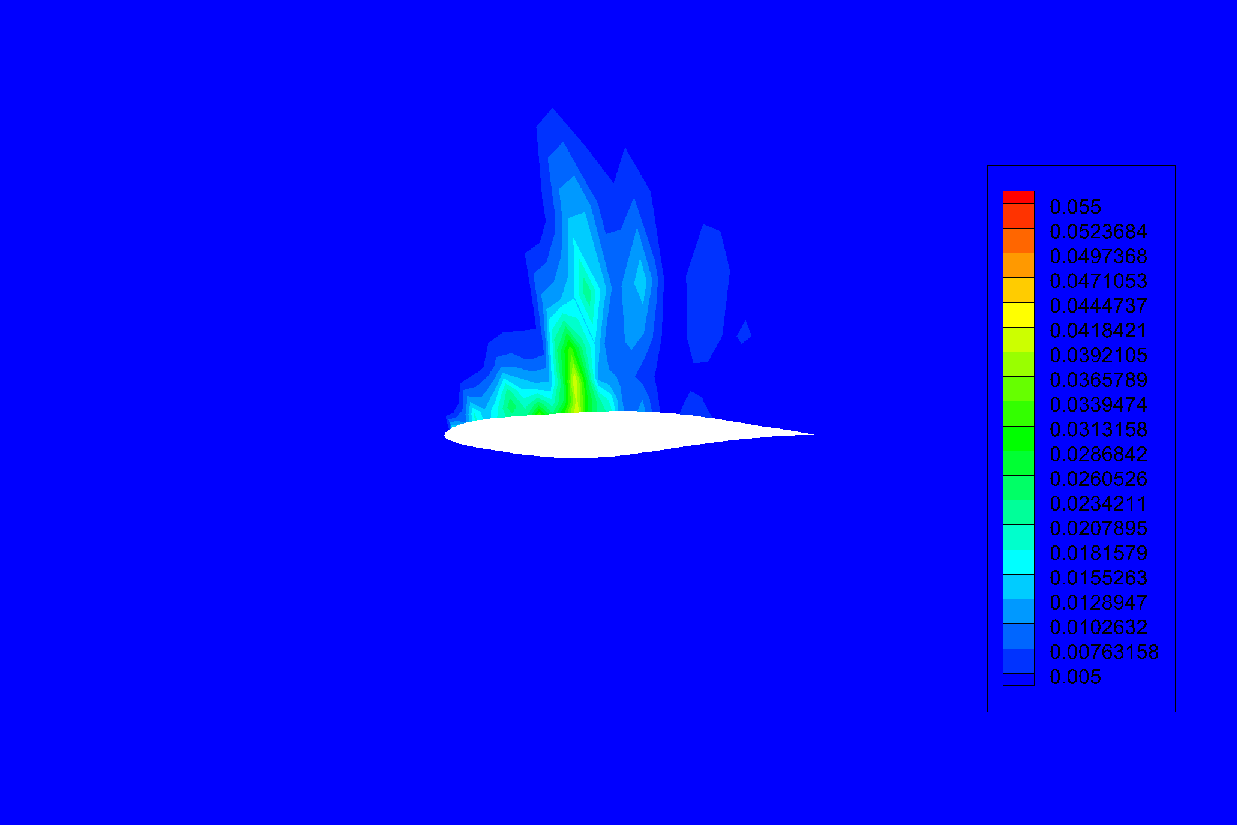}}
 \subfigure[model prediction of No.2]{
 \includegraphics[width=4.8cm]{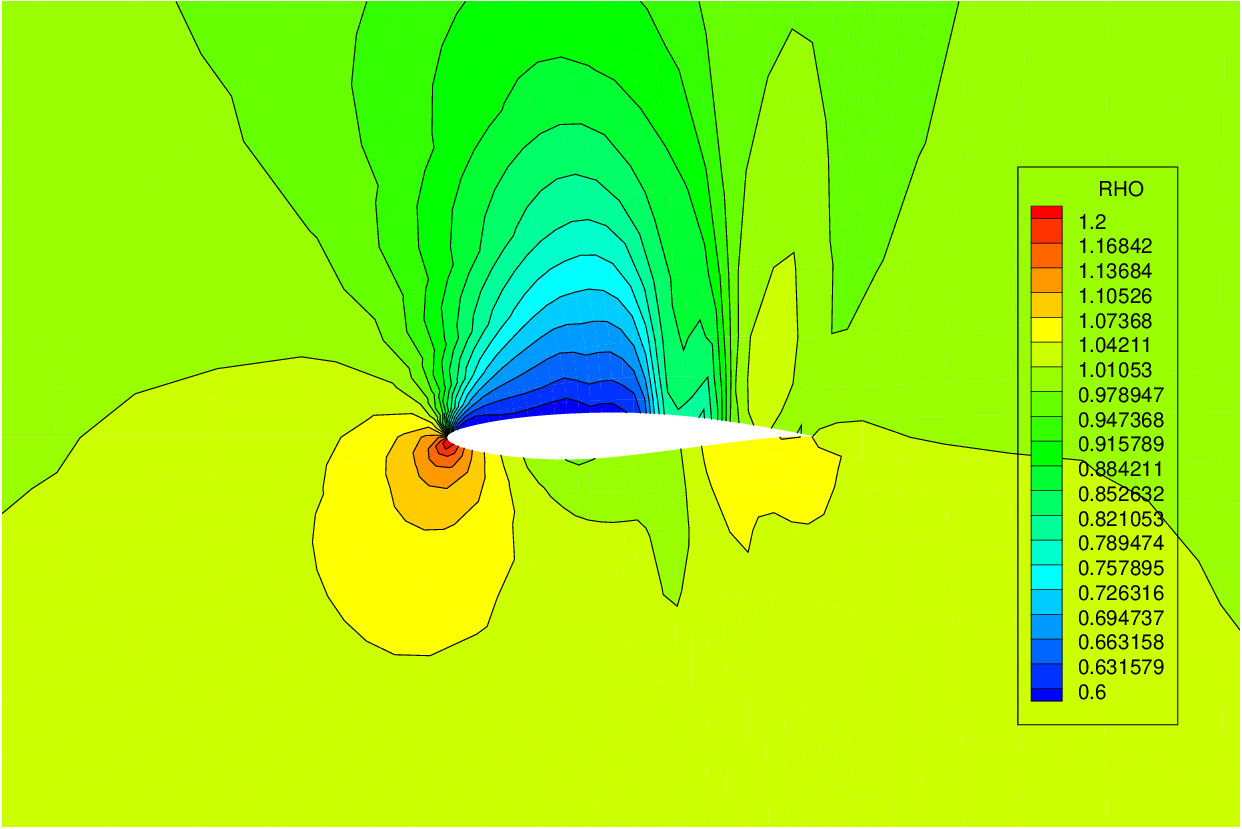}}
 \subfigure[CFD correction of No.2]{
 \includegraphics[width=4.8cm]{figure/flow-CFD-7.eps}}
 \subfigure[relative error of No.2]{
 \includegraphics[width=4.8cm]{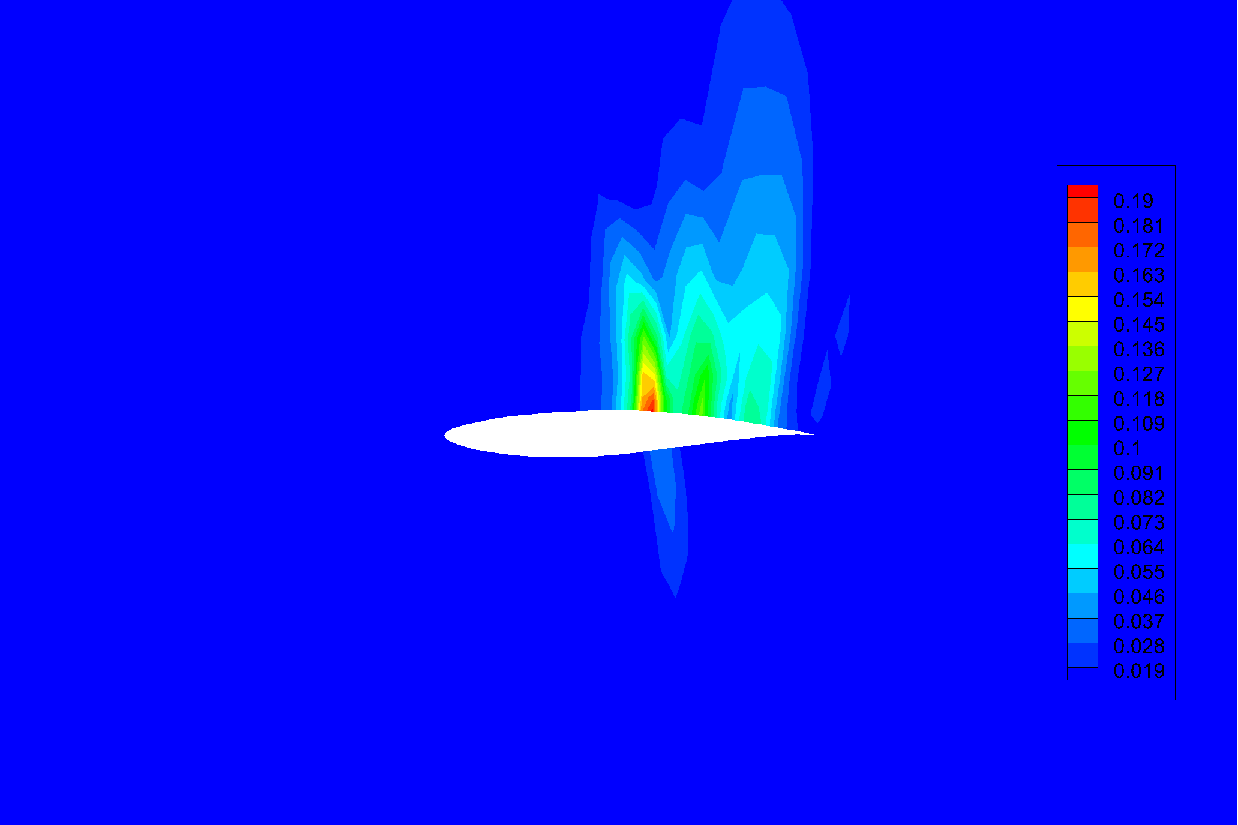}}
 \subfigure[model prediction of No.3]{
 \includegraphics[width=4.8cm]{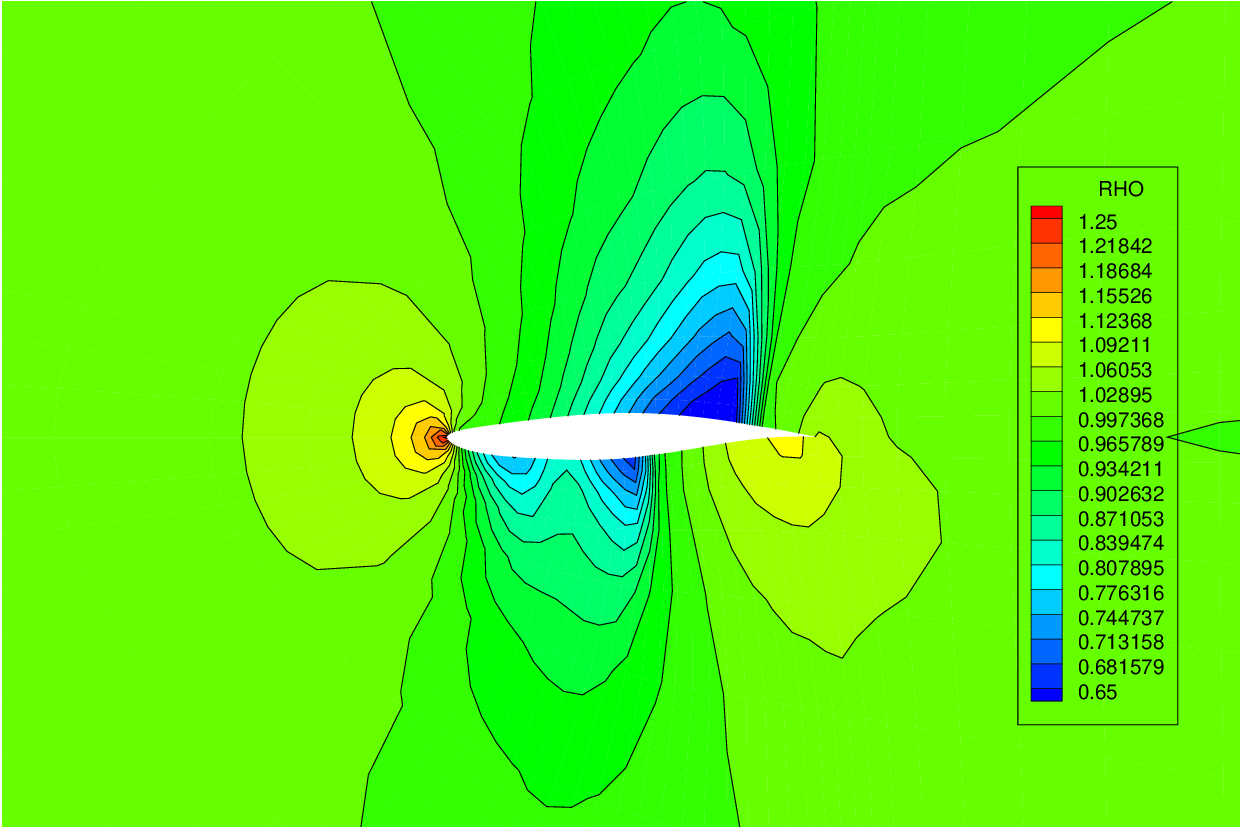}}
 \subfigure[CFD correction of No.3]{
 \includegraphics[width=4.8cm]{figure/flow-CFD-8.eps}}
 \subfigure[relative error of No.3]{
 \includegraphics[width=4.8cm]{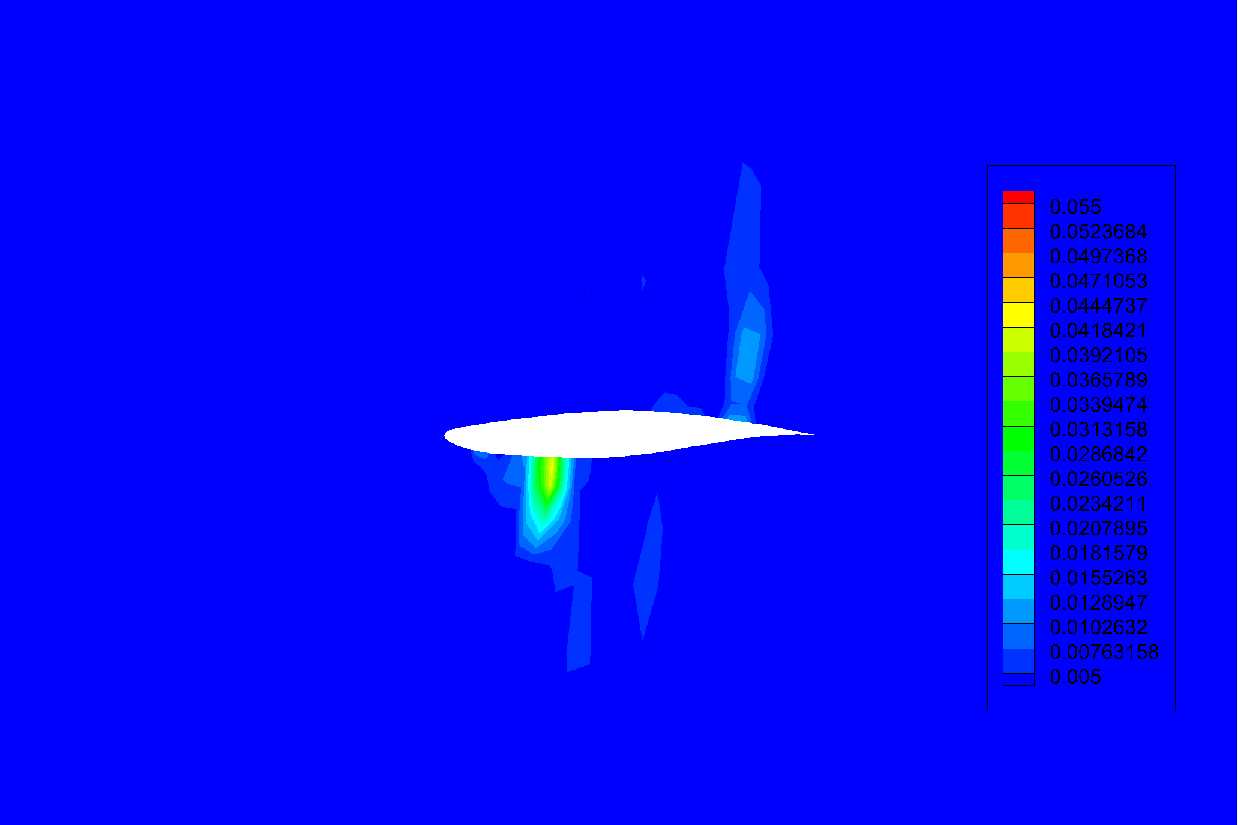}}
 \subfigure[model prediction of No.4]{
 \includegraphics[width=4.8cm]{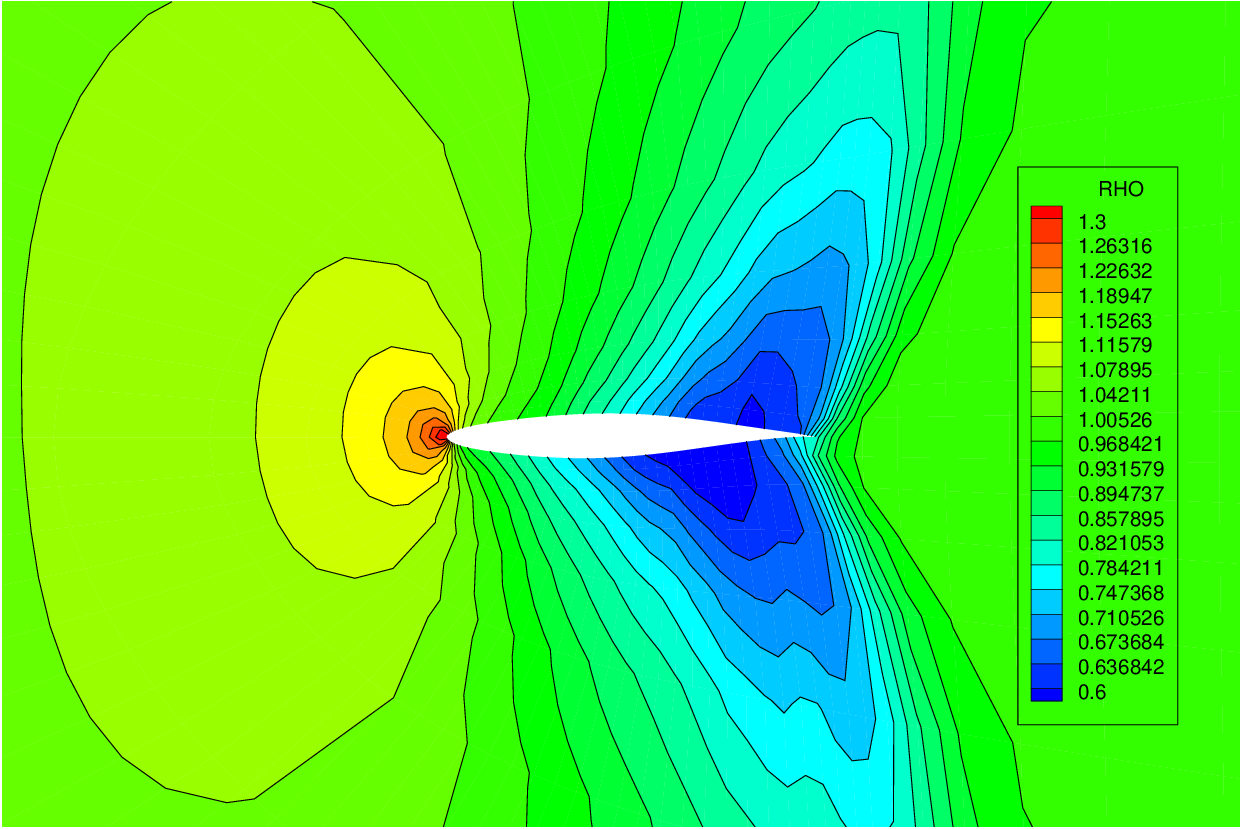}}
 \subfigure[CFD correction of No.4]{
 \includegraphics[width=4.8cm]{figure/flow-CFD-9.eps}}
 \subfigure[relative error of No.4]{
 \includegraphics[width=4.8cm]{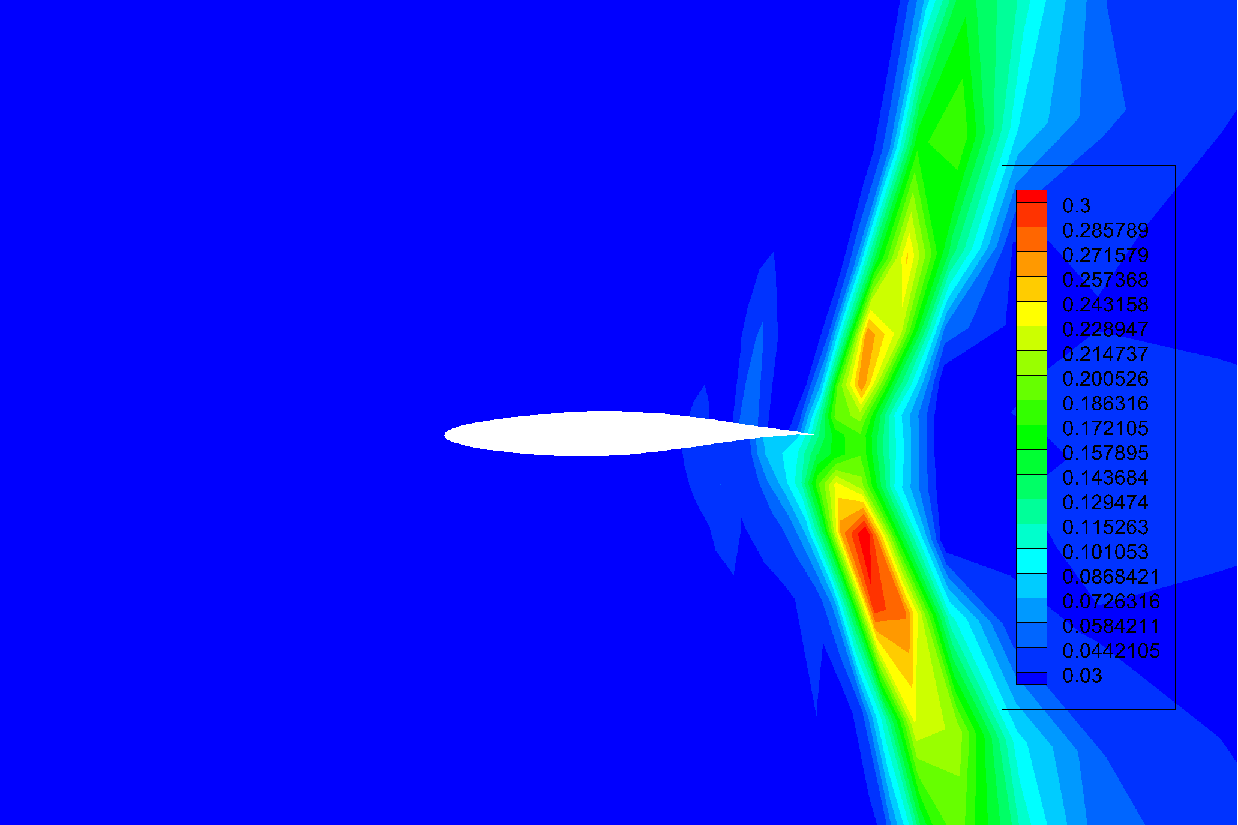}}
 \caption{Flow-field produced by Zonal POD + Mod RBF model and CFD in validation set.}
 \label{fig:rae2822flow5}
\end{figure}

\begin{figure}[htbp]
 \centering
 \subfigure[$C_p$ curves of No.1]{
 \includegraphics[width=7cm]{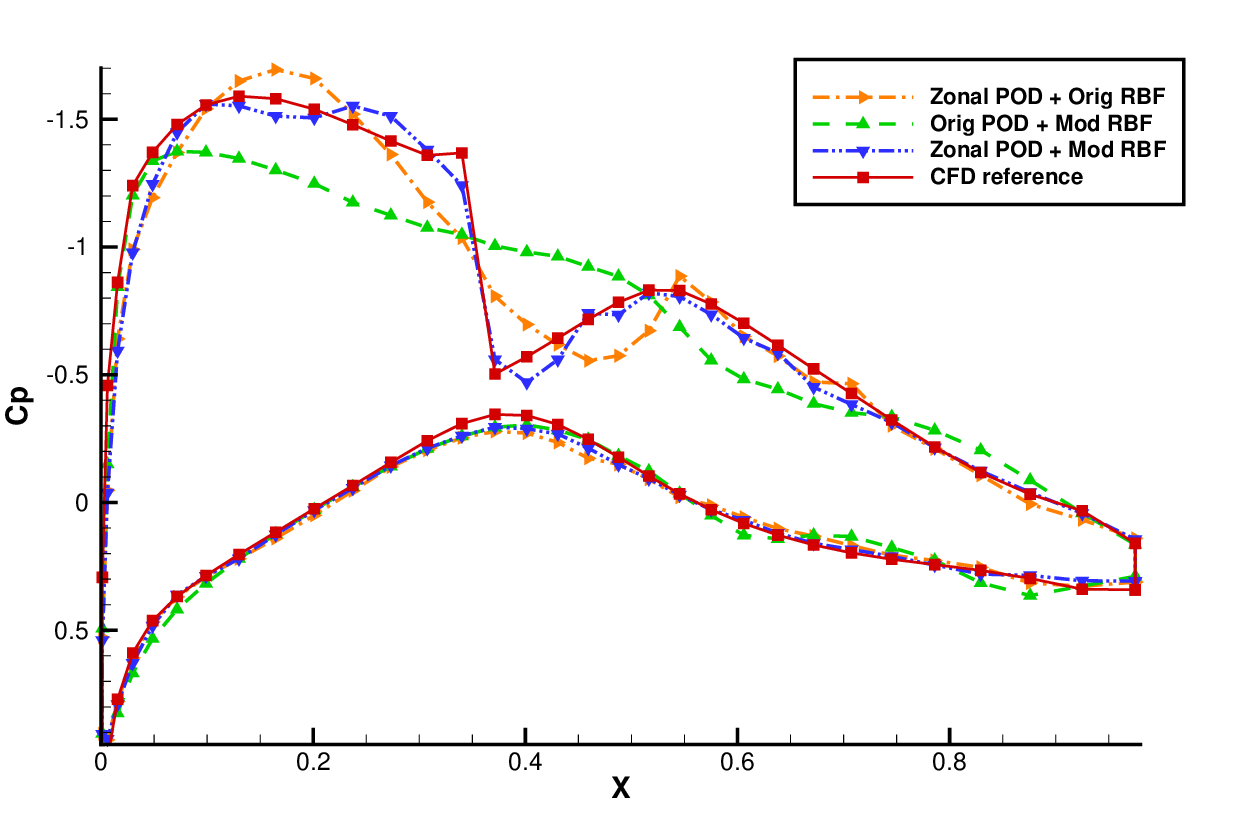}}
 \subfigure[$C_p$ curves of No.2]{
 \includegraphics[width=7cm]{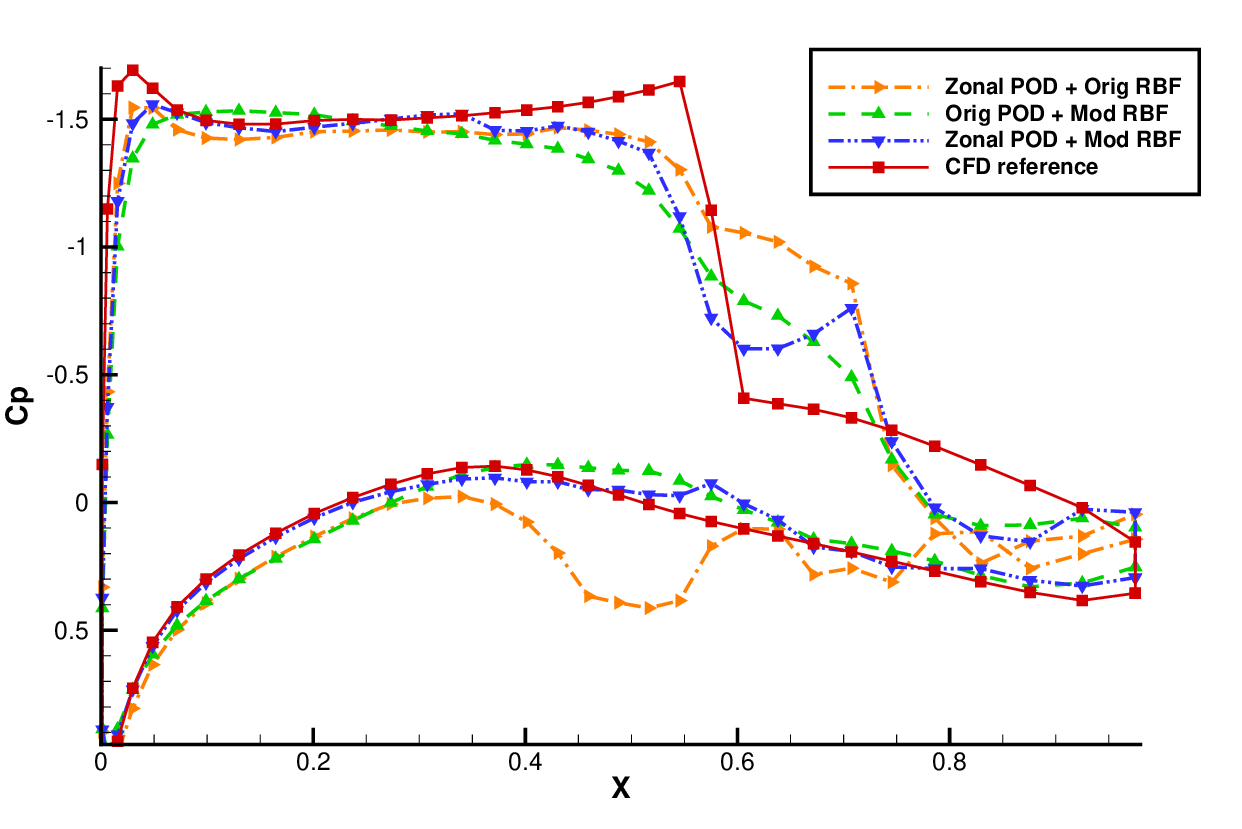}}
 \subfigure[$C_p$ curves of No.3]{
 \includegraphics[width=7cm]{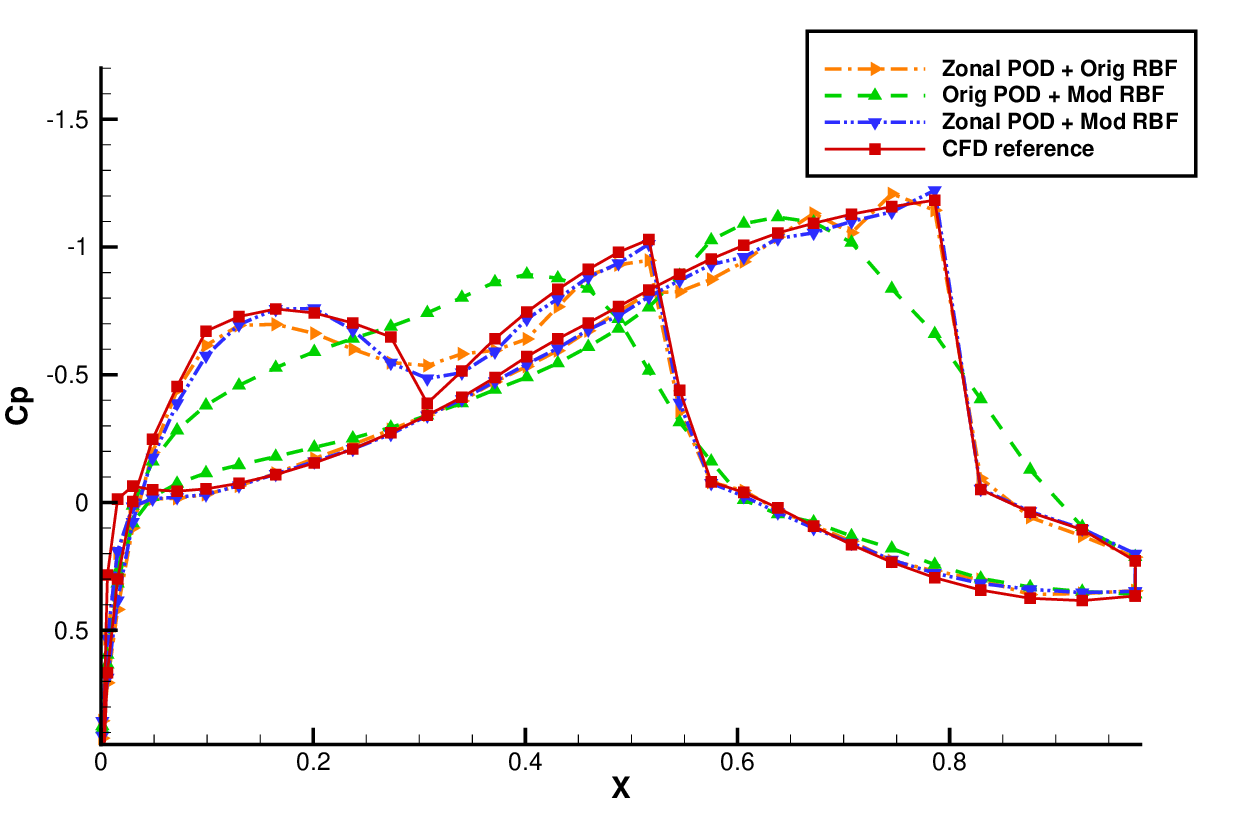}}
 \subfigure[$C_p$ curves of No.4]{
 \includegraphics[width=7cm]{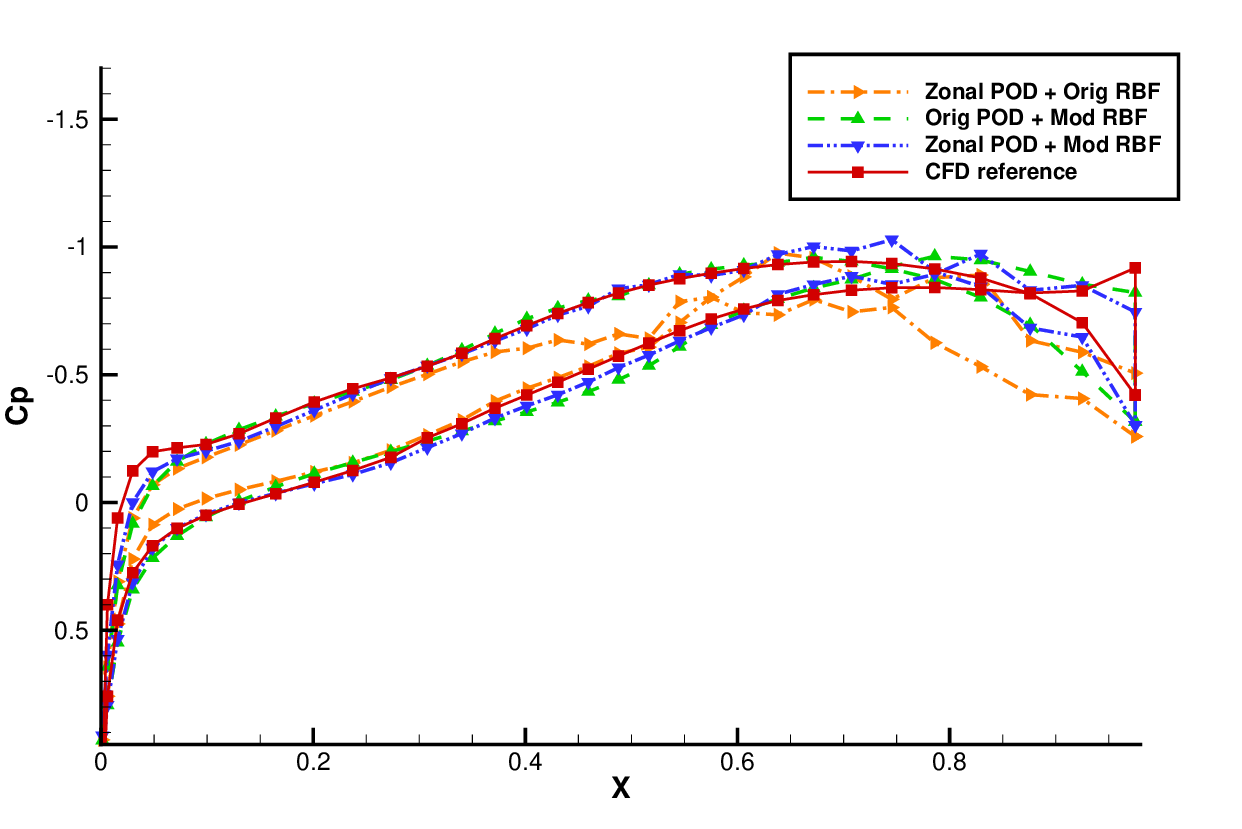}}
 \caption{$C_p$ distribution produced by data-driven models and CFD in validation set.}
 \label{fig:rae2822Cp2}
\end{figure}

Figure~\ref{fig:rae2822flow3}-\ref{fig:rae2822flow5} show flow-field predicted by different data-driven models and CFD, together with their relative errors. Figure~\ref{fig:rae2822Cp2} compares the difference of $C_p$ distribution among different models and CFD.

As can be observed from Figure~\ref{fig:rae2822flow3}, although the Zonal POD can capture more detailed flow features, the Zonal POD + Orig RBF model shows poor generalization capability on the validation set due to the utilization of the Orig RBF model. Despite the overall flow structure produced by Zonal POD + Orig RBF model is roughly matched to that produced by CFD, there exist considerable discrepancies in the position and finer details of flow structures, such as shock waves, as compared to the CFD reference. Figure~\ref{rae2822flow4} illustrates the effectiveness of the Mod RBF model in correcting the positional information of flow structures, particularly shock waves. Nevertheless, the limitation of the Orig POD in capturing flow details leads to a more dissipative shock wave structure. Furthermore, Figure~\ref{fig:rae2822flow5} highlights the prowess of the Zonal POD + Mod RBF model in predicting shock positions and enhancing the sharpness of shock wave structures, thereby significantly reducing the relative error in the flow-field as compared to the other strategy combinations.

Similar behaviors can be observed from the $C_p$ curves shown in Figure~\ref{fig:rae2822Cp2} that (a) Orig POD + Mod RBF model brings high dissipation to the $C_p$ curve, making it difficult to reach the pressure extreme points; (b) Zonal POD + Orig RBF model produces relatively large positional deviations and oscillations in the $C_p$ curve; (c) Zonal POD + Mod RBF model can effectively improve accuracy of both shock positions and detailed flow features, but its predicted flow-field still show visible errors compared to CFD at certain sample points.

\subsection{Accelaration performance for steady flow simulation}
\qquad This part presents a comparison of various strategies in terms of accelaration performance. The number of iteration steps for the convergence of steady flows is used to evaluate the acceleration performance of the initial DoFs assignment strategies, and the averaged converged steps of various strategy combinations in both training and validation set are provided in Table~\ref{tab:accelaration-performace}.

\begin{table}[htbp]
\centering \caption{Accelaration performance (averaged converged steps) of various strategies in dataset.}
\begin{tabular}{m{2cm}<{\centering}|m{2cm}<{\centering}|m{2cm}<{\centering}|m{2cm}<{\centering}|m{2cm}<{\centering}|m{2cm}<{\centering}}
\toprule
  & \small{Freestream} & \scriptsize{Orig POD + Orig RBF} & \scriptsize{Orig POD + Mod RBF} & \scriptsize{Zonal POD + Orig RBF} & \scriptsize{Zonal POD + Mod RBF}  \\ \hline
  \footnotesize{Training set}  & 169 & 42 & 42 & 35 & 35 \\ \hline
  \footnotesize{Validation set}  & 152  & 109 & 60 & 96 & 51 \\
\bottomrule
\end{tabular}
\label{tab:accelaration-performace}
\end{table}

It can be summarized from Table~\ref{tab:accelaration-performace} that (a) the initial DoFs assignment strategy assisted by data-driven models can effectively enhance the efficiency of CFD simulation, with the converged steps reduced to around $1/3\sim2/3$ in average; (b) regardless of the specific RBF model employed, utilizing Zonal POD method consistently yields slightly better acceleration performance compared to Orig POD method; (c) Orig RBF model exhibits poor generalization capability on the current flow-field prediction task, manifesting in significantly degraded acceleration effects on the validation set, even though it exhibits excellent acceleration performance on training set; (d) Zonal POD + Mod RBF model ensures consistently good acceleration performance on both the training and validation sets.

\section{Practical applications of the ML-enhanced DGM}
\label{sec:5}
\qquad This section presents practical applications of the ML-enhanced DGM in viscous flow simulation on arbitrary shapes/grids and in SBO for rapid aerodynamic evaluation. It aims to demonstrate the superiority of the ML-enhanced DGM compared to classical DGMs that assign initial DoFs via freestream or solution remapping in terms of improving computational efficiency and simulation robustness.

\subsection{Transonic flow around arbitrary airfoils}
\qquad The ML-enhanced DGM can be utilized to simulate transonic flow around a specific airfoil, even if it falls outside the range of the original dataset. Given the airfoil data on upper and lower surface,
\begin{equation*}
\begin{aligned}
 &\bm{x}^{\text{low}}=\left( x^{\text{low}}_{1}, y^{\text{low}}_{1},...,x^{\text{low}}_{n}, y^{\text{low}}_{n}\right),\\
 &\bm{x}^{\text{up}} =\left(x^{\text{up}}_{1}, y^{\text{up}}_{1},...,x^{\text{up}}_{n}, y^{\text{low}}_{n}\right).
\end{aligned}
\end{equation*}
There is an additional step to solve the corresponding design variable $\bm{D}^*$ with respective to the data $\bm{X}^{*}_{\text{surf}}=\left(\bm{x}^{\text{low}},\bm{x}^{\text{up}}\right)$ through optimization,
\begin{equation}
 \bm{D}^*=\underset{\bm{D}\in\mathcal{D}}{\text{argmin}}~\big(\bm{X}_{\text{surf}}(\bm{D})-\bm{X}^{*}_{\text{surf}}\big)^2.
\label{eq:D-star}
\end{equation}

\begin{figure}[htbp]
  \centering
  \includegraphics[width=8cm]{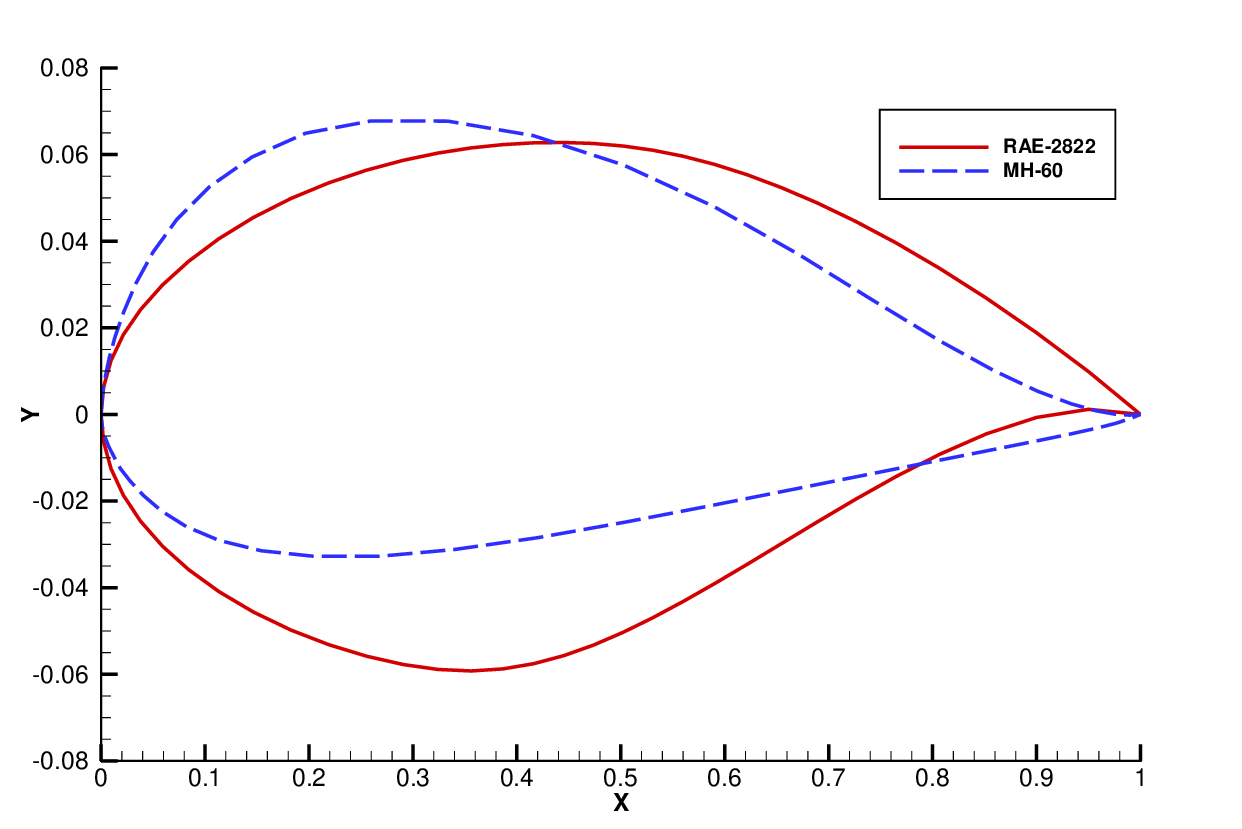}
  \caption{Shape comparison of the RAE-2822 and MH-60 airfoils.}
  \label{fig:mh60shape}
\end{figure}

\begin{figure}[htbp]
 \centering
 \subfigure[inviscid flow via ML]{
 \includegraphics[width=4.8cm]{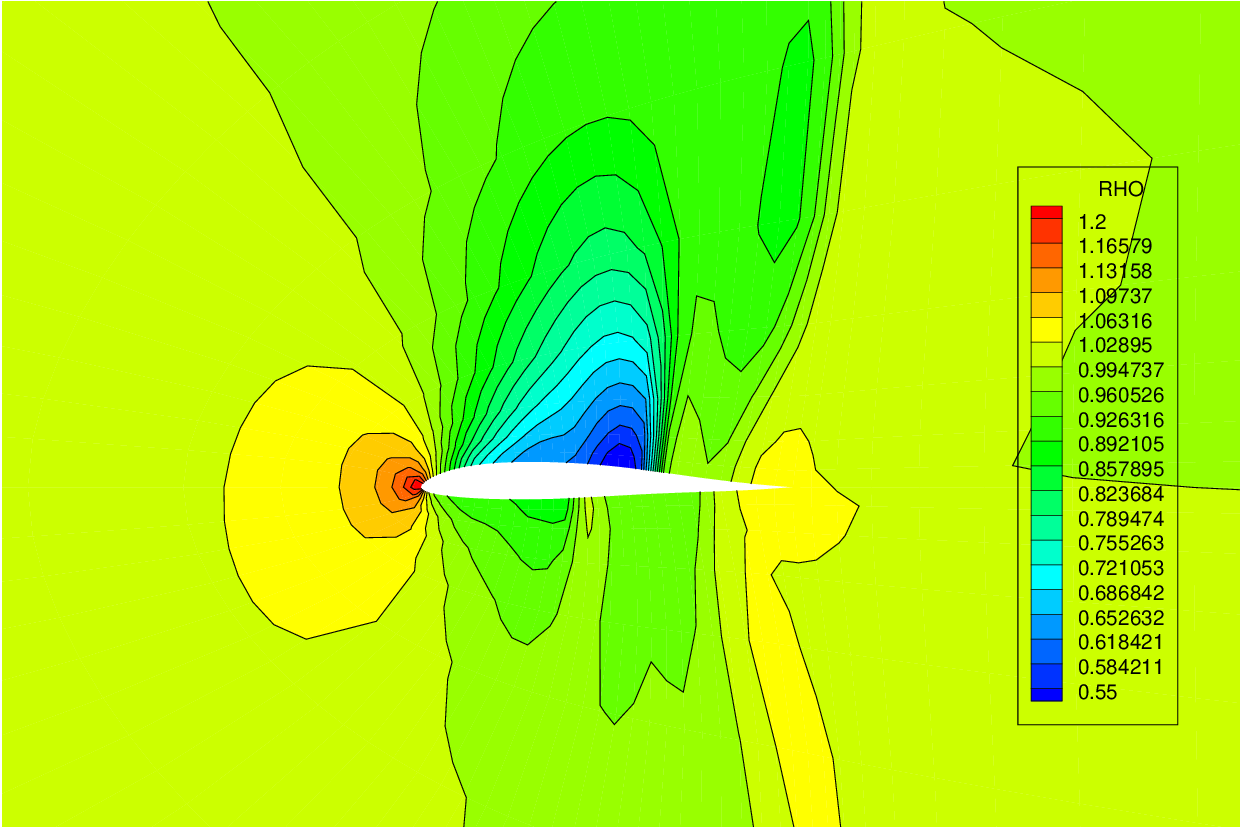}}
 \subfigure[inviscid flow via CFD solver]{
 \includegraphics[width=4.8cm]{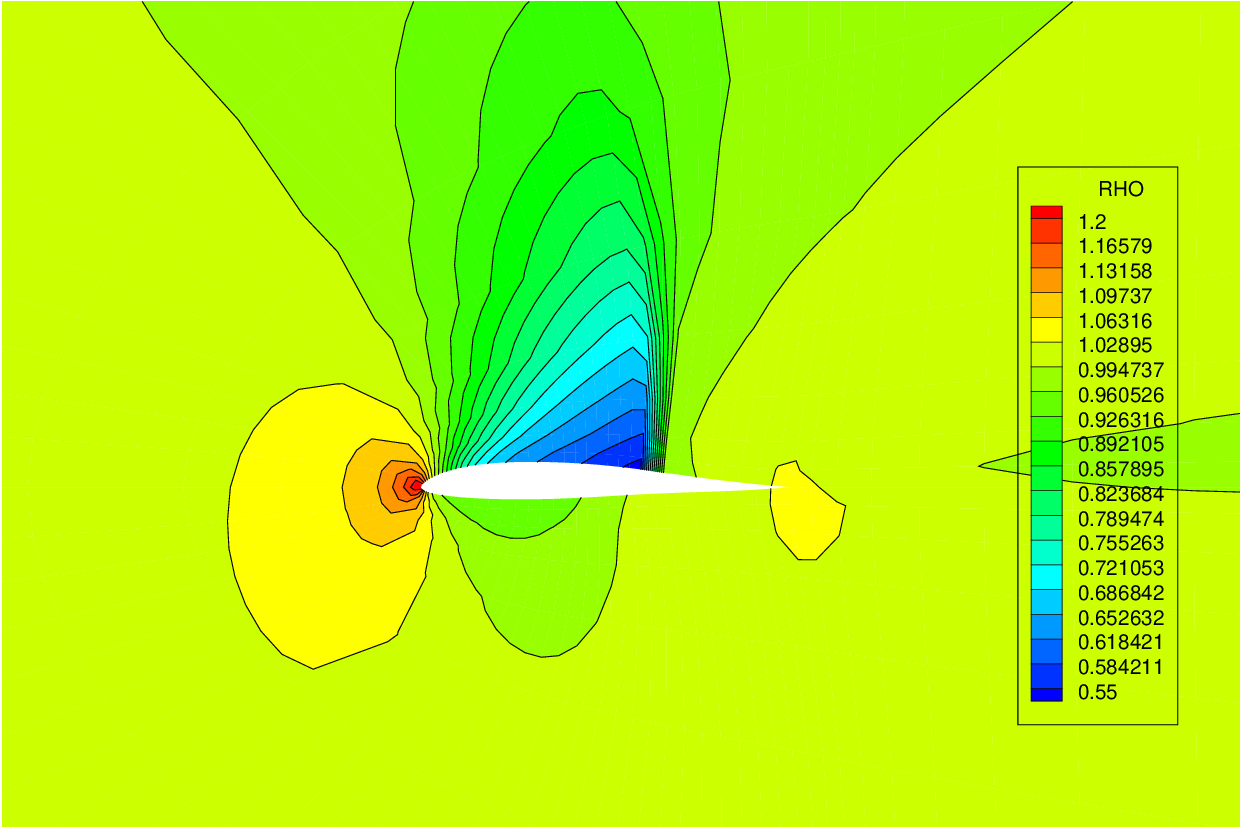}}
 \subfigure[relative error]{
 \includegraphics[width=4.8cm]{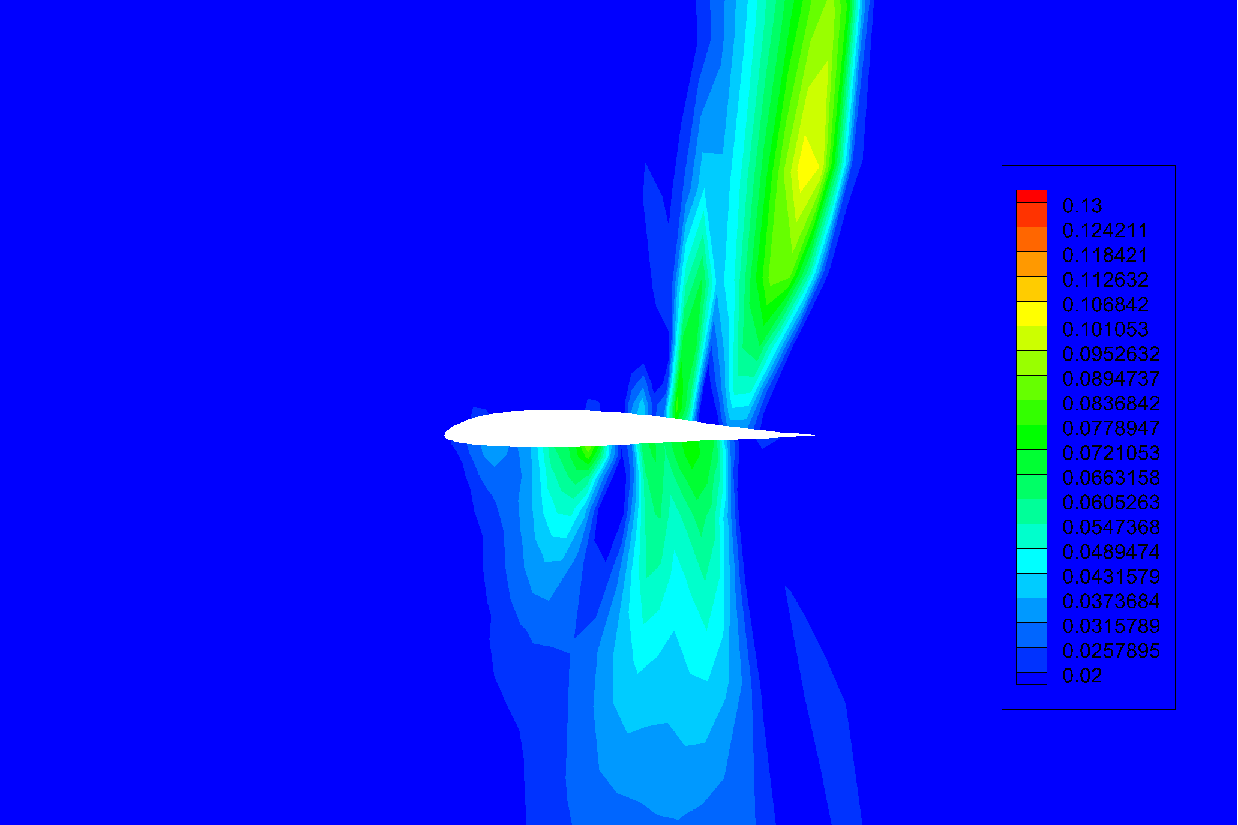}}
 \caption{Flow-field produced by the data-driven model and CFD solver for inviscid transonic flow around the MH-60 airfoil.}
 \label{fig:mh60flow}
\end{figure}

\begin{figure}[htbp]
  \centering
  \includegraphics[width=8cm]{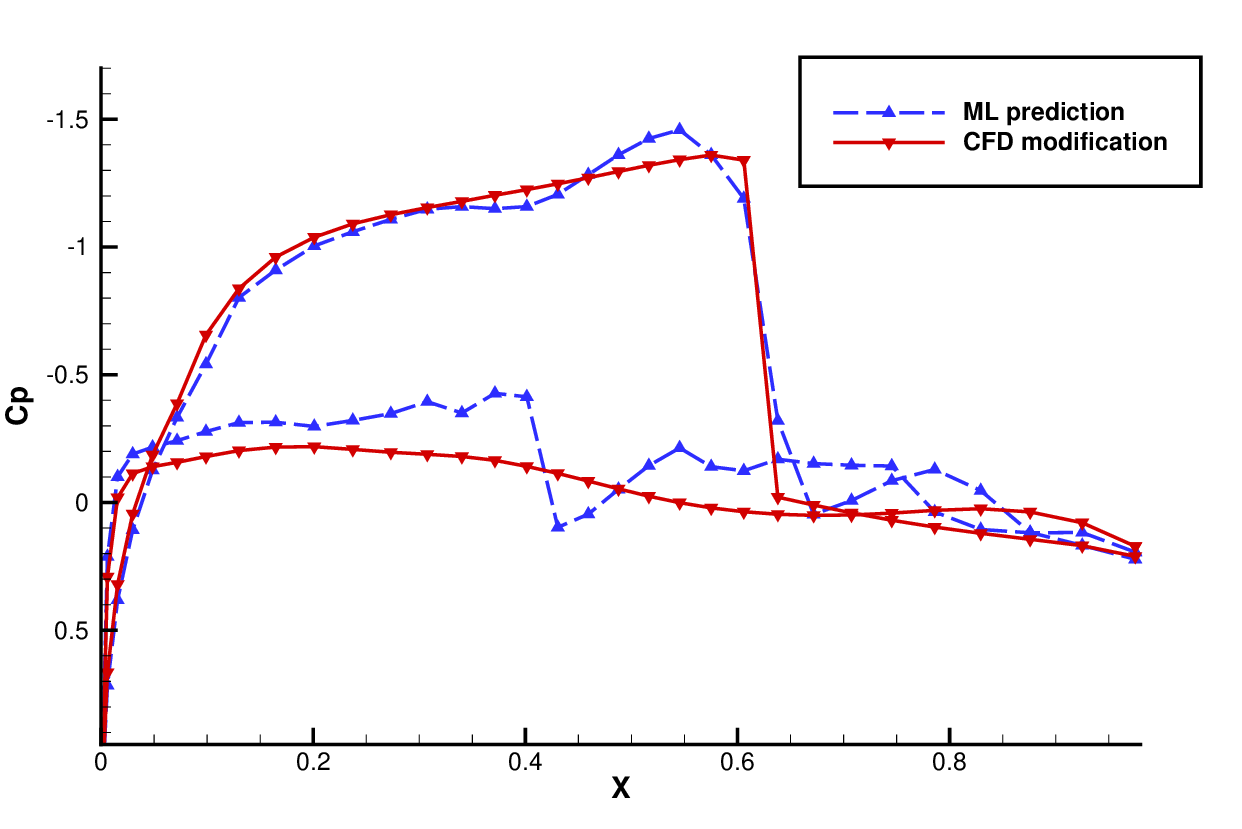}
  \caption{Comparison of $C_p$ curves produced by data-driven model and CFD solver.}
  \label{fig:mh60Cp}
\end{figure}

Once obtain the geometric design variable $\bm{D}^*$, the ML-enhanced DGM in Section \ref{sec:3} can be used directly. This part tests a case of the transonic inviscid flow around the MH-60 airfoil with $\text{Ma=0.78, AoA=}1.25^{\circ}$, and the results are presented in Figure~\ref{fig:mh60shape}-\ref{fig:mh60Cp}.

The results demonstrate that the data-driven model is able to predict relatively accurate airfoil flow-field and $C_p$ curves even outside the range of the dataset. The DGM correction can facilitate the expansion of the applicable scope of data-driven model in the flow-field prediction task. In addition, the ML-enhanced DGM can effectively reduce the converged steps for steady-state flow (66 steps) as compared to the DGM with freestream initialization (requiring 148 steps).

\subsection{Transonic viscous flow on arbitrary grids}
\qquad The ML-enhanced DGM can be utilized to simulate transonic viscous flow on arbitrary computational grids. Given a computational grid with boundary layers for simulation of viscous flow, the initial DoFs on the new grid can be assigned through the projection of the ML predicted flow-field onto the current grid.

\begin{figure}[htbp]
 \centering
 \subfigure[mesh for inviscid flow]{
 \includegraphics[width=6cm]{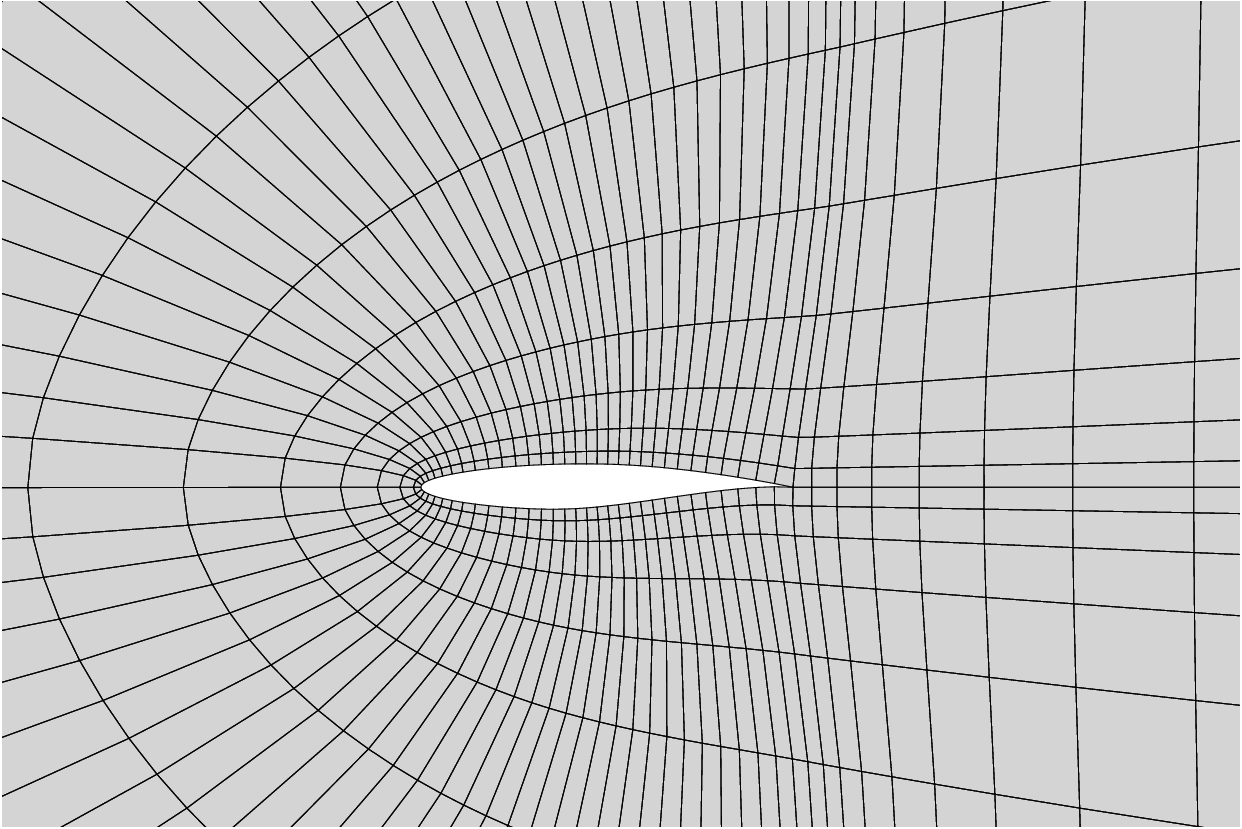}}
 \subfigure[mesh for viscous flow]{
 \includegraphics[width=6cm]{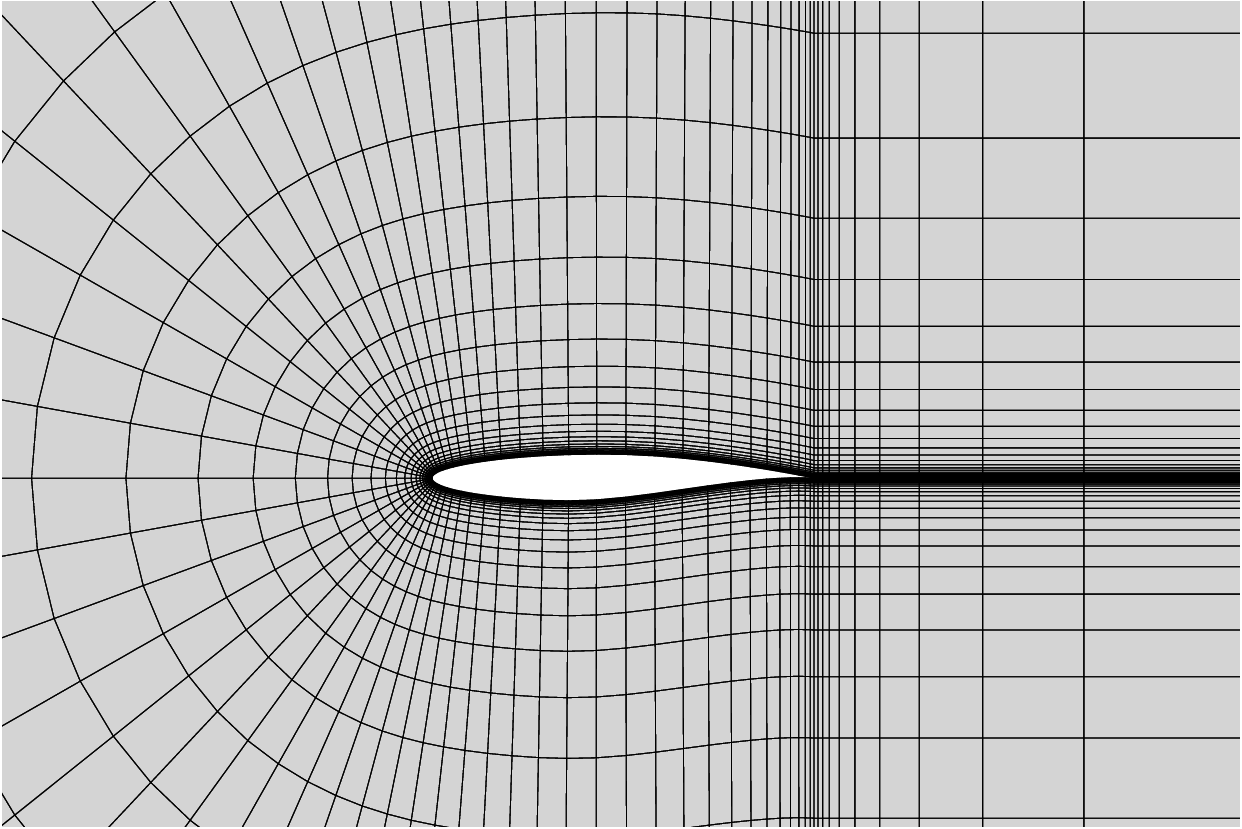}}
 \caption{Computational grids for inviscid and viscous flows.}
 \label{fig:rae2822mesh}
\end{figure}

\begin{figure}[htbp]
 \centering
 \subfigure[inviscid flow via data-driven model]{
 \includegraphics[width=6cm]{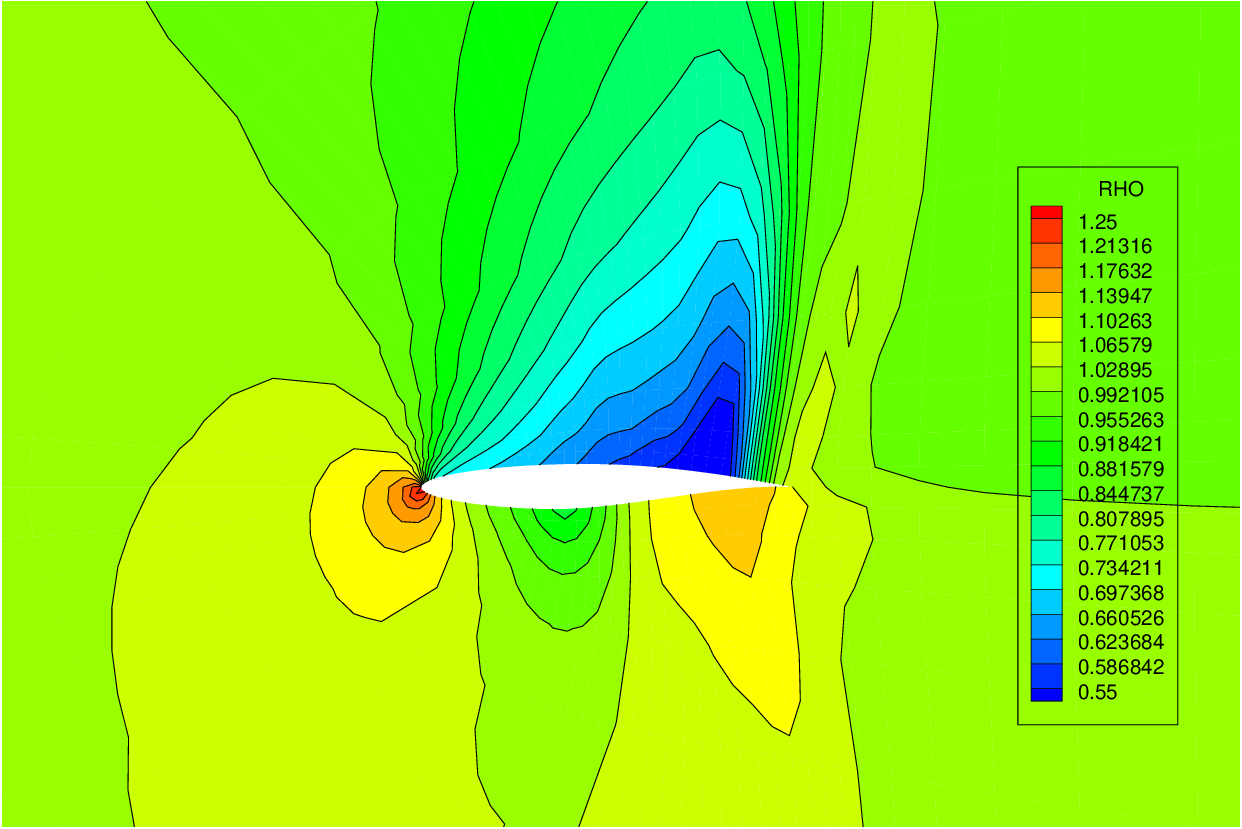}}
 \subfigure[inviscid flow via CFD solver]{
 \includegraphics[width=6cm]{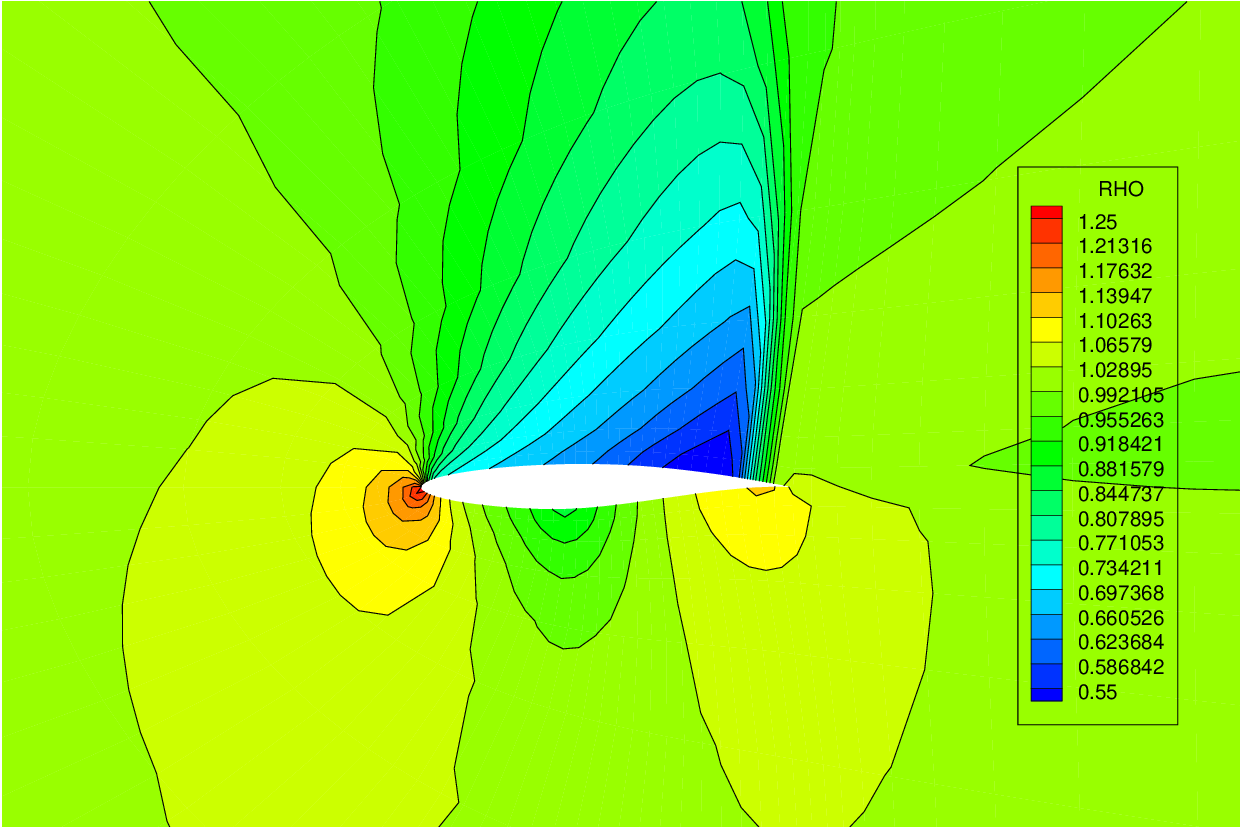}}
 \subfigure[viscous flow (Re=1e7) via CFD solver]{
 \includegraphics[width=6cm]{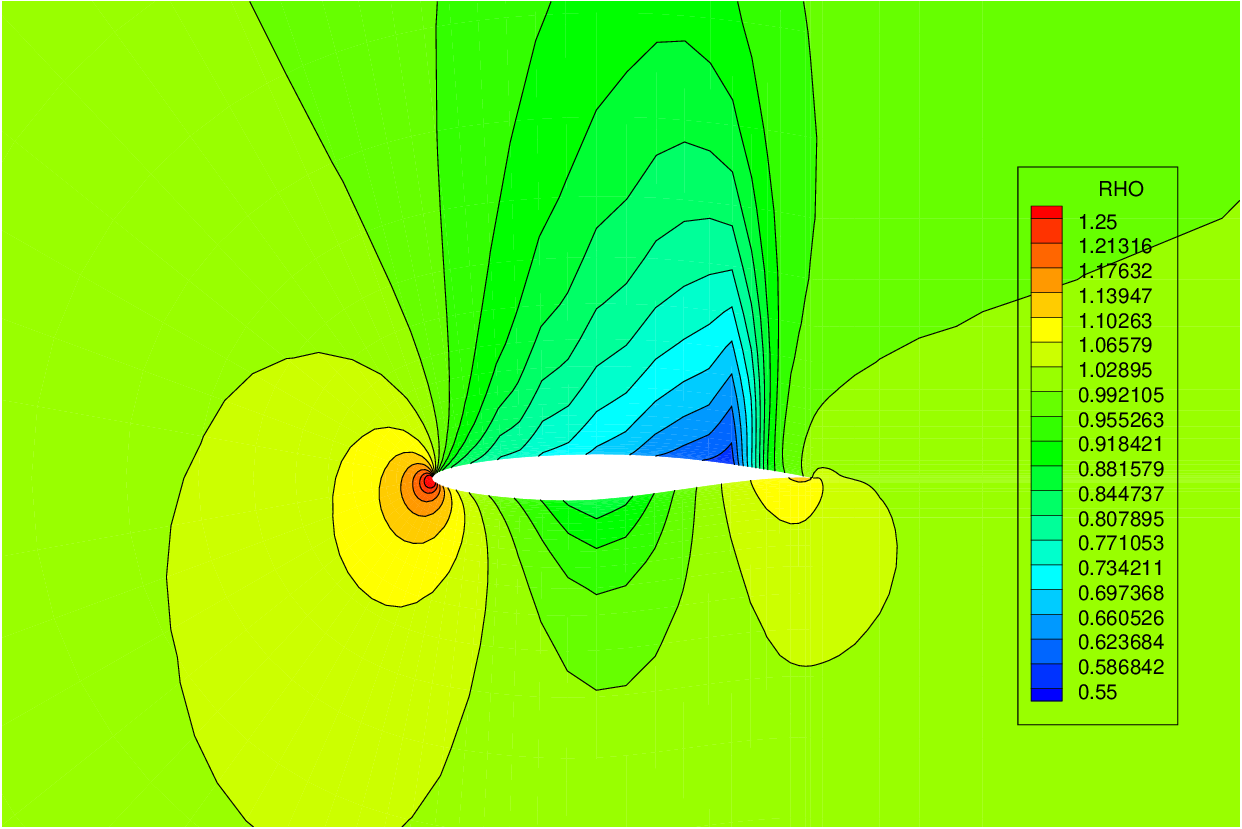}}
 \subfigure[viscous flow (Re=1e6) via CFD solver]{
 \includegraphics[width=6cm]{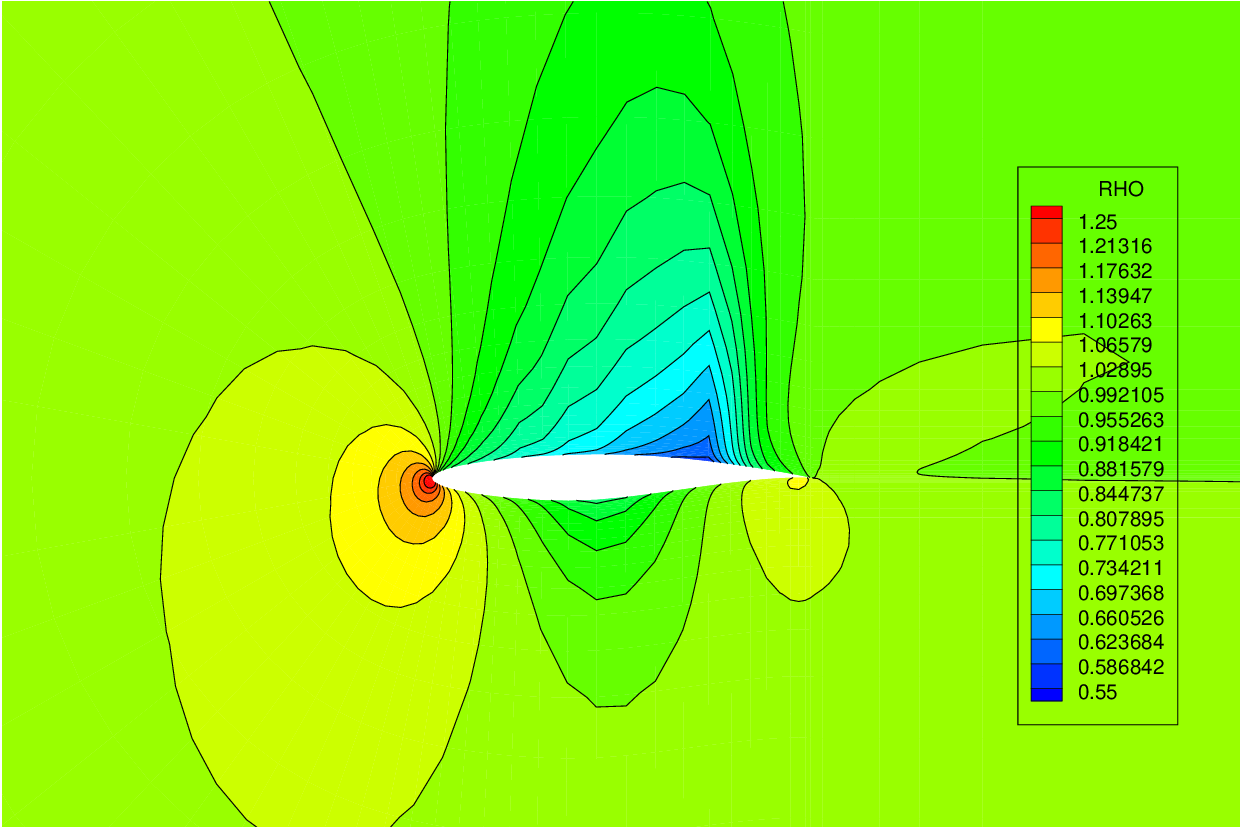}}
 \caption{Flow-field produced by various methods for inviscid and viscous flows.}
 \label{fig:rae2822visflow}
\end{figure}

This part presents a test case of transonic viscous flows around the RAE-2822 airfoil with various Reynolds numbers. The computational grids used for ML prediction and for viscous flow simulation are presented in Figure~\ref{fig:rae2822mesh}, and the numerical results are provided in Figure~\ref{fig:rae2822visflow} and Table~\ref{tab:rae2822visflow}.

\begin{table}[htbp]
\centering \caption{Accelaration performance (averaged converged steps) for simulating inviscid and viscous flows with different Reynolds numbers.}
\begin{tabular}{m{3cm}<{\centering}|m{3cm}<{\centering}|m{3cm}<{\centering}|m{3cm}<{\centering}}
\toprule
  & Inviscid flow & Viscous flow, Re=1e7 & Viscous flow, Re=1e6  \\ \hline
  Freestream      & 134  & 235   & 189 \\ \hline
  ML prediction   & 30   & 54    & 56 \\
\bottomrule
\end{tabular}
\label{tab:rae2822visflow}
\end{table}

Figure~\ref{fig:rae2822visflow} demonstrates that due to the similarity between inviscid flows and viscous flows with high Reynolds numbers, the ML-enhanced DGM can be extended to simulate viscous airfoil flow-field on given grids, and Table~\ref{tab:rae2822visflow} illustrates that the ML-enhanced DGM still exhibits remarkable acceleration effects when extended for simulating viscous flows at high Reynolds numbers.

\subsection{Flow-field simulation in surrogate-based aerodynamic optimization}
\qquad This part shows a practical application of the data-driven model to surrogate-based aerodynamic optimization \cite{feng2024sbo}. During the design of experiment (DoE) stage, the CFD solver is required to evaluated the aerodynamic objectives and constraints at numerous sample points within the design space. In the test case of aerodynamic optimization of RAE-2822 airfoil, the flight conditions (Mach number and angle of attack) are included as the design variable, thereby expanding the design space $\widetilde{\mathcal{D}}$ to
\begin{equation*}
 \widetilde{\mathcal{D}}=\left\{(\bm{D},\bm{S})\Big|~\Delta D_i\leq 0.25,~\text{Ma}\in[0.7,0.95],\text{AoA}\in[-5^o,5^o]\right\}.
\end{equation*}

SLHS is utilized to generate 40 sample points within $\widetilde{\mathcal{D}}$, and the related data (average converged steps and serial CPU time) of different DoFs assignment strategies are provided in Table~\ref{tab:opt-efficiency}. Additionally, Table~\ref{tab:opt-efficiency} also includes the solution remapping technique~\cite{wang2021local} which is commonly used for accelerating aerodynamic shape optimization within the HODG framework.

\begin{table}[htbp]
\centering \caption{Accelaration performance (averaged converged steps and serial CPU time) of different strategies during the sample generation.}
\begin{tabular}{m{4cm}<{\centering}|m{4cm}<{\centering}|m{4cm}<{\centering}}
\toprule
  & Converged steps & CPU time (sec)  \\ \hline
  Freestream  & 184 & 129.3 \\ \hline
  Solution remapping  & 294  & 251.5 \\ \hline
  ML prediction  & 59  & 44.1 \\
\bottomrule
\end{tabular}
\label{tab:opt-efficiency}
\end{table}

It needs to be mentioned that in numerical experiments with fixed Mach number and angle of attack, the solution remapping technique can effectively accelerates the convergence of steady flows. However, when flow conditions are taken into consideration and there are significant changes in geometry and flow conditions between consecutive sample points, the initial DoFs assigned by the solution remapping technique deviates greatly from that of the final steady solution. That might lead to CFD simulation broken down, or make simulation hard to achieve convergence.

In terms of simulation robustness, the DGMs with initial DoFs assigned by freestream and the ML prediction can successfully simulate the flow-field for all 40 samples in the current optimization test case, while solution remapping technique results in 2 out of the 40 samples to break down (due to negativity of pressure) and 1 sample failing to converge. In terms of computational efficiency, it can be observed from Table~\ref{tab:opt-efficiency} that the solution remapping strategy exhibits a deceleration compared to the freestream strategy. Nevertheless, the DoFs assigned via ML prediction demonstrates a stable accelaration in the convergence of steady flow-field.

\section{Conclusion and perspectives}
\label{sec:6}
\qquad In this work, based on our recently open-sourced HODG platform, we developed a ML-enhanced DGM for rapidly simulating transonic airfoil flow-field. The combination of ML prediction with CFD modification can effectively enhance the generalization capability of data-driven models and improve the computational efficiency of CFD methods. The data-driven model, utilizing the zonal POD method and weighted-distance RBF interpolation, possesses excellent properties of lightweightness and updatability, and is applicable to a wide range of airfoils at Mach numbers ranging from $0.7$ to $0.95$ and angles of attack from $-5^{\circ}$ to $5^{\circ}$. Numerical experiments have validated that the proposed data-driven model maintains reliable predictive performance on both the training and validation sets, and is capable of rapidly providing predictions that closely resemble actual flow-field distribution. Furthermore, the ML-enhanced DGM can be extended to viscous flow simulations and SBO, and significantly improving the efficiency and robustness of normal DGMs. Future work will focus on the 3D practical ML-enhanced DGM and its applications in aerodynamic optimization designs.

\section*{Acknowledgments}
This work is supported by National Nature Science Foundation of China (No.12302380).

\bibliographystyle{reference}
\bibliography{references}

\end{document}